\documentstyle[floats,prd,aps,epsf]{revtex}

\draft
\begin{document}
%@twocolumnfalse\endcsname

\newcommand{\beq}{\begin{equation}}
\newcommand{\eeq}{\end{equation}}
\newcommand{\beqn}{\begin{eqnarray}}
\newcommand{\eeqn}{\end{eqnarray}}
\newcommand{\pa}{\partial}
\newcommand{\vp}{\varphi}
\newcommand{\varep}{\varepsilon}
\def\zero{\hbox{$_{(0)}$}}
\def\bL{\hbox{$\,{\cal L}\!\!\!\!\!-$}}

\begin{center}
{\large\bf{Axisymmetric general relativistic hydrodynamics: 
Long-term evolution of neutron stars and 
stellar collapse to neutron stars and black holes}}
~\\
~\\
Masaru Shibata\\
{\em Graduate School of Arts and 
Sciences, University of Tokyo, Tokyo, 153-8902, Japan}\\
\end{center}

\begin{abstract}
We report a new implementation for axisymmetric simulation 
in full general relativity. 
In this implementation, the Einstein equations 
are solved using the Nakamura-Shibata 
formulation with the so-called cartoon method to impose 
an axisymmetric boundary condition, and 
the general relativistic hydrodynamic equations are
solved using a high-resolution 
shock-capturing scheme based on an approximate Riemann solver. 
As tests, we performed the following simulations: 
(i) long-term evolution of non-rotating and rapidly rotating neutron stars, 
(ii) long-term evolution of neutron stars of 
a high-amplitude damping oscillation accompanied with shock formation,
(iii) collapse of unstable neutron stars to black holes, 
and (iv) stellar collapses to neutron stars. 
The tests (i)--(iii) were 
carried out with the $\Gamma$-law equation of state, and
the test (iv) with a more realistic parametric 
equation of state for high-density matter. 
We found that this new implementation works very well: 
It is possible to perform the simulations for 
stable neutron stars for more than 10 dynamical time scales, 
to capture strong shocks formed at stellar core collapses, 
and to accurately compute the mass of black holes 
formed after the collapse and subsequent accretion. 
In conclusion, this implementation
is robust enough to apply to astrophysical problems such as 
stellar core collapse of massive stars to a neutron star and 
black hole, phase transition of a neutron star to 
a high-density star, and accretion-induced collapse of a neutron star
to a black hole. The result for the first simulation
of stellar core collapse 
to a neutron star started from a realistic initial condition 
is also presented. 
\end{abstract}
%\vskip 2mm
%\vskip2pc]

\section{Introduction}

In the 1980s, one of the most important issues in the 
field of numerical relativity 
involved performing simulations of rotating stellar collapse 
with the assumption of axial symmetry.
Simulations of rotating stellar collapse in full general relativity 
were first performed by Nakamura and collaborators \cite{Nakamura,NOK}.  
Using the (2+1)+1 formalism developed by Maeda {\it et al}. \cite{Maeda}, 
they succeeded in performing simulations of a 
rotating collapse of massive stars to black holes. 
They used cylindrical coordinates $(\varpi, z)$ with 
a nonuniform grid spacing and with 
at most (42,42) grid resolution for $(\varpi, z)$ 
because of restricted computational resources. 

To compute gravitational waves emitted during gravitational collapse to 
black holes, Stark and Piran \cite{SP} subsequently 
performed simulations similar to those of Nakamura {\it et al}., 
adopting spherical polar coordinates with 
a typical grid size $(100,16)$ for $(r, \theta)$. The distinguishing 
feature of their work is that they adopted the Bardeen-Piran 
formalism \cite{BP}, 
which is well suited for computation of gravitational waves 
in the wave zone. As a result of this 
choice of formalism, they succeeded in computing 
gravitational wave forms, and clarified that the wave forms  
are characterized by the quasinormal mode of rotating black holes 
formed after gravitational collapse and that the total radiated 
energy of gravitational waves is at most 0.1 \% of the 
gravitational mass of the system \cite{SP}. 

Since the completion of their works, 
no new work in this field was done for the next 15 years \cite{acst}.  
Although several questions that they originally wished to answer 
have been answered by their simulations,
it was not feasible to perform sophisticated
astrophysical simulations in the 1980s.
This is likely due to the fact that 
the computational resources were severely restricted, 
and, in addition, that techniques in numerical relativity 
such as methods to provide realistic
initial conditions and to perform the 
long-term simulations were not sufficiently developed. 
As a result, there still remain many unsolved issues in astrophysics and 
general relativity that can be studied with 
axisymmetric hydrodynamic simulations in full general relativity. 
Among them, realistic simulations of rotating core collapse of massive stars, 
which thereby become black holes or protoneutron stars 
in full general relativity, have not yet been performed. 
Stellar collapse is a common phenomenon in the universe, and 
hence, understanding the formation mechanism of 
black holes and neutron stars in nature is one of 
the most important issues in astrophysics. 
Actually, the study of the formation of rapidly rotating black holes with 
surrounding accretion disks in stellar core collapse 
is currently one of the hot topics in connection with 
a hypothetical scenario for the central 
engine of $\gamma$-ray bursts \cite{hyper}. 
To date, simulations of a rotating collapse 
of a massive stellar core have been done 
in the Newtonian gravity \cite{Newton,Newton2,Newton3,YS,Muller,fryer}
or in an approximate general relativistic 
gravity \cite{HD} using the so-called conformal flatness 
approximation (or the Isenberg-Wilson-Mathews approximation). 
In rotating stellar core collapses, 
general relativity plays an important role. 
As demonstrated in \cite{HD}, general relativistic effects modify 
the collapse, bounce, and amplitude of 
gravitational waves emitted significantly, even in the formation of
neutron stars. Of course, 
general relativity plays a crucial role in the formation of black holes. 
Thus, general relativistic simulation is inevitable to 
precisely understand the nature of stellar core collapses. 

One long-standing issue for axisymmetric 
simulations in full general relativity has been to develop 
methods in which the accuracy and stability 
for a long-term simulation can be preserved. 
In axisymmetric simulations, we have in general used 
cylindrical and/or spherical polar coordinate 
systems, which have coordinate singularities at the origin and 
along the symmetric axis $\varpi=0$. 
At such coordinate singularities, the finite differencing scheme 
has to be changed, resulting often in numerical instabilities. 
To stabilize computation, 
we have often been required to add artificial viscosities around the 
coordinate singularities to stabilize the numerical system \cite{David}. 

Recently, the Potsdam numerical relativity group has proposed 
the so-called cartoon method by which 
a robust numerical relativity implementation 
for axisymmetric systems can be made \cite{alcu}.  
The essence of their idea is that the Cartesian coordinates
$(x, y, z)$ 
could be used even for simulations of axisymmetric systems if 
the Einstein field equations are solved only for the 
$y=0$ (or $x=0$) plane, using the boundary condition 
at $y=\pm \Delta y$ (or $x=\pm \Delta x$) provided by the axial symmetry.
(Here, $\Delta x$, $\Delta y$, and $\Delta z$ denote the
grid spacing.) 
Since the field equations are written in the Cartesian 
coordinate system, we neither have singular terms nor do we have to change 
the finite differencing scheme anywhere, except at the outer boundaries. 
Thus, it is possible to perform a stable and accurate 
long-term simulation without any prescription or 
artificial viscosities, but only by a minor modification  
of a three-dimensional implementation that has already been developed
\cite{gw3p2,gr3d,bina,bina2}. 

Other important progress has been made regarding computational resources.
Current large-scale supercomputers that we can use are 
typically of several hundred Gbytes memory. 
Necessary memory in an axisymmetric simulation 
with double precision, with $N^2$ 
grid points, and with $N_v$ variables is 
\beq
\sim 2~{\rm Gbytes}
\biggl( {N \over 10^3}\biggr)^2\biggl({N_v \over 250}\biggr),
\eeq
where $N_v$ is $\sim 200$ in our general relativistic implementation. 
This implies that the memory of current supercomputers 
is large enough to carry out
an axisymmetric numerical simulation with $N \sim$ several thousands. 
Using $N$ of order $10^3$, it is feasible to carry out a well-resolved 
simulation and a careful convergence test, changing the grid resolution 
for a wide range from $N \sim 100$ to $1000$.
This situation is in contrast with that of 
three-dimensional numerical relativity, 
since it is still very difficult to carry out 
a three-dimensional simulation with $N \sim 10^3$, for which
the required computational memory is of order TByte.  

Motivated by the status mentioned above, we recently started 
a project in axisymmetric numerical relativity. 
In \cite{gr2d}, we reported a numerical hydrodynamic
implementation for axisymmetric spacetimes 
that is made using the cartoon method and incorporating a
hydrodynamic implementation in the cylindrical coordinates. 
In that paper, we presented numerical results for simulations of 
rotating stellar collapse adopting simple initial conditions and 
simple equations of state to investigate the effects of rotation 
on the criteria for prompt collapse to black holes
in an idealized setting. We demonstrated 
that the axisymmetric implementation and current computational 
resources are well suited to systematically 
perform stable and well-resolved hydrodynamic simulations 
in axisymmetric numerical relativity. 
That implementation has also been applied to a study of collapse of 
rotating supermassive stars to supermassive black holes \cite{SS}. 

We have recently remade a hydrodynamic implementation 
using a high-resolution shock-capturing scheme based on
a Godunov-type scheme \cite{Font,Val,Val2,Iba,fontrev,other}. 
Although the previous one \cite{gr2d} works 
well for problems in which shocks are weak, such as evolution of 
single rotating stars and collapse of neutron stars and
supermassive stars to a black hole, 
it is expected that such implementation
cannot produce an accurate numerical result for 
problems in which shocks are strong. 
During rotating core collapses to a neutron star 
or a black hole, strong shocks are likely to be accompanied. Therefore, 
implementing a high-resolution shock-capturing scheme, 
such as that adopted in \cite{other}, 
is a promising strategy. To check that 
the new implementation works well,
we have performed a wide variety of test simulations. 
In this paper, we present the numerical results, 
paying particular attention to long-term numerical simulations 
of neutron stars as done in, e.g., \cite{gr3d,other}, 
and to stellar collapse in which strong shocks are accompanied. 
Finally, we present the first numerical
results of stellar core collapse of a neutron star
for which the simulation is started from a realistic initial condition. 
In addition to presenting the successful numerical results, 
we address the advantage of axisymmetric simulations 
in testing a new general relativistic hydrodynamic implementation, 
since we can study in detail the convergence of numerical results 
in the test simulations changing 
the grid number for a wide range to a 
well-resolved level (e.g., $N \sim$ several hundreds), 
which is still difficult in three-dimensional 
simulations because of restricted computational resources. 

The paper is organized as follows. 
In Sec. II, we describe the formulation that we adopt. 
In Sec. III, we define global quantities of the system and 
describe the calibration method for the numerical results. 
In Sec. IV, we present the numerical results. 
Section V is devoted to a summary and discussion. 
Throughout this paper, we use the geometrical
units in which $G=1=c$, where $G$ and $c$ 
denote the gravitational constant and speed of light.  
We use Cartesian coordinates, $x^k=(x, y, z)$, 
as the spatial coordinates, with $r=\sqrt{x^2+y^2+z^2}$, 
$\varpi=\sqrt{x^2+y^2}$, and $\varphi=\tan^{-1}(y/x)$. $t$ denotes
the coordinate time. Greek indices $\mu, \nu, \cdots$ 
denote $x, y, z$, and $t$, 
small Latin indices $i, j, \cdots$ denote $x, y$, and $z$, and 
capital Latin indices $A, B, \cdots$ denote $x$ and $z$. 

\section{Formulation}

\subsection{Formulation for the Einstein equation} 

The Einstein equation is solved in the (3+1) formulation, in which 
the line element is written in the form 
\beqn
ds^2 = (-\alpha^2 + \beta_k\beta^k)dt^2 + 2 \beta_k dx^k dt
+ \gamma_{ij}dx^i dx^j,
\eeqn
where $\alpha$, $\beta^k$, and $\gamma_{ij}$ are the lapse 
function, shift vector, and three-metric. The three-metric 
is defined by 
\beq
\gamma_{\mu\nu}=g_{\mu\nu}+n_{\mu}n_{\nu}, 
\eeq
where $n^{\mu}$ is a timelike, unit-normal vector that is
orthogonal to a spacelike hypersurface, 
and its components are written as $(1/\alpha, -\beta^k/\alpha)$. 
The extrinsic curvature is defined as
\beqn
K_{ij}=-{1 \over 2}\bL_n \gamma_{ij}=
-{1 \over 2\alpha}\biggl(\pa_t \gamma_{ij}
-D_i\beta_j - D_j \beta_i \biggr),
\eeqn
where $\bL_n$ is the Lie derivative with respect to $n^{\mu}$, and
$D_{i}$ is the covariant derivative with respect to $\gamma_{jk}$. 

The Einstein field equations are solved using 
the same formulation, gauge conditions, and outer boundary conditions
as in previous papers \cite{gw3p2,gr3d,bina,bina2,SBS,gr3drot,gr2d}: 
We adopt the so-called Nakamura-Shibata formulation 
\cite{NOK,SN} with some modification
from the original version (see \cite{bina2}, 
to which the reader may refer for basic equations and gauge conditions
in the latest version). In this formalism, we evolve
the following geometric variables using a free evolution code: 
\beqn
\phi~ && \equiv {1 \over 12}\ln[{\rm det}(\gamma_{ij})], \\
\tilde \gamma_{ij}~ && \equiv e^{-4\phi}\gamma_{ij}, \\
K &&\equiv K_{ij}\gamma^{ij},\\
\tilde A_{ij} && \equiv e^{-4\phi}(K_{ij}-\gamma_{ij}K/3), \\
F_i &&\equiv \delta^{jk} \pa_k \tilde \gamma_{ij}. 
\eeqn
The Hamiltonian and momentum 
constraint equations are solved at $t=0$, and used 
to check the accuracy of numerical solutions during computation. 

The slicing and spatial gauge conditions
for determining $\alpha$ and $\beta^k$ are 
basically the same as those adopted in our previous series of 
papers \cite{gw3p2,gr3d,bina,bina2,SBS,gr3drot,gr2d}, {i.e.}, we 
impose an ``approximate'' maximal slicing condition ($K \approx 0$)
and an ``approximate'' minimum distortion (AMD) gauge condition 
[$\tilde D_i (\pa_t \tilde \gamma^{ij}) \approx 0$, 
where $\tilde D_i$ is the 
covariant derivative with respect to $\tilde \gamma_{ij}$]. 
However, in contrast with previous papers \cite{gw3p2,SBS,bina,bina2}, 
we do not modify the spatial gauge condition 
even in the formation of black holes, 
since in the axisymmetric simulation, a sufficient number of grid points  
can be taken to resolve black hole formation and subsequent evolution
even using the AMD gauge condition without modification. 

We impose the axially symmetric condition to the geometric
variables using the so-called cartoon method
proposed by Alcubierre {\it et al}. \cite{alcu}. 
First, we define the computational domain as $0 \leq x,z \leq L$ and 
$-\Delta y \leq y \leq \Delta y$, 
where $L$ denotes the location of the outer boundaries, and 
reflection symmetry with respect to the $z=0$ plane is assumed. 
With this computational domain, we need only three points 
in the $y$ direction, $0$ and $\pm \Delta y$. 
We determine here that the Einstein equation 
is solved only in the $y=0$ plane.  
Then, the boundary conditions at $y=\pm \Delta y$ that 
are necessary in evaluating $y$ derivatives 
are supplied from the assumption of axial symmetry as 
\beqn
Q_{AB} && = \Lambda_{A}^{~C} \Lambda_B^{~D} Q_{CD}^{(0)}, \nonumber \\
Q_{Az} && = \Lambda_{A}^{~C} Q_{Cz}^{(0)}, \hskip 1cm 
Q_{A}  = \Lambda_{A}^{~C} Q_{C}^{(0)}, \nonumber \\
Q_{zz} && =Q_{zz}^{(0)},\hskip 1cm
Q_{z} =Q_{z}^{(0)},\hskip 1cm   Q  = Q^{(0)},
\eeqn
where
\beqn
\Lambda_A^{~B}=
\left(
\begin{array}{ll}
\cos\varphi(x) & -\sin\varphi(x) \\
\sin\varphi(x) & ~~\cos\varphi(x) \\
\end{array}
\right),
\eeqn
and $\varphi(x)=\tan^{-1}[\pm\Delta y/\sqrt{x^2+(\Delta y)^2}]$. 
$Q_{ij}$, $Q_i$, and $Q$ denote 
($\tilde \gamma_{ij}, \tilde A_{ij}$), ($F_i, \beta^i$), and 
($\phi, K, \alpha$), respectively, 
and $Q_{ij}^{(0)}, Q_i^{(0)}$ and $Q^{(0)}$ 
are the values of $Q_{ij}, Q_i$, and $Q$ at 
$(\sqrt{x^2+(\Delta y)^2},0,z)$, which are interpolated using 
Lagrange's formula \cite{recipe} 
with three nearby grid points along the $x$ direction 
({i.e.}, $x \pm \Delta x$ and $x$). At $x=L$, we use only two points, 
$x-\Delta x$ and $x$, for the extrapolation. 

To impose the gauge conditions, as well as to solve the constraint 
equations in preparing the initial conditions, we 
solve scalar and vector elliptic-type equations of 
the form \cite{gr3d}
\beqn
&& \Delta_{\rm flat} Q = S, \label{Poissons}\\
&& \Delta_{\rm flat} Q_i=S_i,\label{Poissonv}
\eeqn
where $\Delta_{\rm flat}$
denotes the Laplacian in the flat three-dimensional space, and 
$S$ and $S_i$ denote the source terms. Using the interpolation 
mentioned above, $\pa_{yy} Q$ and $\pa_{yy} Q_i$ 
are evaluated in the finite differencing as 
\beqn
&&\pa_{yy} Q = 2{ Q^{(0)}-Q(x,0,z) \over (\Delta y)^2}, \hskip 1.8cm
\pa_{yy} Q_z = 2{ Q_z^{(0)}-Q_z(x,0,z) \over (\Delta y)^2},\nonumber \\
&&\pa_{yy} Q_x = 2{ Q_x^{(0)}|\cos\varphi(x)|-Q_x(x,0,z) \over (\Delta y)^2},~~
\pa_{yy} Q_y = 2{ Q_y^{(0)}|\cos\varphi(x)|-Q_y(x,0,z) \over (\Delta y)^2}.
\nonumber
\eeqn
On the other hand, the finite differencing in the $x$ and $z$ 
directions, $\pa_{xx} Q_i$ and $\pa_{zz} Q_i$, is 
written in the standard form as 
\beqn
&&{Q_i(x+\Delta x, 0, z)-2 Q_i(x,0,z)+Q_i(x-\Delta x,0,z) 
\over (\Delta x)^2}, \nonumber \\
&&~ \nonumber \\
&&{Q_i(x, 0, z+\Delta z)-2 Q_i(x,0,z)+Q_i(x,0,z-\Delta z) 
\over (\Delta z)^2}.\nonumber 
\eeqn
Thus, in the finite differencing form for each component of 
Eq.~(\ref{Poissonv}), 
only one component of $Q_i$ is included, 
implying that each component of the vector elliptic-type 
equation is solved independently, 
as in the case of the scalar elliptic equation. 

Finally, we note a necessary modification 
for numerically handling Einstein's 
evolution equations in the axisymmetric case. 
In our formalism, the evolution equations are written in
the form \cite{SN,gw3p2,bina2}
\beq
\pa_t Q + \beta^k \pa_k Q = {\rm right-hand~side}, 
\eeq
where $Q$ denotes one of the geometric variables $\tilde \gamma_{ij}$,
$\tilde A_{ij}$, $\phi$, $K$, and $F_k$.
In the three-dimensional case, we apply an upwind scheme to numerically
handle
the transport term $\beta^k \pa_k Q$ for all the components \cite{gw3p2}.
In the axisymmetric case, the same method is used for 
$\beta^x \pa_x Q$ and $\beta^z \pa_z Q$, but is not for 
$\beta^y \pa_y Q$, since it is not appropriate. 
(Remember that in the hydrodynamic equations in the axisymmetric case, 
there is no transport term for the rotational direction.) 
For $\beta^y \pa_y Q$, we use the following schemes:
For $Q=\phi$, $K$, $\tilde \gamma_{zz}$, and $\tilde A_{zz}$,
we set $\beta^y \pa_y Q=0$ because of symmetry.
For other variables, we simply use the
cell-centered (second-order) finite difference. 

%%%%%%%%%%%%%%%%%%%%%%%%%%%%%%%%%

\subsection{Formulation for the
hydrodynamic equations in general relativity}

The hydrodynamic equations in general relativity are written as 
\beqn
&&\nabla_{\mu} (\rho u^{\mu})=0,\label{eq1} \\
&&\nabla_{\mu} T^{\mu}_{~\nu}=0, \label{eq2}
\eeqn
where $\nabla_{\mu}$ is the covariant derivative with respect to
the spacetime metric $g_{\mu\nu}$, $\rho$ is the baryon rest-mass density,
$u^{\mu}$ is the four-velocity, and
\beqn
T^{\mu\nu}=\rho h u^{\mu} u^{\nu} + P g^{\mu\nu}. 
\eeqn
Here, $P$ is the pressure, $h \equiv 1+ \varepsilon + P/\rho$
is the enthalpy, 
and $\varepsilon$ is the specific internal energy. 
Equations (\ref{eq1}) and (\ref{eq2}) are the continuity equation
and the equations of motion, respectively. 

We adopt the so-called high-resolution shock-capturing
scheme in numerically handling the transport terms of
hydrodynamic equations. To use such a scheme, 
the hydrodynamic equations should be of a
conservative form as 
\beqn
&& \pa_t (\rho_*\sqrt{\eta}) + \pa_i (\rho_*\sqrt{\eta} v^i)=0,\label{cont}\\
&& \pa_t (\rho_* \sqrt{\eta} \hat u_j)
+ \pa_i (\rho_* \sqrt{\eta} v^i \hat u_j
+ P\alpha e^{6\phi}\sqrt{\eta}\delta^i_{~j}) \nonumber \\
&&~~~~~~~=P\pa_j(\alpha e^{6\phi}\sqrt{\eta})-\rho_* \sqrt{\eta}
[w h\pa_j \alpha - \hat u_i
\pa_j \beta^{i}+{1 \over 2 u^t h}\hat u_k \hat u_l \pa_j \gamma^{kl}]
\label{euler}\\
&& \pa_t (\rho_* \hat e\sqrt{\eta})+ \pa_i [\rho_* \sqrt{\eta}
\hat e v^i + P e^{6\phi} \sqrt{\eta} (v^i+\beta^i)] \nonumber \\
&&~~~~~~~=\alpha e^{6\phi} \sqrt{\eta}P K
+{\rho_* \sqrt{\eta}\over u^t h}\hat u_i \hat u_j K^{ij}
-\rho_* \sqrt{\eta}\hat u_i \gamma^{ij} D_j \alpha, \label{energy}
\eeqn
where 
\beqn
&& \rho_* \equiv \rho w  e^{6\phi},\\
&& v^i \equiv {u^i \over u^t}
=-\beta^i+\alpha \gamma^{ij}{\hat u_j \over h w},\\
&& \hat u_i \equiv h u_i,\\
&& \hat e \equiv {e^{6\phi} \over \rho_*}
T_{\mu\nu} n^{\mu}n{^\nu} =  h w-{P \over \rho w},\\
&& w \equiv \alpha u^t,
\eeqn
and $\eta$ is a determinant in curvilinear coordinates; 
in the cylindrical coordinates, $\eta=\varpi$. 
We note that subscripts $i, j, \cdots$ here
denote the components in curvilinear spatial coordinates. 
Equations (\ref{cont}), (\ref{euler}), and (\ref{energy}) are 
the continuity, Euler, and energy equations.
The Euler and energy equations are derived from
$\gamma^{\nu}_{~j} \nabla_{\mu} T^{\mu}_{~\nu}=0$ and 
$n^{\nu} \nabla_{\mu} T^{\mu}_{~\nu}=0$, respectively. 

We solve the hydrodynamic equations in the assumption of
axial symmetry. Thus, we first write equations in the
cylindrical coordinates $(\varpi, \varphi, z)$. 
However, the Einstein equations are 
solved in the $y=0$ plane with the Cartesian coordinates.
Hence, 
we rewrite the hydrodynamic equations in the Cartesian coordinates
using relations such as $\varpi=x$ and $u_{\varphi}=x u_y$ for $y=0$. 
Then, the explicit forms of the equations can be written as
\beqn
&&\pa_t \rho_* + \pa_x(\rho_* v^x)+\pa_z(\rho_* v^z)
=-{\rho_* v^x \over x},\label{continuity1}\\
&&\pa_t (\rho_* \hat u_A)
+ \pa_x [\rho_* \hat u_A v^x + P\alpha e^{6\phi} \delta^x_{~A}]
+ \pa_z [\rho_* \hat u_A v^z + P\alpha e^{6\phi} \delta^z_{~A}] 
\nonumber \\
&&~~=-{\rho_* \hat u_A v^x \over x}
+{\rho_* \hat u_y v^y \over x}\delta_{Ax}
+P \pa_A (\alpha e^{6\phi}) \nonumber \\
&&~~~~-\rho_* \biggl[w h \pa_A \alpha - \hat u_j\pa_A \beta^j
+{\alpha e^{-4\phi} \hat u_i \hat u_j \over 2 w h} 
\pa_A \tilde \gamma^{ij}
-{2\alpha h (w^2-1) \over w} \pa_A \phi \biggr],\label{eulerA}\\
&&\pa_t (\rho_* \hat u_y) 
+ \pa_x(\rho_* \hat u_y v^x)+\pa_z(\rho_* \hat u_y v^z)
=-{2\rho_* \hat u_y v^x \over x},\label{eulery}\\
&& \pa_t (\rho_* \hat e)
+\pa_x[\rho_* \hat e v^x + P e^{6\phi} (v^x + \beta^x)]
+\pa_z[\rho_* \hat e v^z + P e^{6\phi} (v^z + \beta^z)] \nonumber \\
&&~~~~=
-{\rho_* \hat e v^x + P e^{6\phi} (v^x + \beta^x) \over x}
+\alpha e^{6\phi} P K + {\rho_* \over u^t h} \hat u_i \hat u_j K^{ij}
-\rho_* \hat u_i \gamma^{ij}D_j \alpha,\label{energy1}
\eeqn
where a subscript $A$ denotes $x$ or $z$, and $i, j, \cdots$ here 
denote $x, y$, and $z$. 
For numerically handling the transport terms as 
$\pa_x (\cdots)$ and $\pa_z (\cdots)$, we apply an approximate 
Riemann solver with third-order (piecewise parabolic) spatial interpolation. 
Other terms are regarded as the source terms.
No artificial viscosity is added, in contrast with our
previous axisymmetric implementation \cite{gr2d}. 
The time integration is done with the second-order Runge-Kutta method
as explained in \cite{gr3d}. 
Detailed numerical methods with respect to the treatment of the
transport terms are also described in Appendix A. 

We note that from Eqs. (\ref{continuity1}) and (\ref{eulery}), 
conservation of baryon rest-mass and angular momentum is
derived. However, we write these equations as nonconservative forms
and, hence, these conserved quantities are not precisely conserved 
in numerical computation. To suppress the growth of 
violation of the conservations in an acceptable level (e.g., within 1\%), 
we should be careful in the grid resolution (see Sec. IV). 

In every time step of computation, $w$ at each grid point 
is obtained by solving the following equation 
which is derived from the normalization relation of the four-velocity as
\beq
w^2 = 1 +\gamma^{ij} u_i u_j 
=1 +\gamma^{ij} \hat u_i \hat u_j \biggl({\hat e \over w}+{P \over \rho w^2}
\biggr)^{-2}.\label{wweq}
\eeq
Here, $P=P(\rho,\varepsilon)=P[\rho_*/(w e^{6\phi}),\hat e]$
(see Sec. II C) and
$\rho=\rho_*/(w e^{6\phi})$. Thus for a given $\tilde \gamma_{ij}$,
$\phi$, $\hat u_i$, $\hat e$, and $\rho_*$, 
Eq. (\ref{wweq}) constitutes an algebraic equation for $w$, which
can be solved by standard numerical techniques \cite{recipe}. 
After $w$ is obtained, $\rho$,
$P$, $\varepsilon$, $h$, and $v^i$ can be updated. 
We note that this procedure is essentially the same as that
used in our previous papers (see \cite{gr3d} for details). 

\subsection{Equations of state} 

We adopt two equations of state. 
One is the so-called $\Gamma$-law equation of state of the form 
\beq
P=(\Gamma-1)\rho \varep, \label{GEOS} 
\eeq
where $\Gamma$ is an adiabatic constant. 
In using Eq. (\ref{GEOS}), we always give an initial condition 
using the polytropic equation of state $P=K_{\rm P} \rho^{\Gamma}$
of the identical $\Gamma$, where $K_{\rm P}$ is a polytropic constant. 
In this paper, we set the adiabatic constant as $\Gamma=2$ 
[{i.e.}, the polytropic index $n$ is given by 
$n=1/(\Gamma-1)=1$] as a qualitative 
approximation of moderately stiff equations of state for neutron stars.
We note that if we prepare a polytropic star as an initial condition 
and the system evolves in an adiabatic manner (with no shock, cooling,  
and heating), 
the equation of state is preserved in the polytropic form 
even using Eq.~(\ref{GEOS}); {i.e.}, 
the value $P/\rho^{\Gamma} \equiv K_{\rm P}(x^{\mu})$ for any fluid 
element remains a constant ($=K_{\rm P}$). 

The second one is a parametric equation of
state that has been used by Yamada and Sato \cite{YS}, and 
by M\"uller and his collaborators \cite{Muller,HD} for the simulation of
a rotating stellar core collapse. 
In this equation of state, we assume that
the pressure consists of the sum of polytropic and thermal parts as
\beq
P=P_{\rm P}+P_{\rm th}. \label{EOSII}
\eeq
The polytropic part is in general given as
$P_{\rm P}=K_{\rm P}(\rho) \rho^{\Gamma(\rho)}$, where
$K_{\rm P}$ and $\Gamma$ are not constants but
functions of density $\rho$.
In this paper, we follow \cite{HD} for the choice of
$K_{\rm P}(\rho)$ and $\Gamma(\rho)$: 
For density smaller than the nuclear density
$\rho_{\rm nuc} \equiv 2\times 10^{14}~{\rm g/cm^3}$,
$\Gamma=\Gamma_1(=$const) is set to be $\alt 4/3$, and
for $\rho \geq \rho_{\rm nuc}$,
$\Gamma=\Gamma_2(={\rm const}) \geq 2$. Thus,
\beqn
P_{\rm P}=
\left\{
\begin{array}{ll}
K_1 \rho^{\Gamma_1}, & \rho \leq \rho_{\rm nuc}, \\
K_2 \rho^{\Gamma_2}, & \rho \geq \rho_{\rm nuc}, \\
\end{array}
\right.\label{P12EOS}
\eeqn
where $K_1$ and $K_2$ are constants. 
Since $P_{\rm P}$ should be continuous, we demand that 
the relation, $K_2=K_1\rho^{\Gamma_1-\Gamma_2}$, should be satisfied.
Following \cite{Muller,HD}, we set $K_1=5\times 10^{14}$ cgs, 
because we can well
approximate the polytropic part of the equation of state for
$\rho < \rho_{\rm nuc}$ in which the degenerate pressure of electrons is
dominant. Taking into account that the specific internal energy
should also be continuous at $\rho=\rho_{\rm nuc}$,
the polytropic specific internal energy 
$\varepsilon_{\rm P}$ is written as 
\beqn
\varepsilon_{\rm P}=
\left\{
\begin{array}{ll}
\displaystyle
{K_1 \over \Gamma_1-1} \rho^{\Gamma_1}, & \rho \leq \rho_{\rm nuc}, \\
\displaystyle 
{K_2 \over \Gamma_2-1} \rho^{\Gamma_2}
+{(\Gamma_2-\Gamma_1)K_1 \rho_{\rm nuc}^{\Gamma_1-1}
\over (\Gamma_1-1)(\Gamma_2-1)},  & \rho \geq \rho_{\rm nuc}. \\
\end{array}
\right.
\eeqn
With these settings, we mimic a realistic equation of state
for high-density, cold nuclear matter. 

The thermal part of pressure plays a role in the case that shocks
are generated. Here, we write it as 
\beq
P_{\rm th}=(\Gamma_t-1)\rho\varepsilon_{\rm th},
\eeq
where $\varepsilon_{\rm th}\equiv \varepsilon-\varepsilon_{\rm P}$. 
Following \cite{Muller,HD}, we set $\Gamma_{\rm th}=1.5$
in this paper. 

We performed simulations of rotating stellar collapses 
using this parametric equation of state. In choosing this, we always give
equilibrium stars as 
initial conditions using the polytropic equation of state 
\beq
P=K_0 \rho^{4/3},\label{EOS43}
\eeq
where $K_0$ is a constant. Following \cite{Muller,HD},
we set $K_0=5 \times 10^{14}~{\rm cm^3/s^2/gr^{1/3}}$, 
with which a soft equation of state governed by 
the electron degenerate pressure is well approximated \cite{ST}.
Here, $K_0$ and $K_1$ are related by
$K_1=K_0\rho_0^{4/3-\Gamma_1}$, where $\rho_0=1~{\rm g/cm^3}$.

\subsection{Adding atmosphere} 

In using high-resolution shock-capturing schemes, 
we have to add an atmosphere of small density outside
stars, since $\rho$ and $P$ have to be nonzero.
At $t=0$, we put an atmosphere of uniform density and specific 
internal energy in the computational domain of $\rho=0$, 
according to the following methods.
For the $\Gamma$-law equation of state with $\Gamma=2$, 
the uniform density of the atmosphere is set as 
$\rho_a=10^{-6}\rho_{\rm max}$, 
where $\rho_{\rm max}$ denotes the maximum density of a star.
The specific internal energy is given using the polytropic 
constant as $K_{\rm P}/4$.
For parametric equations of state (\ref{EOSII}),
the uniform density of the atmosphere is set as 
$\rho_a \approx 1~{\rm g/cm^3}$.
In this case, the specific internal energy is
given using the polytropic constant as $K_0$. 

For the $\Gamma$-law equation of state with $\Gamma=2$, 
the density decreases steeply around the surface of a neutron star. 
In such a case, numerical instability could often turn on around the stellar 
surface, if 
the density of the atmosphere is too low. This is the reason that 
we attach the atmosphere of relatively high density.
On the other hand, a small value of $\rho_a$ is acceptable 
in parametric equations of state.

\section{Global quantities and method for calibration}

We monitor the 
conservation of the total baryon rest-mass $M_*$,
ADM mass $M$, and angular momentum $J$, 
which are computed in the $y=0$ plane  as 
\beqn
&&M_*=4\pi \int_0^{L} x dx \int_0^{L} d z \rho_*, \\ 
&&M=-2\int_0^{L} x dx \int_0^{L} dz \biggl[ -2\pi E e^{5\phi}
+{e^{\phi} \over 8} \tilde R 
-{e^{5\phi}\over 8}\Bigl\{\tilde A_{ij}\tilde A^{ij}-{2 \over 3}K^2\Bigr\}
\biggr],~~~~~\\
&&J=4\pi \int_0^{L} x^2 dx \int_0^{L} dz \rho_* \tilde u_y,
\eeqn
where $E=\rho h w^2-P$ and $\tilde R$ is the Ricci scalar 
with respect to $\tilde \gamma_{ij}$. 
$M_*$ should be conserved in any system. 
Because of the axial symmetry, $J$ should also be conserved. 
On the other hand, $M$ is not conserved in general 
because of gravitational radiation. 
However, the total radiated energy 
of gravitational waves is likely to be quite 
small in the axisymmetric spacetime, so that we can consider 
$M$ as an approximately conserved quantity.
In our axisymmetric hydrodynamic implementation, $M_*$ and $J$ 
are not guaranteed to be conserved precisely. Thus, monitoring
the conservation of them is a good check of numerical accuracy. 

In addition to the mass and angular momentum, 
we also check the conservation of the 
specific angular momentum spectrum \cite{SP2}, 
\beq
M_*(j_0)=4\pi \int_{j \leq j_0} x dx dz \rho_*,
\eeq
where $j$ is the specific angular momentum computed as 
$x\hat u_y(=h u_{\varphi})$ and $j_0$ denotes a particular value for $j$.

The numerical accuracy is also checked monitoring 
the violation of the Hamiltonian constraint, which is written as
\beq
H=-8 \psi^{-5} \biggl[
\tilde \Delta \psi - {\psi \over 8}\tilde R +2\pi E \psi^5
+{\psi^5 \over 8}\tilde A_{ij}\tilde A^{ij}-{\psi^ 5 \over 12}K^2\biggr],
\eeq 
where $\psi \equiv e^{\phi}$, and
$\tilde \Delta$ denotes the Laplacian with respect to $\tilde \gamma_{ij}$. 
In this paper, we define the averaged violation according to
\beq
{\rm ERROR}={1 \over M_*} \int \rho_* |V| d^3x,
\eeq
where
\beq
V={\displaystyle \tilde \Delta \psi - {\psi \over 8}\tilde R +2\pi E \psi^5
+{\psi^5 \over 8}\tilde A_{ij}\tilde A^{ij}-{\psi^ 5 \over 12}K^2
\over \displaystyle
|\tilde \Delta \psi| + \Big|{\psi \over 8}\tilde R \Big| +2\pi E \psi^5
+{\psi^5 \over 8}\tilde A_{ij}\tilde A^{ij}+{\psi^ 5 \over 12}K^2}. 
\eeq
Namely, we use $\rho_*$ as the weight factor for the average.
The reason that we introduce this average is as follows. 
In using high-resolution shock-capturing schemes, 
we add an atmosphere of small density 
outside neutron stars and/or collapsing stars.  
In the atmosphere, a small error in the metric results in a large 
violation of the Hamiltonian constraint because $E$ is 
a very small value. Furthermore, the volume fraction occupied
by the atmosphere in the whole computational domain
is larger than that 
for main bodies. Thus, if we simply compute the volume integral of $|V|$, 
it is close to unity irrespective of the grid resolution. 
However, the numerical accuracy in the atmosphere is not very important for
evolution of the main bodies and for global evolution of the system 
in which we are interested. Therefore, to monitor whether the main bodies 
(neutron stars and collapsing stars) are accurately 
computed or not, this type of weight factor is necessary.

%%%%%%%%%%%%%%%%%%%%%%%%%%%%%%%%%%%%%%%%%%%
\section{Numerical results}

In the numerical simulations reported in Secs. IV A--C below, 
we adopted a fixed uniform grid, in which the grid spacing 
$\Delta x=\Delta y=\Delta z$ is constant, with grid size 
$(N+1, N+1)$ for $(x, z)$ to cover a computational domain 
as $0 \leq x,~z \leq L$, where $L=N\Delta x$.
In the simulation reported in Sec. IV D, we varied the
grid spacing during the computation, but still used the uniform grid
in which $\Delta x=\Delta y=\Delta z$. 
To check the
convergence of the numerical results for $\Delta x \rightarrow 0$, 
numerical computations were carried out with 
three levels of the grid resolution while fixing $L$. 
All the computations were done on the FACOM VPP5000 machine in 
the data processing center of the 
National Astronomical Observatory of Japan. 
The memory and CPU time in one run with a grid size 
$N=600$ and with four processors 
are about 1 GByte and $12$ CPU hours for $\sim 30000$ time steps. 

We note that for a simulation with $N=600$ in three spatial dimensions, we 
would need $\sim 300$ Gbytes memory and it takes $\sim 1000$ CPU hours for 
$30000$ time steps using 32 processors 
\cite{bina2}. Such simulation is pragmatically 
impossible, because the computational costs are too high 
(note that under normal circumstances, we can use at most $\sim 1000$
CPU hours per year). 
However, taking $N=600$ in an axisymmetric simulation is an easy task 
with the current computational resources.

\subsection{Spherical neutron stars}

In this subsection, we focus on the long-term
evolution of spherical neutron stars. 
The initial conditions are given using the polytropic equation of state 
with $\Gamma=2$, 
and during the time evolution, the $\Gamma$-law equation of state (\ref{GEOS}) 
is used. Although we did the same test simulations for the previous 
hydrodynamic implementation and obtained successful outputs
from it \cite{gr3d}, we repeated the tests again in the 
present new implementation to demonstrate that it also works well. 
A difference in the present tests from the previous ones is that 
we have performed much longer-term simulations than those in the 
previous tests, since it is computationally inexpensive and 
pragmatically possible to do in the axisymmetric case. 

In the polytropic equations of state, the polynomial relation 
$c^{3-n}K_{\rm P}^{n/2}G^{-3/2}$ has dimension of mass. With this property, 
all the quantities can be scaled to be nondimensional, if we 
multiply an appropriate combination of $c$, $G$, and $K_{\rm P}$. 
Thus, we will only show the nondimensional quantities 
in using this equation of state.
In other words, we adopt the units of $c=G=K_{\rm P}=1$. 
In these units, the maximum ADM mass and baryon rest-mass 
of spherical neutron stars are $\approx 0.164$ and 0.180, respectively, 
with the central density $\rho_c \approx 0.318$. 
Recovering $K_{\rm P}$, the mass and density in dimensional units
can be written as 
\beqn
&& M_{*{\rm phys}}=2.11 M_{\odot}
\biggr({K_{\rm P} \over 2\times 10^5~{\rm cgs}}\biggr)^{1/2}
\biggl({M_* \over 0.180}\biggr),\\
&& \rho_{\rm phys}=1.35 \times 10^{15}~{\rm g/cm^3} 
\biggr({K_{\rm P} \over 2\times 10^5~{\rm cgs}}\biggr)^{-1}
\biggl({\rho \over 0.3}\biggr). 
\eeqn
In Table I, we list several quantities for five models of spherical 
neutron stars that we pick up in this subsection. Below, 
we refer to these models as models (S1) -- (S5). 
Models (S1) -- (S4) are stable against gravitational collapse, while 
(S5) is marginally stable. 

\begin{table}[t]
\begin{center}
\begin{tabular}{|c|c|c|c|c|c|} \hline
& $\rho_c$ & $M_*$ & $M$ & $M/R$ & $P_{\rm osc} \rho_c^{1/2}$\\ \hline
S1 & 0.0637 & 0.105 & 0.100 & 0.0932 & \\ \hline
S2 & 0.127  & 0.150 & 0.140 & 0.146 & 5.0\\ \hline
S3 & 0.191  & 0.170 & 0.156 & 0.178 & 6.9\\ \hline
S4 & 0.255  & 0.178 & 0.162 & 0.200 & 11\\ \hline
S5$^{\dagger}$ & 0.318 & 0.180 & 0.164 & 0.214 & \\ \hline
\end{tabular}
\caption{The central density $\rho_c$,
baryon rest-mass $M_*$, ADM mass $M$, and compactness $M/R$ of
spherical neutron stars with $\Gamma=2$ that we pick up in this paper.
Here, $R$ denotes the circumference radius.  
All the quantities are shown in units of $c=G=K_{\rm P}=1$.
The star denoted with $\dagger$ mark is of the maximum allowed mass and
hence marginally stable against gravitational collapse.
In the last column, numerical results of 
the radial oscillation period for the $f$ mode, $P_{\rm osc}$, 
are presented in units of $\rho_c^{-1/2}$.
}
\end{center}
\vspace{-5mm}
\end{table}

\begin{figure}[htb]
\vspace*{-4mm}
\begin{center}
\epsfxsize=2.6in
\leavevmode
(a)\epsffile{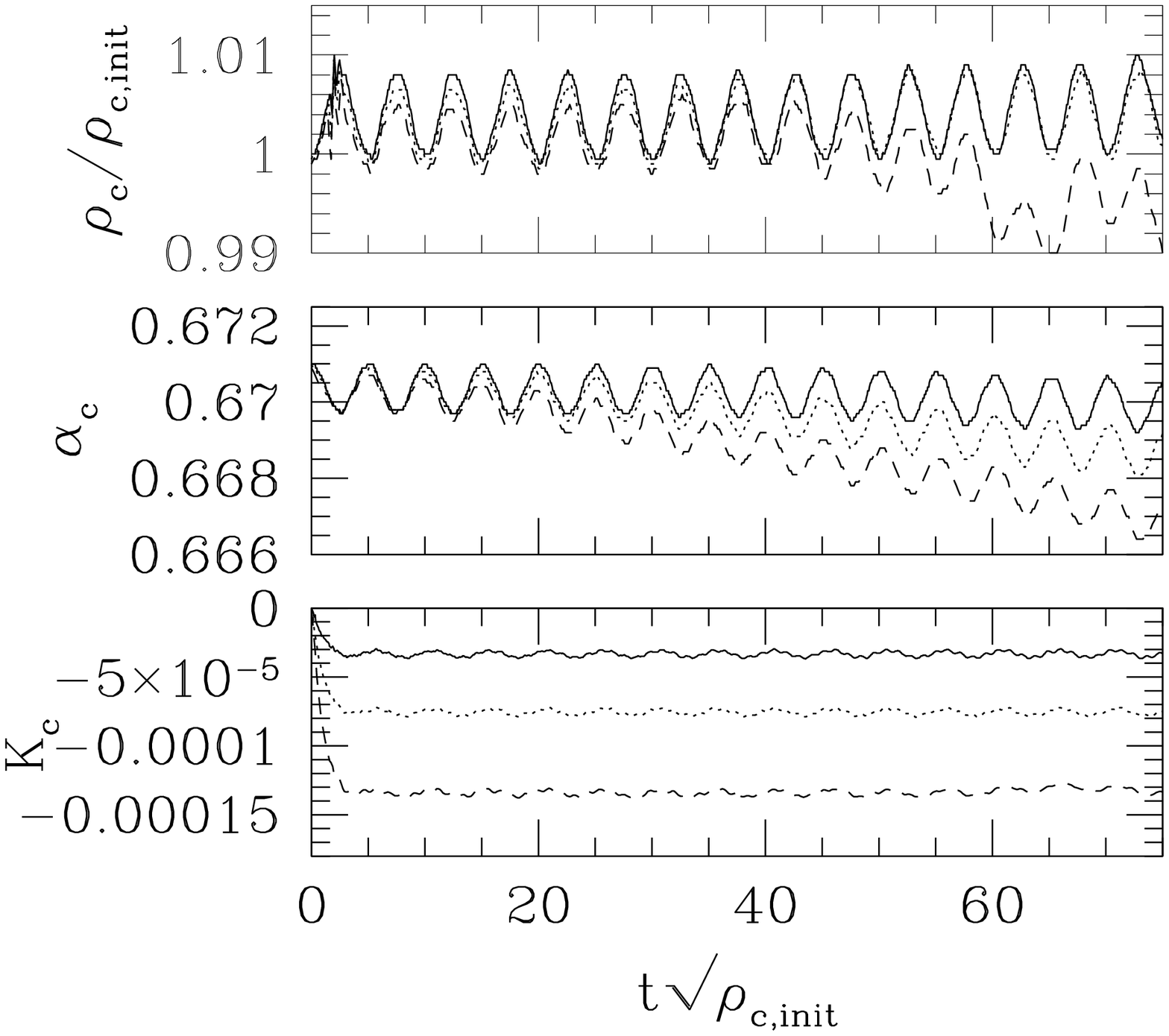}
\epsfxsize=2.6in
\leavevmode
~~(b)\epsffile{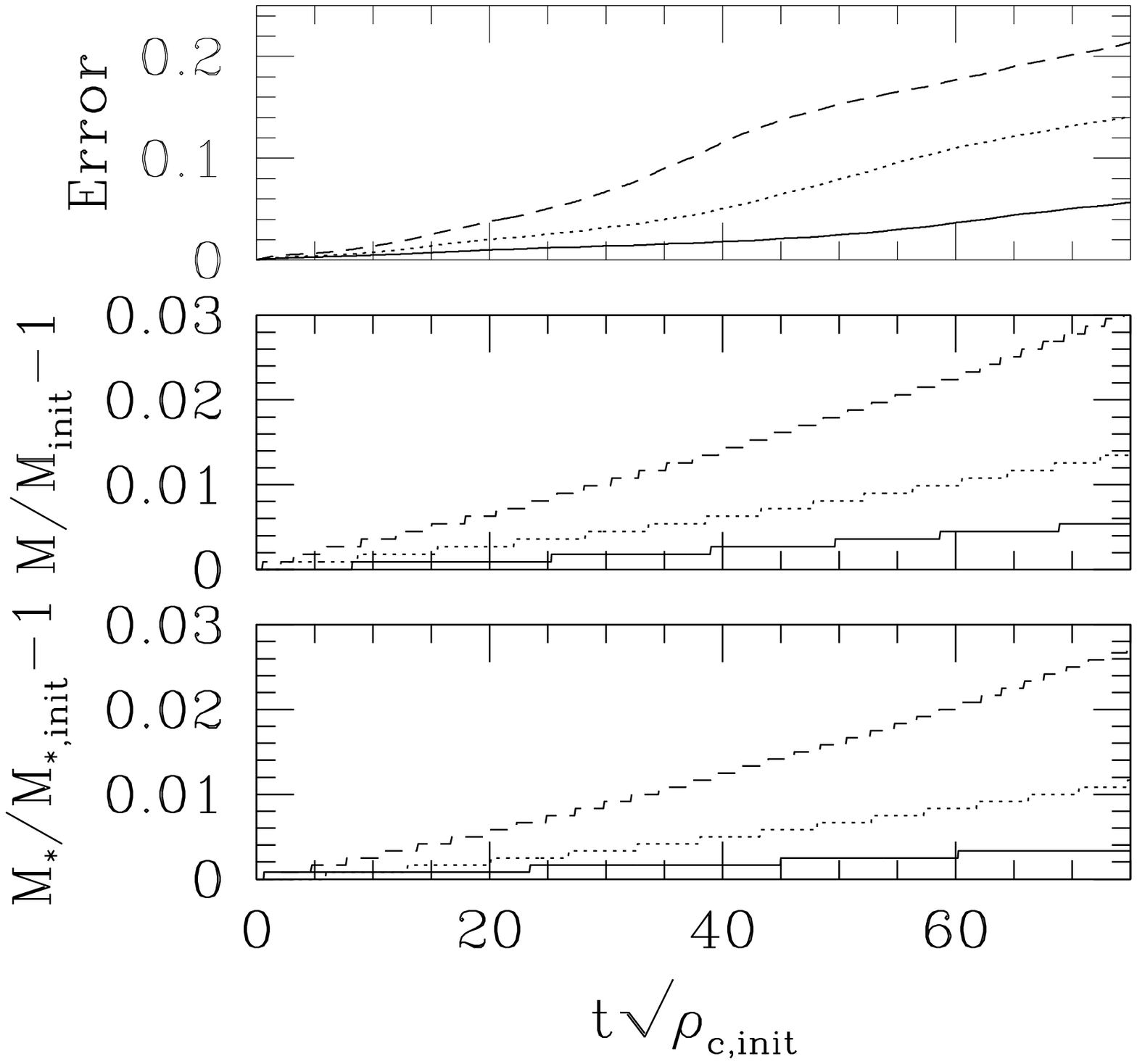}
\caption{(a)
Time evolution of central density, central value of lapse function, 
and extrinsic curvature at origin, and
(b) time evolution of averaged 
violation of the Hamiltonian constraint,
violation of the ADM mass conservation, and
violation of the baryon rest-mass conservation 
for a stable spherical neutron star (S2).
In both figures, the solid, dotted, and dashed 
curves denote the results with $N=180$, 120, and 90.
With these grid numbers, the radius is initially covered by about 
65, 44, and 33 grid numbers. $\rho_{c,{\rm init}}$ denotes
the central density at $t=0$, and the time is shown in units of
$\rho_{c,{\rm init}}^{-1/2}$. 
\label{FIG1}}
\end{center}
\end{figure}

In Figs. \ref{FIG1} (a) and (b), we display the time evolution of
the central density, central value of the lapse function
(hereafter $\alpha_c$), central value of $K$ ($K_c$),
averaged violation of the Hamiltonian constraint, and 
violation of ADM mass and baryon rest-mass conservation for model (S2).
Throughout this subsection, the time is shown in units of
$\rho_{c,{\rm init}}^{-1/2}$, where $\rho_{c,{\rm init}}$ denotes
the central density at $t=0$. 
To induce a small oscillation, we initially 
reduce the pressure by 0.2\%. We note that whenever we
superimpose a perturbation to an equilibrium configuration,
we reinforce the Hamiltonian and momentum constraints at $t=0$. 
Numerical results are 
shown for $N=90$, 120, and 180 with a fixed value of $L$, 
to demonstrate that the convergence is achieved. 
The simulations continued for $\sim 30$ dynamical time scales (see
also Fig. 2) until the crash of the run,
irrespective of the grid resolution, 
although the accuracy deteriorates gradually with time. 
Here, we refer to the period of 
the fundamental radial (and quasiradial) oscillation as  
the dynamical time scale.

$L$ is chosen as $\sim 3R_s$, where $R_s$ denotes the coordinate radius 
of the neutron star. For simulating spherical systems, 
we imposed the outer boundary conditions as 
\beqn
\tilde \gamma_{ij}=\delta_{ij}, ~\tilde A_{ij}=0, ~(r\phi)_{,r}=0,~ K=0, 
~{\rm and} ~F_i=0.
\eeqn
(For nonspherical problems, we impose an outgoing boundary condition 
for $\tilde \gamma_{ij}$ and $\tilde A_{ij}$.) 
These boundary conditions are adequate 
but not physically perfect. As a result, 
for the choice of a too small value of $L$ as $\sim R_s$, 
the numerical solution is 
affected by the spurious effects of the outer boundaries, 
resulting in an earlier crash of the run. However, for $L > 2R_s$, 
the results are not significantly modified by the spurious effects.  

As we mentioned in Sec. III, the baryon rest-mass 
is not numerically conserved strictly. However, the violation 
does not seriously affect the numerical results. Indeed, 
the averaged values of the central density and lapse remain constants, 
as they should. 
The numerical results indicate that if we want to suppress the violation of 
the mass conservation within 1\% (2\%) after 10 dynamical time scales, 
the radius of the neutron star should be covered by more than
40 (30) grid points. 

The error of mass conservation 
converges to zero with improving the grid resolution at 
approximately second-order. The averaged 
violation of the Hamiltonian constraint 
also indicates approximate second-order convergence. 
Therefore, we can conclude that the numerical solution converges to 
the exact solution in the limit $\Delta x \rightarrow 0$. 
We note, however, that the 
convergence is only approximately at second-order, because 
near stellar surfaces, the gradients of hydrodynamic variables
are so steep that transport terms are often computed
with first-order accuracy in space. 
The convergence may also become first-order if shocks are generated during 
numerical computations (see Sec. IV C), since 
near the shocks, the hydrodynamic computations are done 
with first-order accuracy. A similar tendency is 
reported by Miller {\it et al}. in the simulations of 
a head-on collision of two neutron stars \cite{miller}. 

In the last figure of Fig. \ref{FIG1}(a), it is shown that 
$K_c$ is not zero exactly but relaxes to a finite value. 
This indicates that even in solving the equation for 
the maximal slicing condition, 
$K$ deviates from zero as long as finite differencing methods are used. 
Indeed, $K_c$ converges to zero at second-order 
with improving the grid resolution. 
Thus the maximal slicing condition $K=0$ cannot be 
precisely imposed in numerical computation 
even in a well-resolved simulation, if we adopt finite differencing schemes. 
This tells us that we should follow the evolution of $K$ and
should not a priori set $K=0$ in numerical computation, 
even in choosing the maximal slicing condition. 

\begin{figure}[t]
\vspace*{-4mm}
\begin{center}
\epsfxsize=2.6in
\leavevmode
(a)\epsffile{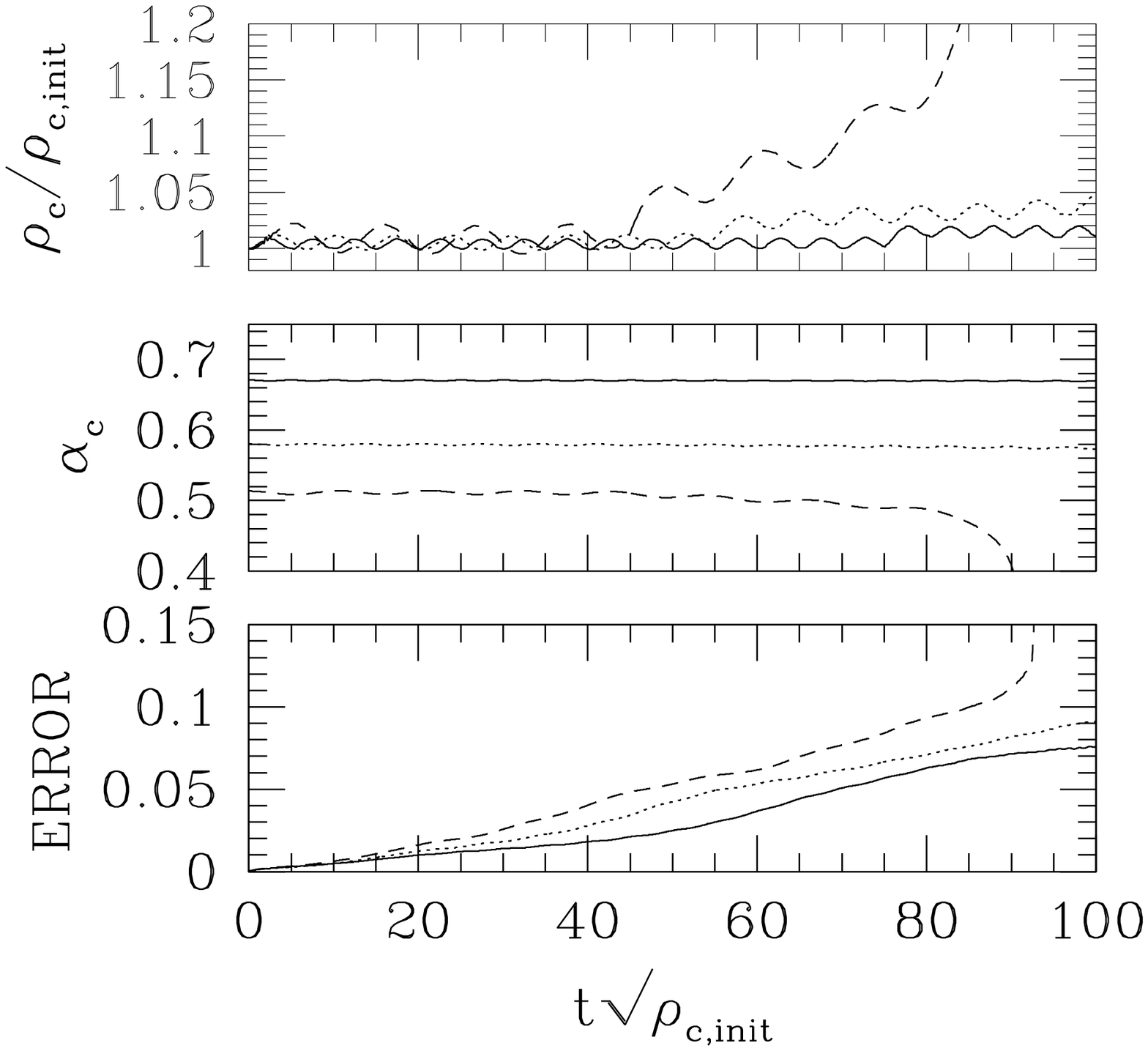}
\epsfxsize=2.6in
\leavevmode
~~~(b)\epsffile{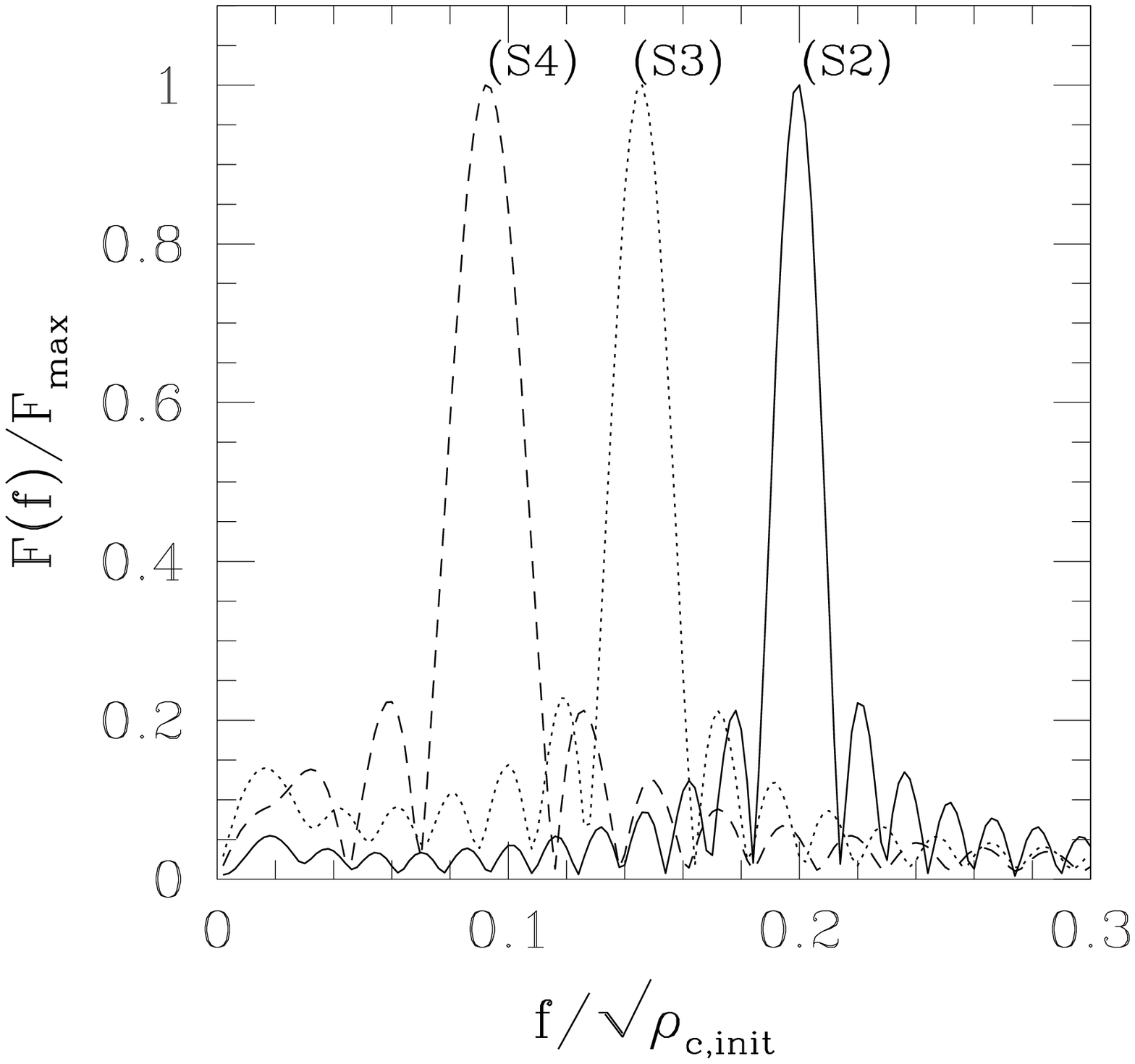}
\caption{(a)
Time evolution of central density, lapse function at origin, 
and averaged violation of the Hamiltonian constraint 
for stable stars (S2) (solid curves), (S3) (dotted curves), 
and (S4) (dashed curves). The simulations were performed with 
$N=180$. With these grid numbers, the radius is covered by about 
65, 58, and 51 grid numbers for (S2), (S3), and (S4), respectively.
(b) Fourier spectra of the central density for (S2) (solid curve),
(S3) (dotted curve), and (S4) (dashed curve). 
\label{FIG2}}
\end{center}
%\vspace{-5mm}
\end{figure}

Long-term simulations for more compact stable stars 
(S3) and (S4) were also carried out.
In Fig. \ref{FIG2}, we display the results for models
(S2), (S3), and (S4) together.
For (S3), the simulation continued for 
$\sim 20$ dynamical time scales until the
crash of the run. However, for (S4), 
the star starts collapsing to a black hole
after about five oscillation periods.
The reason for this consequence is clear. 
The ADM mass of model (S4) is $\approx 99$\%
of the maximum allowed value. Thus, 
with a slight increase of the mass as a result of
the accumulation of numerical error, 
the mass exceeded the maximum allowed value for stable stars, 
resulting in eventual gravitational collapse. 
It is interesting to note that in this case, the computation was 
able to be continued until the formation of a black hole of the apparent 
horizon mass [see Eq. (\ref{MAH}) for definition] $\sim M$, 
where $M$ is the initial ADM mass. Thus, the increase of 
the central density and the decrease of $\alpha_c$ 
[see Fig. \ref{FIG2} (a)] do not imply 
that the computation crashed. To avoid the collapse to 
a black hole and to make the oscillating time longer, we need to 
take more grid numbers to improve the grid resolution.
Actually, we have checked that we can increase 
the oscillation cycles in numerical computation
with improvement of the grid resolution. 

The duration of the simulation to crash
for model (S3) was shorter than 
that of (S2). This fact indicates that 
with increasing the compactness of 
neutron stars, the computations crash earlier. 
We note here that the duration of the 
simulation for these models depends only weakly on the grid resolution. 
Thus, it does not depend on the numerical accuracy, but 
seems to depend on certain factors associated with 
formulation or gauge conditions or outer boundaries.
The same tendency is found 
in the simulations of rotating neutron stars. 
We will argue this point in Sec. IV B again. 

Although the simulations can be continued only for a finite time scale,
the duration of $> 10$ dynamical time scales
seems to be sufficiently long. Indeed, 
we can accurately extract the oscillation frequencies
of the fundamental radial mode from these simulations. 
In Fig. \ref{FIG2}(b), we display the Fourier spectra of
$\rho_c(t)$, which is defined as 
\beq
F(f) \equiv \Big| \int_0^{t_f} [\rho_c(t)-\rho_{c,{\rm av}}]
e^{2\pi i f t} dt \Big|,
\eeq
where $t_f$ is chosen as $\sim 70\rho_c^{-1/2}$, $55\rho_c^{-1/2}$, 
and $40\rho_c^{-1/2}$ for (S2), (S3), and (S4), respectively. 
$\rho_{c,\rm av}$ is computed from 
\beq
\rho_{c,\rm av} \equiv {1 \over t_f} \int_0^{t_f} \rho_c(t) dt. 
\eeq
The Fourier spectra indicate that the oscillation period of
the $f$ mode of the radial oscillation 
is $\approx 5.0\rho_c^{-1/2}$, $6.9\rho_c^{-1/2}$, and
$11\rho_c^{-1/2}$ for (S2), (S3), and (S4). These values 
agree well with those derived from Chandrasekhar's
semianalytic formula \cite{chandra,gr3d}.
Furthermore, the result for (S2) is in good agreement with
that for a spherical star of $\rho_c=0.128$ reported
in \cite{other}. Thus, we conclude that the computation can be continued 
for a sufficiently long time to accurately obtain the
oscillation frequencies of even extremely relativistic neutron stars. 
The simulation would be able to be carried out to study
nonspherical oscillations of neutron stars, as was
done in \cite{gr3d,nick}. 

\begin{figure}[t]
\vspace*{-4mm}
\begin{center}
\epsfxsize=2.6in
\leavevmode
(a)\epsffile{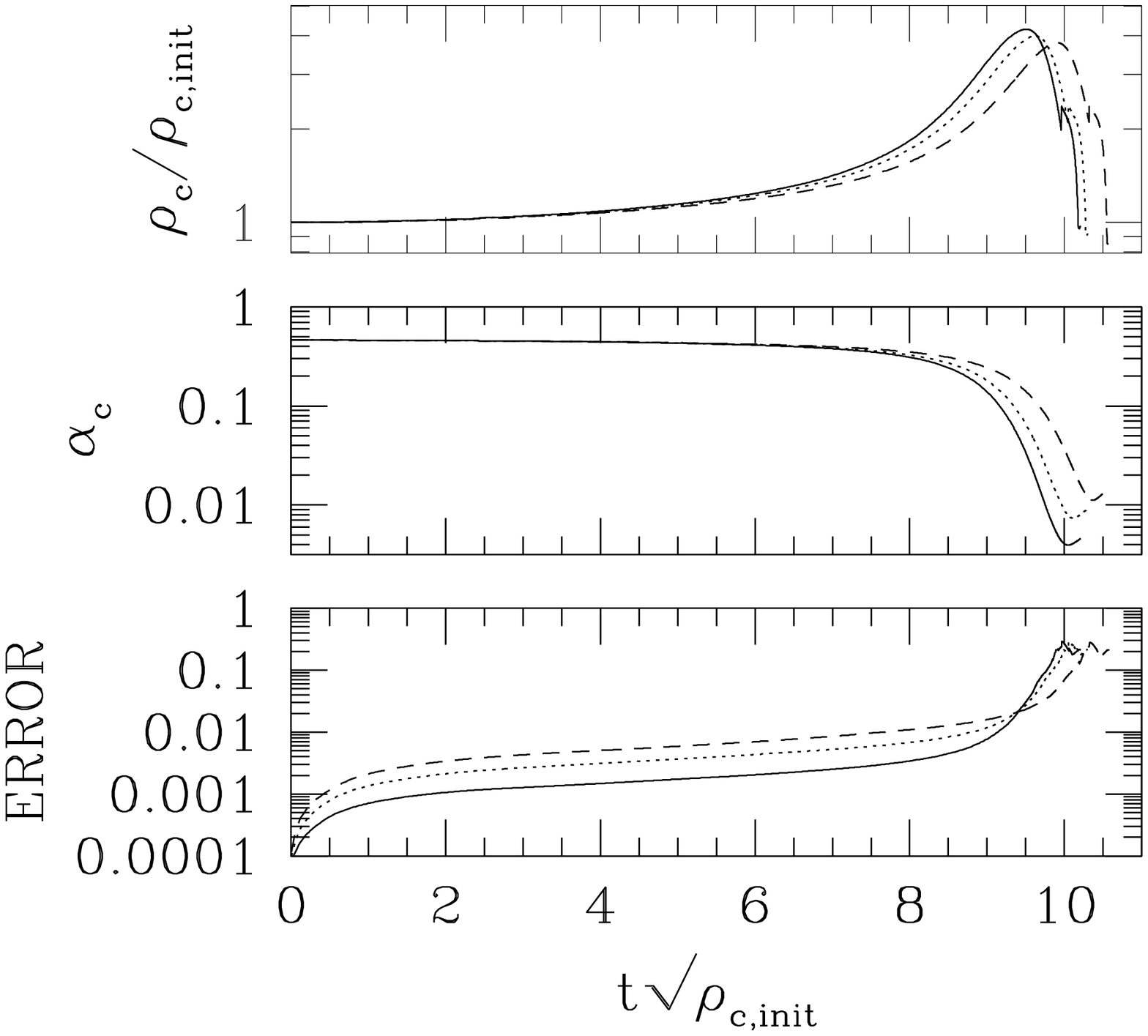}
\epsfxsize=2.6in
\leavevmode
~~~(b)\epsffile{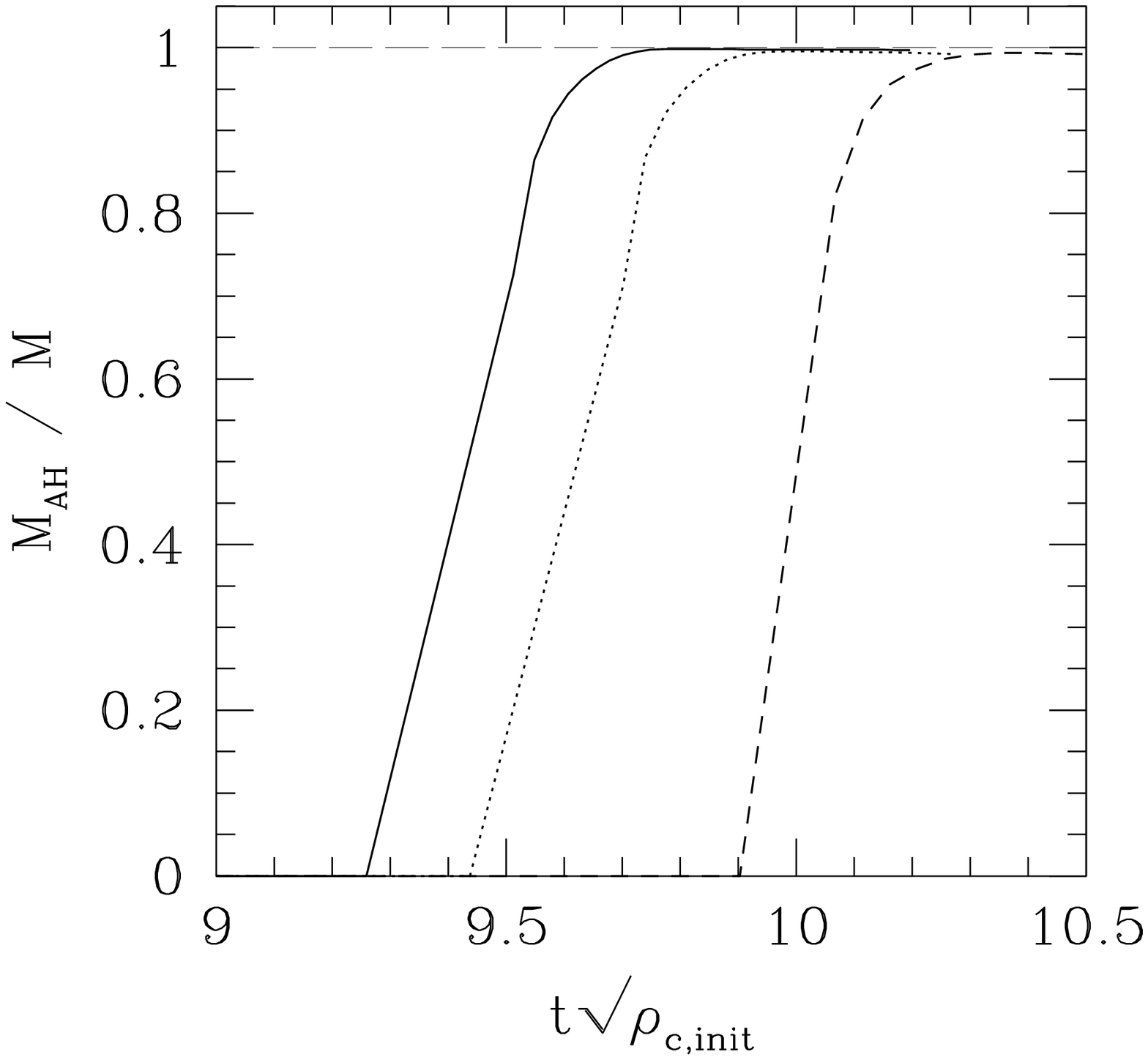}
\caption{(a) 
Time evolution of central density, lapse function at origin, 
and violation of baryon mass conservation, and
(b) time evolution of mass of the apparent horizon $M_{\rm AH}$
in units of the ADM mass of the system for
a marginally stable spherical neutron star (S5). 
To destabilize, we initially reduce the pressure by 0.5\%. 
In both figures, the solid, dotted, and dashed 
curves denote the results with $N=180$, 120, and 90. 
With these grid numbers, the radius is initially covered by about 
71, 47, and 35 grid points. 
\label{FIG3} }
\end{center}
%\vspace{-1mm}
\end{figure}

In Fig. \ref{FIG3}, we display the time evolution of several quantities for 
collapse of a marginally stable spherical neutron star (S5). 
To induce the collapse, we initially reduced the pressure 
by 0.5\%. With this setting, the neutron star 
collapses to a black hole in $\sim 10 \rho_{c,{\rm init}}^{-1/2}$.
We have checked that 
even with a 0.2\% decrease of the pressure, the star collapses 
to a black hole in $\sim 20 \rho_{c,{\rm init}}^{-1/2}$,
which is longer than that for the case of 0.5\%. 
Numerical results are presented for 
$N=90$, 120, and 180 and demonstrate that the convergence 
is achieved. 
Note that with lower grid resolution, it takes a slightly longer time to 
form a black hole. This is because the dissipation 
that prevents the increase of density is larger 
with lower grid resolution. 

In the final phase of the collapse, the grid resolution around the black hole 
forming region became so bad that the computation crashed. 
It is interesting to note that the 
magnitude of the averaged violation of 
the Hamiltonian constraint, ERROR, 
relaxes to $\sim 0.2$ irrespective of the grid 
resolution at $t \sim 10 \rho_{c,{\rm init}}^{-1/2}$. 
This verifies that the computation 
crashes when ERROR is $\sim 0.2$. We can also observe that 
as the accuracy deteriorates, (i) the central 
density stops increasing and instead starts decreasing, and
(ii) exponential decrease of $\alpha_c$ that is
a feature in the maximal slicing condition is modified.
The reason for (i) is as follows. In our implementation, 
$\rho_*$ is a fundamental quantity to evolve, and $\rho$ is computed from  
$\rho_*/(e^{6\phi}w)$. In numerical computations, $\rho_*$ and $\phi$
monotonically increase, but since $\phi$ around the origin
is too large [of $O(1)$]
in the late phase of the
collapse, a small error in $\phi$ leads to a large error
in $\rho$. As a result, $\rho$ decreases in the late phase. 
The reason for (ii) is simply that the computation crashed. 
Indeed, the time at which
the behavior of $\alpha_c$ starts changing agrees with 
that at which the magnitude of the averaged violation of the Hamiltonian 
constraint saturates to $\sim 0.2$. 

Although the accuracy deteriorates in the final phase of the 
collapse, the simulation can be carried out at least 
until the formation of an almost static black hole. To confirm
the black hole formation, 
apparent horizons were located during the simulations. 
In Fig. \ref{FIG3}(b), we display 
the time evolution of the mass of the apparent horizon 
in units of the ADM mass of the system. Here, the 
mass of the apparent horizon is defined as 
\beqn
M_{\rm AH}=\sqrt{{S \over 16\pi}},\label{MAH}
\eeqn
where $S$ is the area of the apparent horizon. 
The figure indicates that $M_{\rm AH}$ relaxes approximately to $M$ 
in the final phase of the collapse. (For $N=180$, $|M_{\rm AH}/M-1|$
is less than 1\%.) This implies that 
the simulation was carried out up to the time
when a spacetime settles down to a static black hole spacetime.

\begin{figure}[htb]
\vspace*{-4mm}
\begin{center}
\epsfxsize=2.6in
\leavevmode
(a)\epsffile{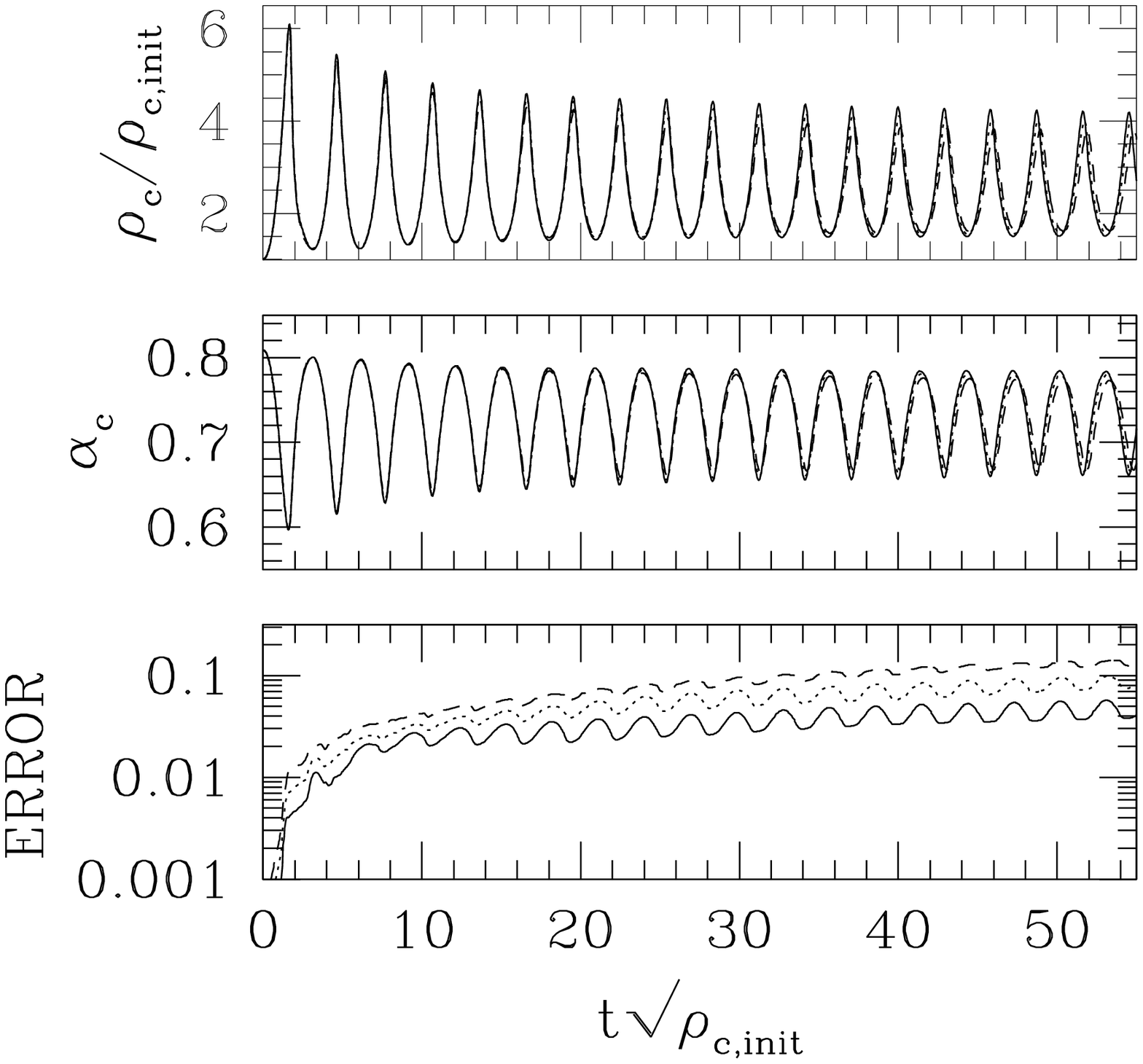}
\epsfxsize=2.6in
\leavevmode
~~~(b)\epsffile{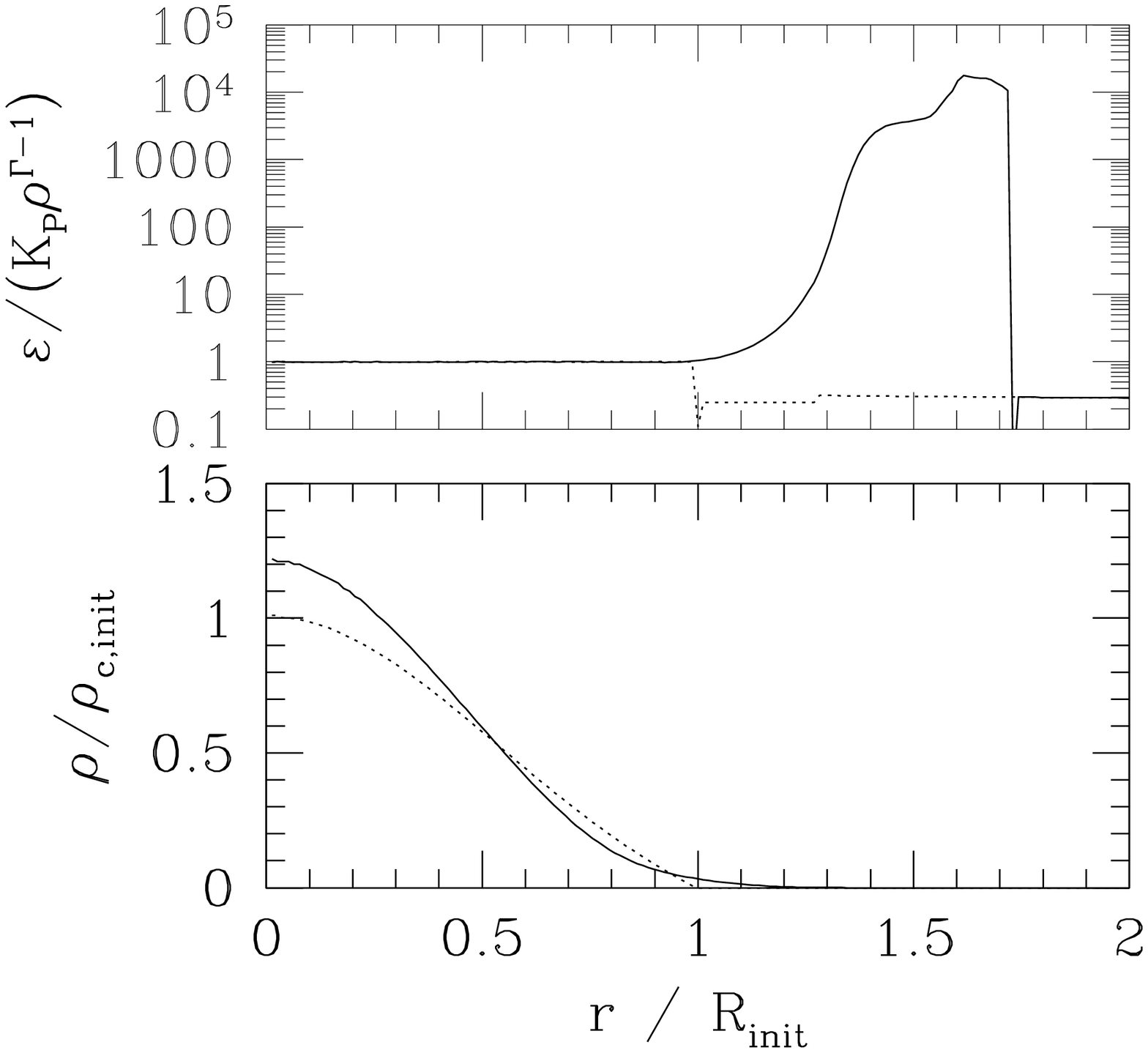}
\caption{
(a) Time evolution of central density, lapse function at origin, 
and averaged violation of Hamiltonian constraint for
an oscillating spherical neutron star.
The solid, dotted, and dashed 
curves denote the results with $N=180$, 120, and 90.
With these grid numbers, the radius is initially covered by about 
78, 52, and 39 grid points. 
(b) Profiles of 
$\varepsilon/(K_{\rm P}\rho^{\Gamma-1})$ and $\rho$
at $t \approx 0$ (dotted curves)
and after about one oscillation period (solid curves).
The unit of the horizontal axis is the initial radius of 
the neutron star. 
\label{FIG4}}
\end{center}
%\vspace{-1mm}
\end{figure}

In Fig. \ref{FIG4}(a), we show the time evolution of several quantities
for an oscillating spherical neutron star of a high amplitude.
In this simulation, we picked up a low-mass spherical star (S1), and 
to induce an oscillation of a high amplitude,
we initially reduced the pressure by 40\%.
The numerical results are shown for 
$N=90$, 120, and 180, and demonstrate that convergence is achieved. 
In each case, the neutron star is initially covered by
39, 52, and 78 grid points, respectively. 

Because of a significant decrease of the pressure, the radius of the 
neutron star decreases 
by a factor of $\sim 2$ soon after the simulation starts. 
However, the magnitude of the pressure decrease is not 
large enough for the star to collapse to a black hole. Instead, 
the star bounces when the central density becomes about six 
times of initial value, and repeats oscillations subsequently.
As the density approaches the maximum, 
shocks are formed around the stellar surface, 
and as a result, outer envelopes explode. 
To illustrate that shock heating indeed occurs, we display 
$\varepsilon/(K_{\rm P}\rho^{\Gamma-1})$ 
at $t \approx 0$ and after about one oscillation
period in Fig. \ref{FIG4}(b). 
As mentioned in Sec. II C, in the absence of shocks, this quantity 
does not change from the initial value, but in the presence of 
shock heating it increases. 
Figure \ref{FIG4}(b) clearly shows that
in the outer envelope, the shock heating is significant. 
On the other hand, a negligible effect of shocks can be seen 
around the central region. 
Since the shocks are generated only in the atmosphere, 
the averaged violation of the Hamiltonian constraint still converges 
approximately at second-order with improving the grid resolution. 

Figure \ref{FIG4}(a) indicates that the
amplitude of the oscillation gradually 
decreases, and after several oscillation periods, 
it settles down approximately to a constant. This illustrates that 
the kinetic energy of the oscillation is dissipated by the shocks 
gradually. Similar results are reported in \cite{other}
for a simulation of a migrating neutron star. 
The lower figure of Fig. \ref{FIG4}(b) shows the density profiles
at $t=0$ and $t \sim$ one oscillation period. 
This indicates that the density profile is modified to a more 
centrally condensed state as a result of the shock dissipation.

We emphasize that this simulation is highly dynamical and general 
relativistic. Even in such a case, the simulation was able to be continued 
for more than 20 oscillation periods. This illustrates the robustness of 
our implementation for dynamical problems in general relativity.

\subsection{Rapidly rotating neutron stars}

We focus here on the long-term 
evolution of rigidly and rapidly rotating neutron stars at mass shedding 
limits for which the angular velocity at the equator 
is equal to the Kepler angular velocity. 
Following Sec. IV A, we adopt the polytropic equation of state 
with $\Gamma=2$ for setting initial conditions and 
evolve neutron stars using the $\Gamma$-law equation of state (\ref{GEOS}). 
The axial ratio of the polar radius to the equatorial
one is $\approx 0.58$ 
for rotating neutron stars at mass shedding limits with $\Gamma=2$.  
As in Sec. IV A, we only present the scaled dimensionless quantities 
with the units of $c=G=K_{\rm P}=1$ throughout this subsection. 
In these units, the maximum ADM mass and baryon rest-mass
of rigidly rotating neutron stars are about 0.188 and 0.207, respectively, 
with the central density $\rho_c \approx 0.27$ \cite{CST}. 
The central density of the marginally stable star against 
gravitational collapse has slightly larger density ($\approx 0.295$) than this 
value \cite{CST,SBS}. 
In Table II, we list several quantities of
six rotating neutron stars that we pick up here.
In the following, we refer to these neutron stars as models 
(R0) -- (R5). Models (R0) -- (R4) are stable against gravitational
collapse, and (R5) is unstable and very close to the marginally stable 
point. 

\begin{table}[t]
\begin{center}
\begin{tabular}{|c|c|c|c|c|c|c|c|c|} \hline
& $\rho_c$ & $M_*$ & $M$ & $M/R$ & $P_{\rm rot}\rho_c^{1/2}$
& $J/M^2$ & $|T/W|$ & $P_{\rm osc} \rho_c^{1/2}$
\\ \hline
R0 & 0.103 & 0.169 & 0.158 & 0.111 & 8.52 &0.667 & 0.0932 &5.7\\ \hline
R1 & 0.136 & 0.186 & 0.172 & 0.129 & 8.53 &0.630 & 0.0909 &6.5\\ \hline
R2 & 0.183 & 0.200 & 0.183 & 0.148 & 8.56 &0.599 & 0.0876 &8.2\\ \hline
R3 & 0.215 & 0.204 & 0.186 & 0.158 & 8.60 &0.585 & 0.0856 &10 \\ \hline
R4 & 0.253 & 0.206 & 0.188 & 0.167 & 8.65 &0.573 & 0.0834 &16 \\ \hline
R5$^{\dagger}$ & 0.296 & 0.206 & 0.188 
& 0.175 & 8.72 & 0.561 & 0.0809& \\ \hline
\end{tabular}
\caption{The central density, baryon rest-mass, ADM mass, $M/R$, 
rotational period ($P_{\rm rot}$) in units of
$\rho_{c,{\rm init}}^{-1/2}$, $J/M^2$, and $|T/W|$ of
rotating neutron stars at mass shedding limits
with $\Gamma=2$ that we pick up in this paper. 
Here $T$ and $W$ are the rotational kinetic energy and
gravitational potential energy, and 
$R$ denotes the circumference radius at the equator. 
All the quantities are shown in units of $c=G=K_{\rm P}=1$.
The star denoted with $\dagger$ is unstable. 
In the last column, numerical results of 
the quasiradial oscillation period for the $f$ mode, $P_{\rm osc}$, 
are presented in units of $\rho_{c,{\rm init}}^{-1/2}$. 
}
\end{center}
\vspace{-5mm}
\end{table} 

\begin{figure}[htb]
\vspace*{-4mm}
\begin{center}
\epsfxsize=2.6in
\leavevmode
(a)\epsffile{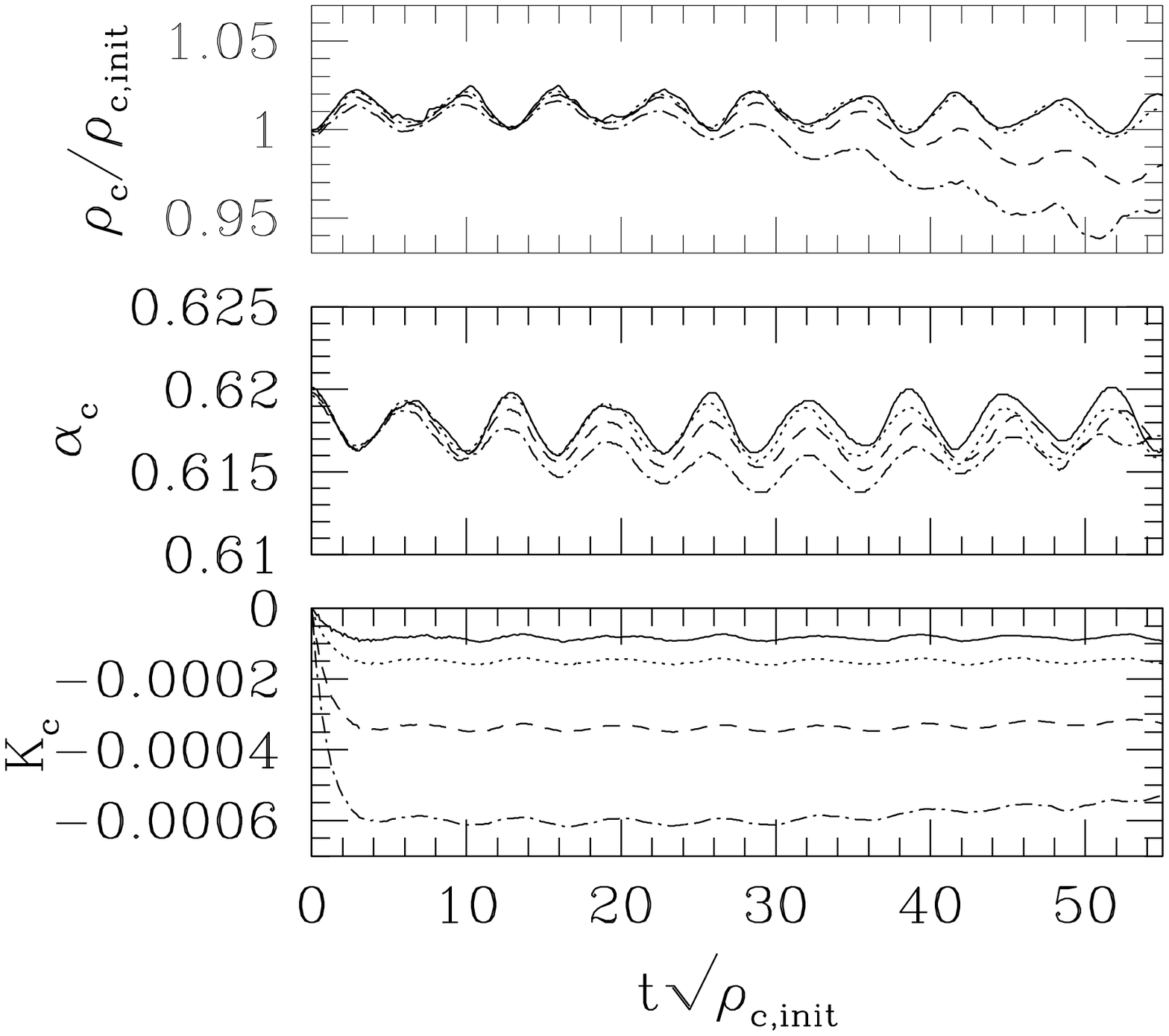}
\epsfxsize=2.6in
\leavevmode
~~~(b)\epsffile{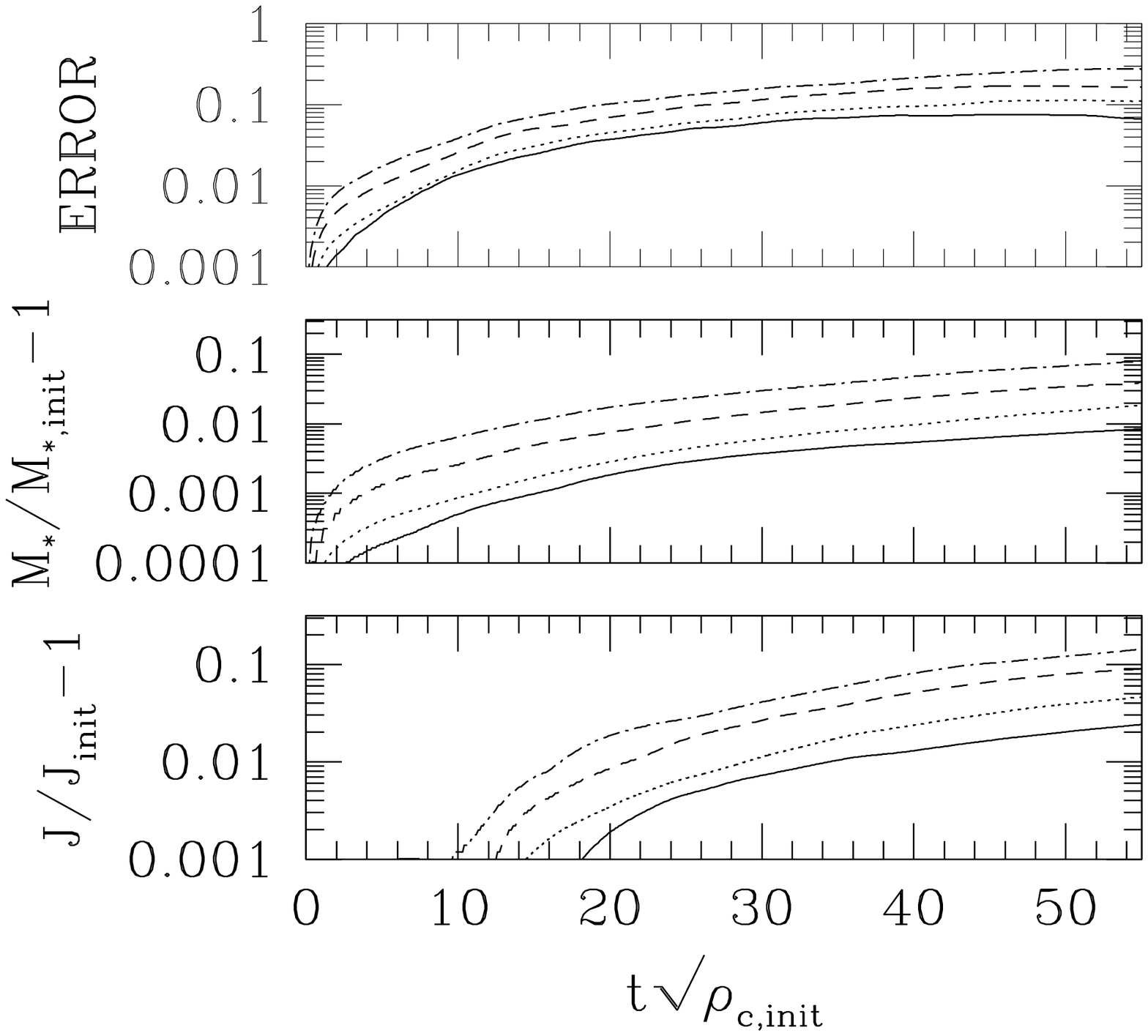}
\caption{
(a) Time evolution of central density, lapse function at origin, 
and extrinsic curvature at origin, and 
(b) time evolution of averaged
violation of the Hamiltonian constraint, 
violation of rest-mass conservation, and 
violation of angular momentum conservation for
a stable and rapidly rotating neutron star at the mass shedding limit (R1). 
In both figures, the solid, dotted, dashed, and dotted-dashed
curves denote the results with $N=240$, 180, 120, and 90.
With these grid numbers, the polar (equatorial)
radius is covered by about
46 (80), 35 (60), 23 (40), and 18 (30) grid points, respectively. 
\label{FIG5} }
\end{center}
%\vspace{-1mm}
\end{figure}

In Fig. \ref{FIG5}, we display the time
evolution of several quantities for model 
(R1). The ADM mass of this model is $\approx 91$\% of the maximum mass 
of rigidly rotating neutron stars with $\Gamma=2$, so that 
it is a sufficiently relativistic model. 
To induce a small oscillation of the fundamental quasiradial mode, 
we initially reduce the pressure by 0.5\%. Numerical results are 
shown for $N=240$, 180, 120, and 90 to demonstrate that convergence is
achieved. With these grid numbers, the polar (equatorial) 
radius is covered by about
46 (80), 35 (60), 23 (40), and 18 (30) grid points, respectively. 

The simulations continued for 
$\sim 10$ dynamical time scales, and eventually crashed. 
The duration of the simulation to crash depends only
weakly on the grid resolution as in the spherical cases. 
This implies that the crash is not triggered by accumulation of 
the numerical error. 
The duration also does not vary much even if 
we change the outer boundary conditions and the 
location of the outer boundary as $L \approx 3R_e$--$4R_e$, 
where $R_e$ denotes the coordinate radius at the equator. 
Furthermore, the duration of the 
simulation for a rotating neutron star is shorter than that for 
a spherical star of identical compactness. 
Therefore, we deduce that the crash of
computations might be associated with our choice of 
the spatial gauge condition or the formulation, 
although we do not understand the reason fully at present. 
There may still be room to improve the spatial gauge condition
and/or the formulation, 
if one wants to perform an extremely long-term simulation 
of the duration $\gg 10$ dynamical time scales. 
However, 10 dynamical time scales are long enough 
to produce scientific results for most problems, so that 
we do not address this problem any longer in this paper.

As in the spherical case, convergence is achieved 
with improvement of the grid resolution. As argued in Sec. III, 
the angular momentum as well as the baryon rest-mass are 
not conserved strictly, although they are conserved quantities.
However, the violation converges to zero nearly at second-order
with improving the grid resolution. The results of this
convergence test indicate that
the polar axis should be covered by at least $\sim 30$ grid points
if one wants to demand that the violation of the conservation
of angular momentum and baryon rest-mass is less than a few \% after
$\sim 10$ dynamical time scales. 
If the polar axis is covered by fewer than 20 grid points, 
the magnitude of the violation becomes larger than 10\% 
after 10 dynamical time scales.

\begin{figure}[t]
\vspace*{-4mm}
\begin{center}
\epsfxsize=2.6in
\leavevmode
(a)\epsffile{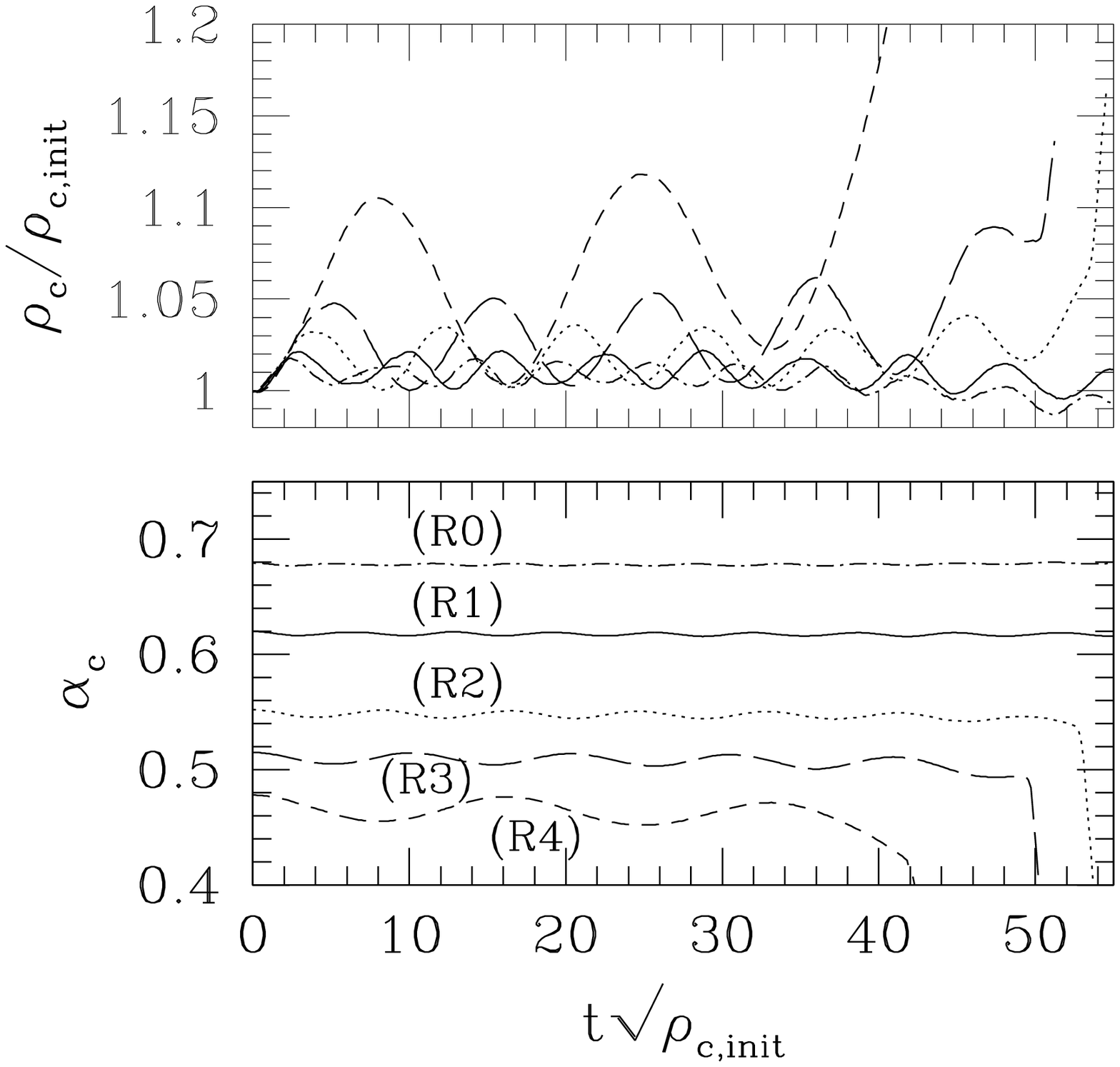}
\epsfxsize=2.6in
\leavevmode
~~~(b)\epsffile{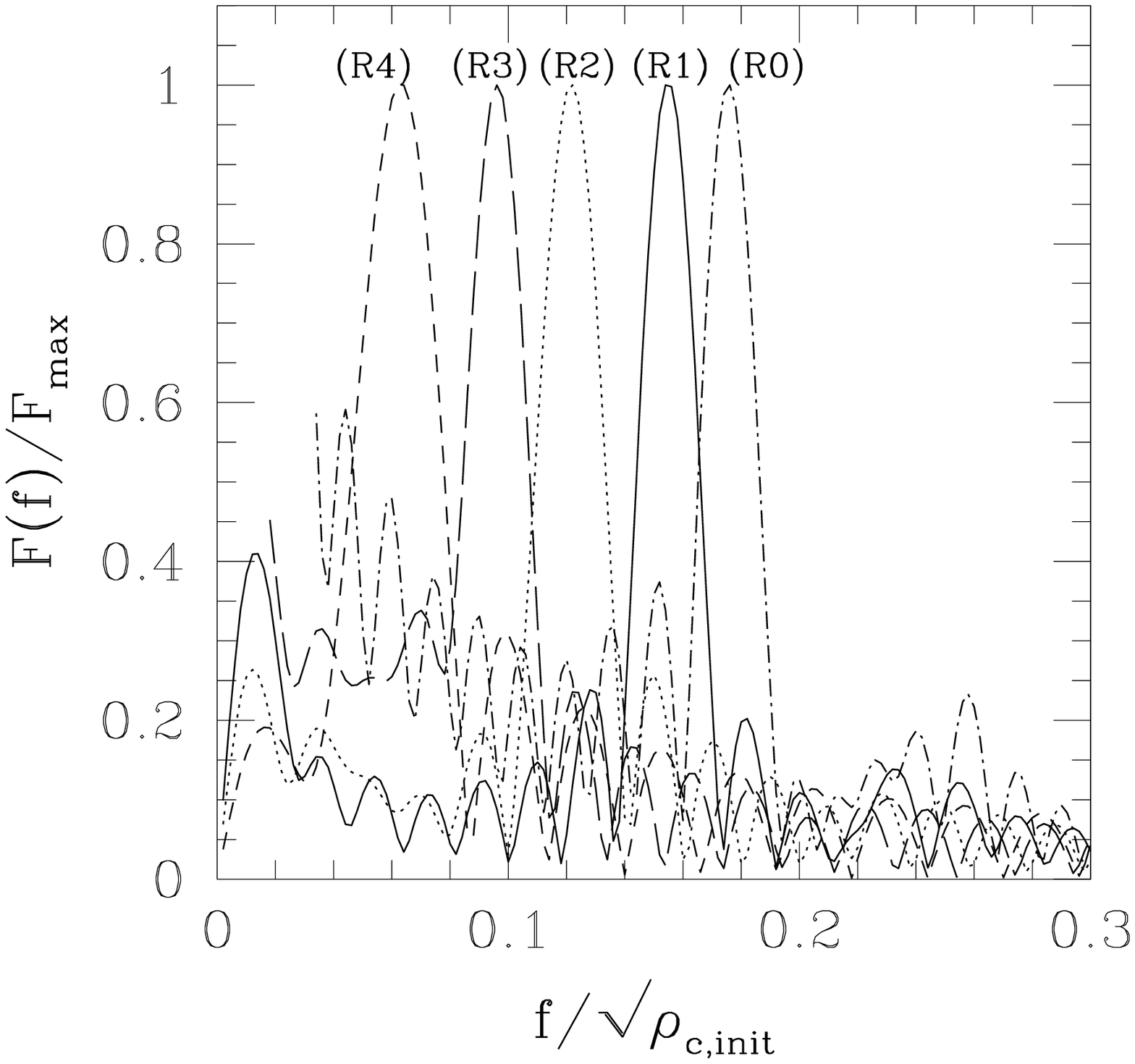}
\caption{(a)
Time evolution of central density and lapse function at origin 
for stable rotating stars
(R0) (dotted-dashed curves), (R1) (solid curves), 
(R2) (dotted curves), (R3) (long-dashed curves), 
and (R4) (dashed curves).
The simulations were performed with 
$N=180$. With these grid numbers, the equatorial
radius is covered by about 60 grid numbers. 
(b) Fourier spectra of $\rho_c(t)$ for 
(R0) (dotted-dashed curves), (R1) (solid curves), 
(R2) (dotted curves), (R3) (long-dashed curves), 
and (R4) (dashed curves).
\label{FIG6} }
\end{center}
%\vspace{-5mm}
\end{figure}

The long-term simulations were also performed for models 
(R0), (R2), (R3), and (R4). In all these simulations,
we initially reduced the pressure by 0.5\% and
take $N=180$. With this grid number, 
the polar (equatorial) radius is covered by 35 (60) grid points. 
The time 
evolution of the central density and central value of the lapse function 
($\alpha_c$) are 
shown together in Fig. \ref{FIG6}(a). As in the spherical case, 
the duration of the simulation is shorter for more compact stars. 
(The moment of the crash of a run is identified with the time 
at which $\alpha_c$ sharply drops.) 
This might be evidence that with our spatial gauge, 
coordinate distortion is accumulated too much  
for the long-term simulation, because 
it is likely to be accumulated more rapidly for more compact stars.
As in the simulation for (S4),
a high-mass rotating star (R4), for which the ADM mass is
$\approx 99.5$\% of the maximum, collapses to a black hole in a 
few dynamical time scales instead of crashing, because of a slight
increase of the baryon rest-mass due to numerical error.
It is necessary to take $N > 180$ to continue computations of 
such high-mass stars for more than two oscillation periods. 
However, as long as the neutron star is not very close to 
the marginally stable point, 
the simulation can be continued for more than five
dynamical time scales with $N \sim 200$, and 
this duration is long enough to produce 
scientific results for most problems.
For example, from these simulations, we can extract the 
frequency of fundamental quasi-radial oscillation modes. 
In Fig. \ref{FIG6}(b), we show the Fourier spectra of the central density 
as in the case of Sec. IV A. The figure indicates clear peaks 
that denote the fundamental frequency of 
the quasiradial oscillation modes. 

\begin{figure}[htb]
\vspace*{-4mm}
\begin{center}
\epsfxsize=2.4in
\leavevmode
(a)\epsffile{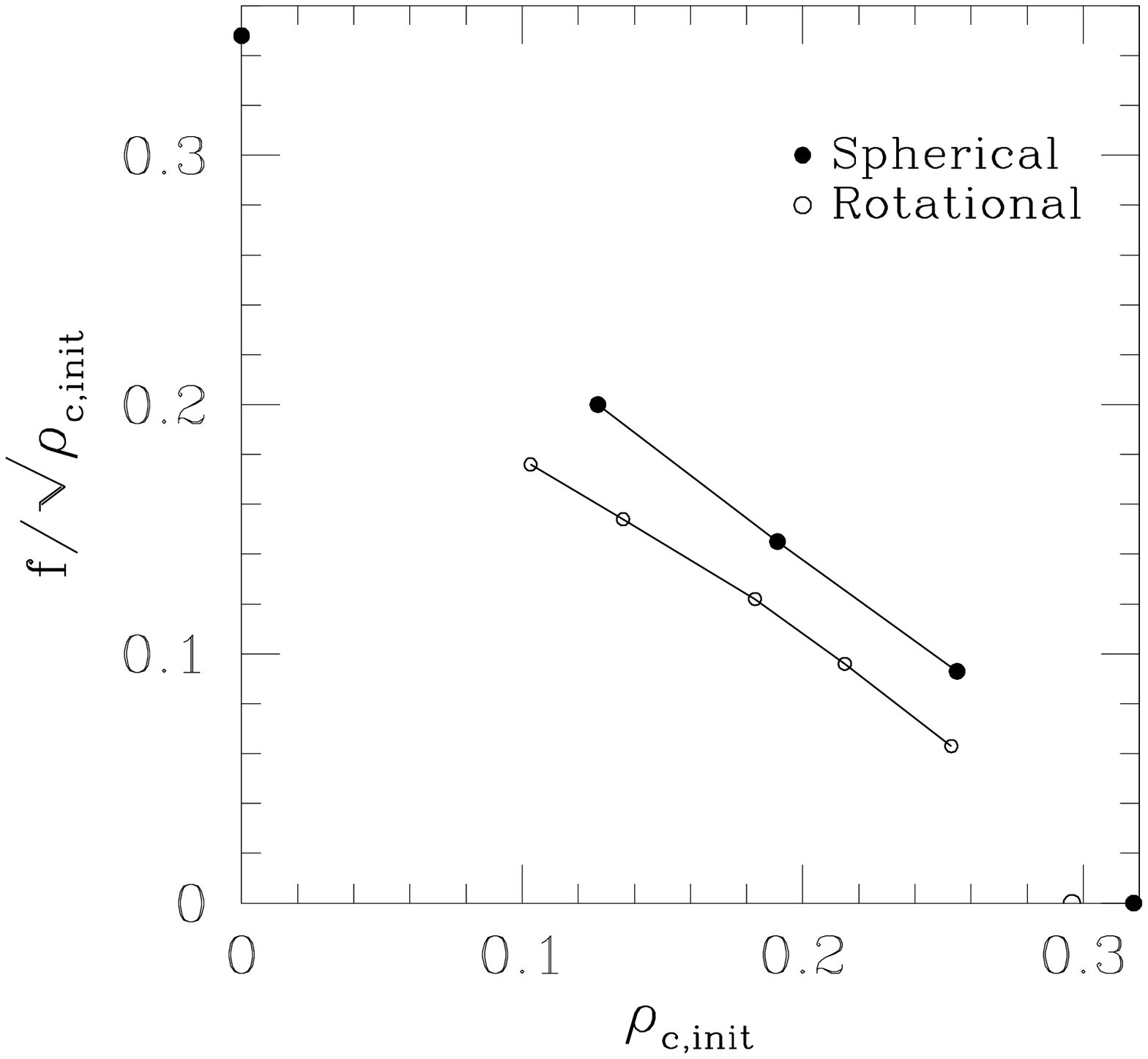}
\epsfxsize=2.4in
\leavevmode
~~~(b)\epsffile{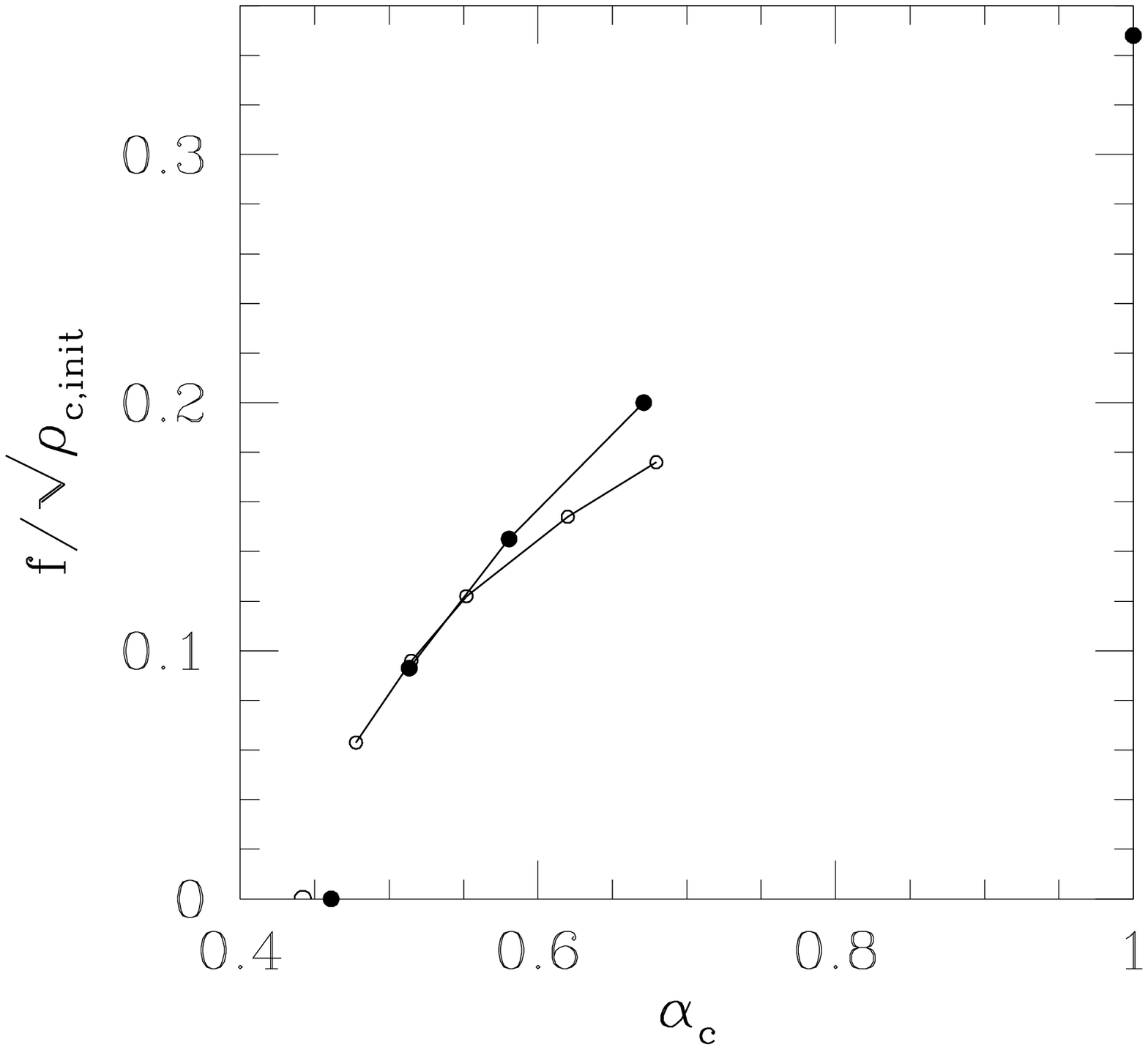}
\caption{Frequency of fundamental radial and quasiradial
oscillation modes for spherical stars (filled circles)
and rotating stars at mass shedding limits (open circles) (a) as 
a function of the central density and (b) as a function of 
the lapse function at origin. 
The frequency at $\rho_{c,{\rm init}}=0$ for the spherical star is 
derived in a Newtonian analysis. 
The filled and open circles along the horizontal axis are plotted 
according to the fact that the oscillation 
frequency of the marginally stable star is zero. 
\label{FIG7}}
\end{center}
%\vspace{-5mm}
\end{figure}

In Fig. \ref{FIG7}, we summarize the frequencies of
the radial and quasiradial
oscillation modes for spherical and rapidly rotating 
neutron stars with $\Gamma=2$. The filled and open circles
denote the numerical results for spherical and rotating
neutron stars, respectively. The filled circle along the vertical axis
is plotted according to
a Newtonian analysis for the spherical polytrope \cite{cox}.
On the other hand, the filled and open circles along the horizontal axis
are plotted by the fact that the frequency of the
fundamental radial and quasiradial oscillations of the marginally stable
stars is zero.
We note that the frequency in dimensional units is computed from
\beq
f \approx 9485~ {\rm Hz} (P_{\rm osc} \rho_{c,{\rm init}}^{1/2} )^{-1}
\biggl( {K_{\rm P} \over 2\times 10^5~{\rm cgs} } \biggr)^{-1/2}
\biggl( {\rho_{c,{\rm init}} \over 0.3} \biggr)^{1/2}, 
\eeq
where $P_{\rm osc}=1/f$ is the oscillation period of the
fundamental quasiradial mode. 

As mentioned in Sec. IV A, the frequencies for spherical neutron stars 
agree well with semianalytical results \cite{chandra}. 
For rotating neutron stars, the frequency is slightly smaller
than that for spherical neutron stars for identical $\rho_c$.
A similar tendency is found in the results of \cite{other}. 
It is interesting to note that 
for the identical value of the central lapse function,
the frequencies approximately coincide for $\alpha_c \alt 0.6$. 

\begin{figure}[hbt]
\vspace*{-4mm}
\begin{center}
\epsfxsize=2.6in
\leavevmode
(a)\epsffile{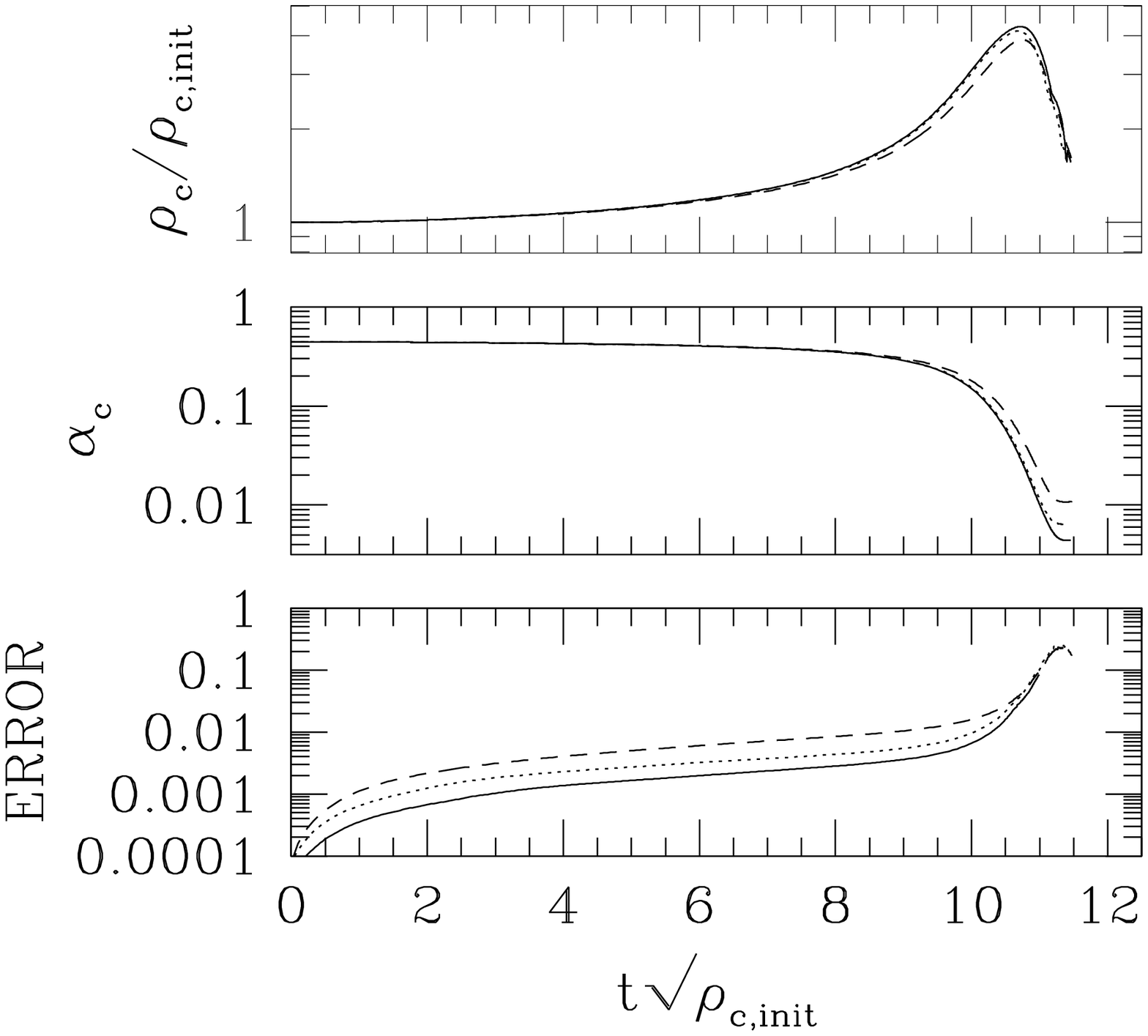}
\epsfxsize=2.6in
\leavevmode
~~~(b)\epsffile{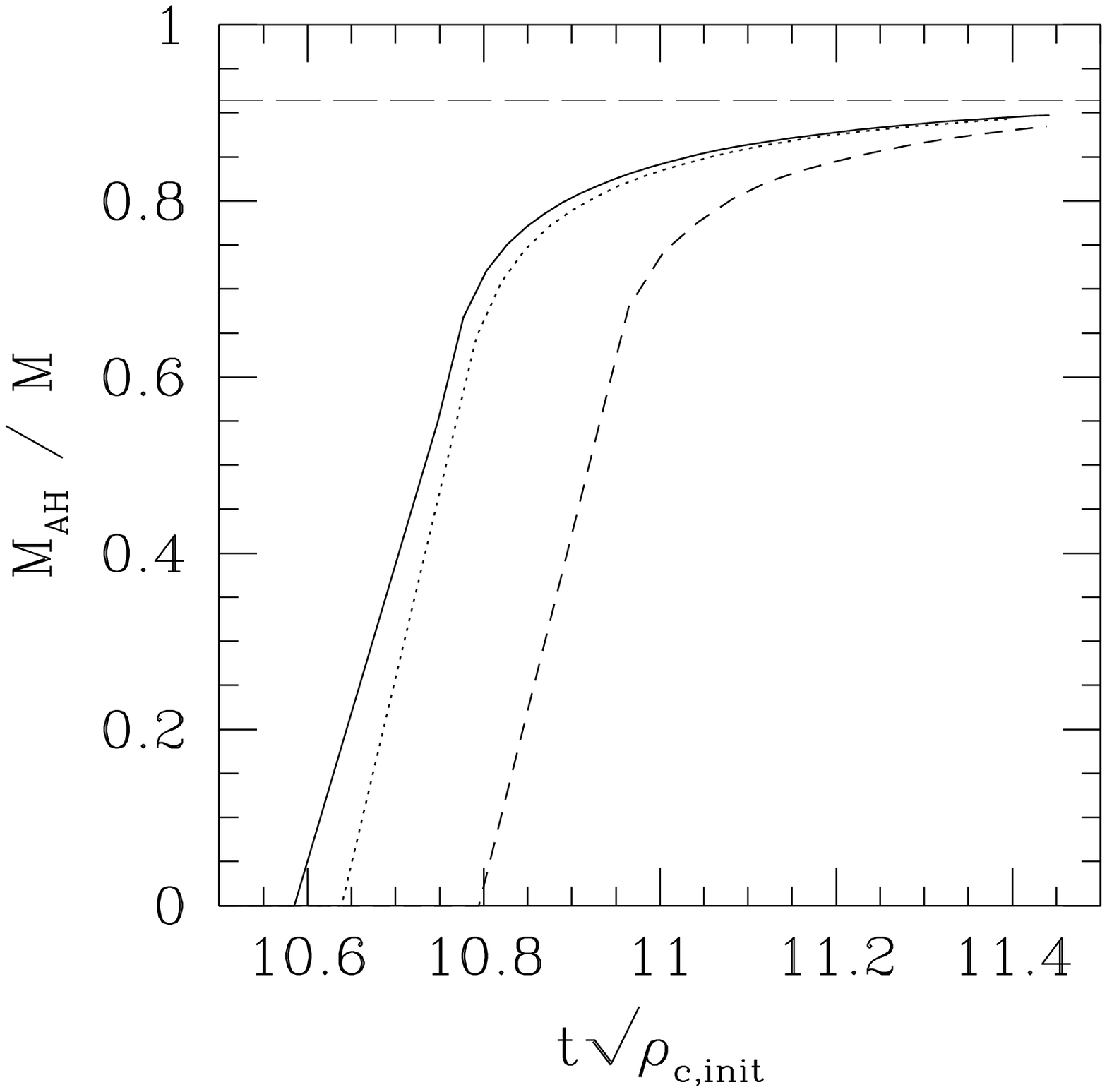}
\caption{
(a) Time evolution of central density, lapse function at origin, 
and averaged violation of the Hamiltonian constraint, and 
(b) mass of the apparent horizon as a function of time 
for collapse of an unstable and rapidly rotating neutron star 
at the mass shedding limit (R5). 
In both figures, the solid, dotted, and dashed 
curves denote the results with $N=240$, 180, and 120. 
With these grid numbers, the polar (equatorial)
radius is initially covered by about 
69 (120), 52 (90), and 35 (60) grid points. 
\label{FIG8} }
\end{center}
%\vspace{-1mm}
\end{figure}

In Fig. \ref{FIG8}(a), 
we display the central density, central value of the 
lapse function, and averaged violation of the
Hamiltonian constraint for collapse of an unstable 
rotating neutron star (R5). To induce the collapse, we initially 
reduced the pressure by 0.5\%. 
We also carried out the simulations with the reduced factor of 0.2\%, 
and have found that the star (R5) collapses also in this case, 
although it takes longer to be a black hole. 
Numerical results are shown for $N=240$, 180, and 120 
and demonstrate that the convergence is achieved well. 
With these grid numbers, the polar (equatorial) 
radius is initially covered by about 
69 (120), 52 (90), and 35 (60), respectively. 

During the collapse, the density (lapse function) monotonically 
increases (decreases) with time, and finally 
a black hole is formed (i.e., the apparent horizon is located)
in the late time when $\alpha_c \alt 0.03$. 
As in the spherical case, the reason that $\rho_c$ decreases in the late
stage of the collapse is that a small error in $\phi$ 
that is of $O(1)$ at the formation of the black hole leads to a large
error in $\rho$ that is computed from $\rho_*/(w e^{6\phi})$. 
It is also found that at the time when the magnitude of 
the averaged violation of 
the Hamiltonian constraint becomes $\sim 0.2$, 
the computation crashed due to the grid stretching 
around the horizon of a black hole, and that 
the magnitude of ERROR at the crash is almost independent of 
the grid resolution. 

In Fig. \ref{FIG8}(b), we show the time evolution of the mass of
the apparent horizon defined by Eq. (\ref{MAH}). 
Here, the long-dashed horizontal line denotes the expected final value 
$\sqrt{S_{\rm EH}/(16\pi M^2)}$, where 
$S_{\rm EH}$ is derived from the formula for the area of the event horizon
of a Kerr black hole as \cite{ST} 
\beq
S_{\rm EH} = 8\pi M^2 \biggl( 1+\sqrt{1-{J^2\over M^4}}\biggr),  
\eeq
where $M$ and $J$ are the ADM mass and angular momentum of
the collapsing neutron star. Since almost all the matter
eventually falls into a black hole in this simulation,
the area of the apparent horizon should settle down to $S_{\rm EH}$. 
The figure indicates that the area of the apparent horizon 
asymptotically approaches the expected value. This demonstrates 
that the spacetime in the final phase of our simulation almost 
relaxes to a stationary, Kerr black hole spacetime.

%%%%%%%%%%%%%%%%%%%%%%%%
%%%%%%%%%%%%%%%%%%%%%%%%

\subsection{Collapse of rotating stars with
parametric equations of state} 

The purpose of this subsection is to demonstrate 
that with our implementation, it is feasible to carry out 
stable and accurate simulations for the collapse of 
rotating stars with parametric equations of state (\ref{EOSII})
that are more realistic for high-density matter than 
the $\Gamma$-law equation of state used 
in the previous two subsections. Since 
the equation of state for high-density matter is still 
not precisely known, the parametric equation of state 
(\ref{EOSII}) is used 
for several choices of $\Gamma_1$ and $\Gamma_2$. 
Initial conditions are set up adopting the polytropic equation of state 
(\ref{EOS43}) with $\Gamma=4/3$. 

In the realistic core collapse of massive stars, 
the central density just before the collapse is of order 
$10^{10}~{\rm g/cm^3}$ \cite{HD,SNinit}.
Since the collapse leads to the formation 
of a neutron star of density of order $10^{15}~{\rm g/cm^3}$ 
or a black hole, the characteristic
length scale changes by a factor of $\sim 100$. 
This implies that we need to take $N$ of $O(10^3)$
for a well-resolved simulation in the fixed uniform grid. 
Although it is possible to take a large value of $N$ as several thousands, 
performing such large-scale simulation is not computationally 
inexpensive even in the axisymmetric case. Since the main purpose 
of this subsection is not to present scientific results, but
both to demonstrate that realistic equations of state can be adopted 
in our implementation and to grasp characteristic behaviors 
associated with new implementation with 
such equations of state, here we pick up more compact stars 
of central density $\approx 6\times 10^{12}~{\rm g/cm^3}$ 
as initial conditions to save the computational costs as a first step. 
In the next subsection, we will 
show a numerical result with a more realistic initial data set
as illustration. 

Velocity profiles of equilibrium rotating stars used 
as initial conditions are given 
according to a popular relation \cite{KEH,Ster}, 
\beq
u^t u_{\varphi} = \varpi_d^2( \Omega_a - \Omega ), 
\eeq
where $\Omega_a$ denotes the angular velocity
along the $z$ axis, and $\varpi_d$ is a constant.
In the Newtonian limit, the rotational profile is written as 
\beq
\Omega = \Omega_a{\varpi_d^2 \over \varpi^2 + \varpi_d^2}. 
\eeq
Thus, $\varpi_d$ indicates the steepness of differential rotation.
In this paper, we pick up the rigidly rotating cases in which
$\varpi_d \rightarrow \infty$ and a differentially rotating case 
in which $\varpi_d/R_e=1/2$, 
where $R_e$ is the equatorial coordinate radius. 
In both cases, we choose the axial ratio of polar radius to
equatorial radius as $2/3$. 
In Table III, we list several quantities for 
the models we adopt in the numerical computation. 
We refer to these models as (C1) and (C2). 

\begin{table}[t]
\begin{center}
\begin{tabular}{|c|c|c|c|c|c|c|c|} \hline
& $\varpi_d/R_e$ & $\rho_c (10^{12} {\rm g/cm^3})$
& $M_*(M_{\odot})$ & $M(M_{\odot})$
& $R_e$ (km) & $J/M^2$ & $|T/W|$ \\ \hline
C1 & $\infty$ & 6.02 & 1.347 & 1.343 & 265 & 0.434 &8.38d-3  \\ \hline 
C2 & 1/2      & 6.24 & 1.465 & 1.465 & 231 & 0.888 &3.51d-2  \\ \hline
\end{tabular}
\caption{The central density, baryon rest-mass, ADM mass, equatorial
radius $R_e$, $J/M^2$, and $|T/W|$ of initial conditions for the simulations 
of stellar collapse in Sec IV C. 
}
\end{center}
\vspace{-5mm}
\end{table}

The simulations were carried out for 
$(\Gamma_1,\Gamma_2)=(1.325,2)$ (a), (1.3,2) (b), (1.325,2.5) (c), 
and (1.3, 2.5) (d). In the following, we specify our choice 
of $\Gamma_1$ and $\Gamma_2$ in terms of (a), (b), (c), and (d).
For example, if we pick up model (C1) as the initial condition and
choose $\Gamma_1=1.325$ and $\Gamma_2=2$, 
we refer to this model as (C1a). 
In Table IV, we list masses and central density 
for spherical neutron stars of maximum mass 
in the parametric, cold equation of states (\ref{P12EOS}) with 
four choices of $\Gamma_1$ and $\Gamma_2$, which are
obtained by numerically solving the TOV equation \cite{ST}.
For $\Gamma_1=1.3$ and $\Gamma_2=2$, the maximum mass is
too small to be a realistic value, but with this model,
we can study qualitative properties of stellar core 
collapses in an extremely soft equation of state. 

\begin{table}[t]
\begin{center}
\begin{tabular}{|c|c|c|c|c|c|c|} \hline
& $\Gamma_1$ & $\Gamma_2$ & $M_*(M_{\odot})$
& $M(M_{\odot})$ & $M_{*{\rm core}}(M_{\odot})$
& $\rho_c({\rm g/cm^3})$ \\ \hline
a  & 1.325 & 2   & 1.425 & 1.363 & 1.362 & 2.68d15  \\ \hline
b  & 1.3   & 2   & 0.991 & 0.925 & 0.980 & 6.01d15  \\ \hline
c  & 1.325 & 2.5 & 2.259 & 2.056 & 2.183 & 1.69d15  \\ \hline
d  & 1.3   & 2.5 & 1.810 & 1.600 & 1.792 & 2.87d15  \\ \hline
\end{tabular}
\caption{Maximum baryon rest-mass, ADM mass,
core rest-mass $M_{*{\rm core}}$, and density at the maximum 
for spherical stars of cold, parametric
equations of state (\ref{P12EOS}),  
with four types of $\Gamma_1$ and $\Gamma_2$. 
}
\end{center}
\vspace{-5mm}
\end{table}

According to a numerical result \cite{BS}, 
rigidly rotating stars in equilibrium with $\Gamma=4/3$ are 
unconditionally unstable against gravitational collapse, 
if the compactness $M/R_e$ is larger than $\sim 1/700$. 
Thus, the equilibrium state of model (C1) is unstable. 
The criterion of the instability has not been established yet
for differentially rotating stars. Thus, the stability is
not clear for (C2). However, 
since the compactness is much larger than 1/700, 
the initial equilibrium of (C2) is also likely to be unstable. 
Hence, by decreasing $\Gamma$ from $4/3$ to $\Gamma_1 < 4/3$ at $t=0$,
the collapse is accelerated. 
To investigate convergence, the simulations were carried out with 
$N=600$, 400, and 300 for all the models. 
In these simulations, the equatorial radius 
was initially covered by $N$ grid points. 

\begin{figure}[htb]
\vspace*{-4mm}
\begin{center}
\epsfxsize=2.2in
\leavevmode
\epsffile{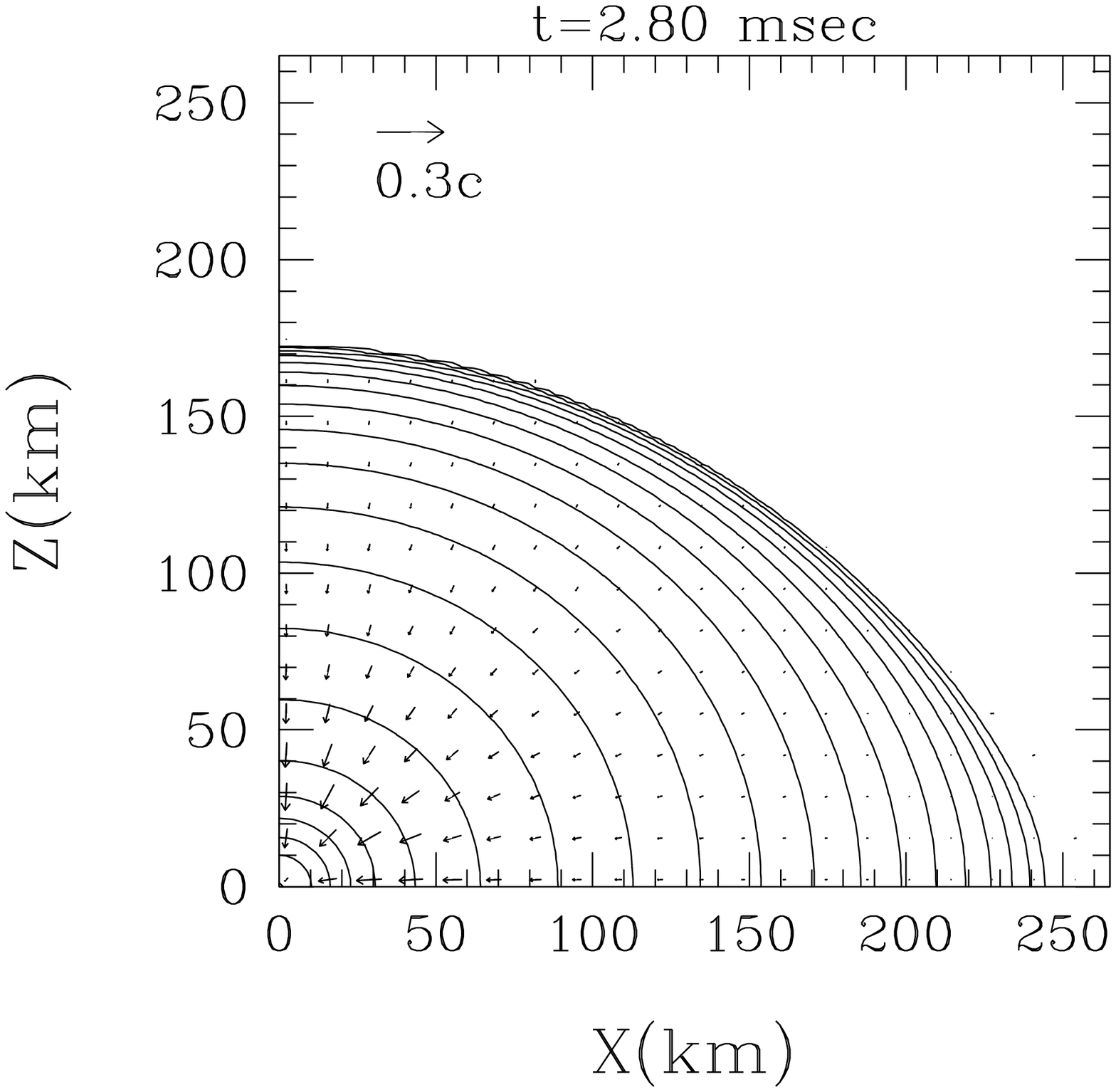}
\epsfxsize=2.2in
\leavevmode
\epsffile{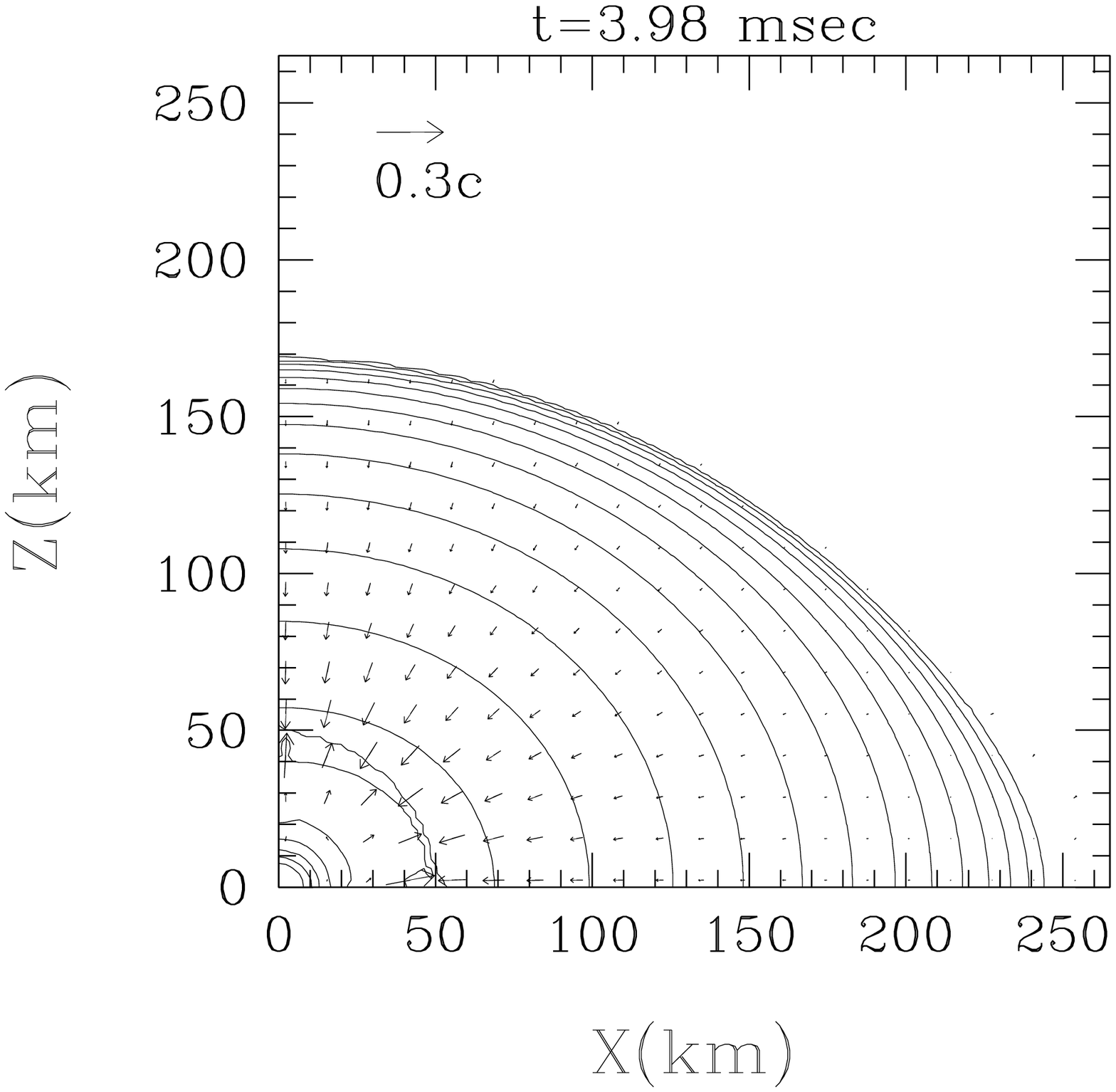}
\epsfxsize=2.2in
\leavevmode
\epsffile{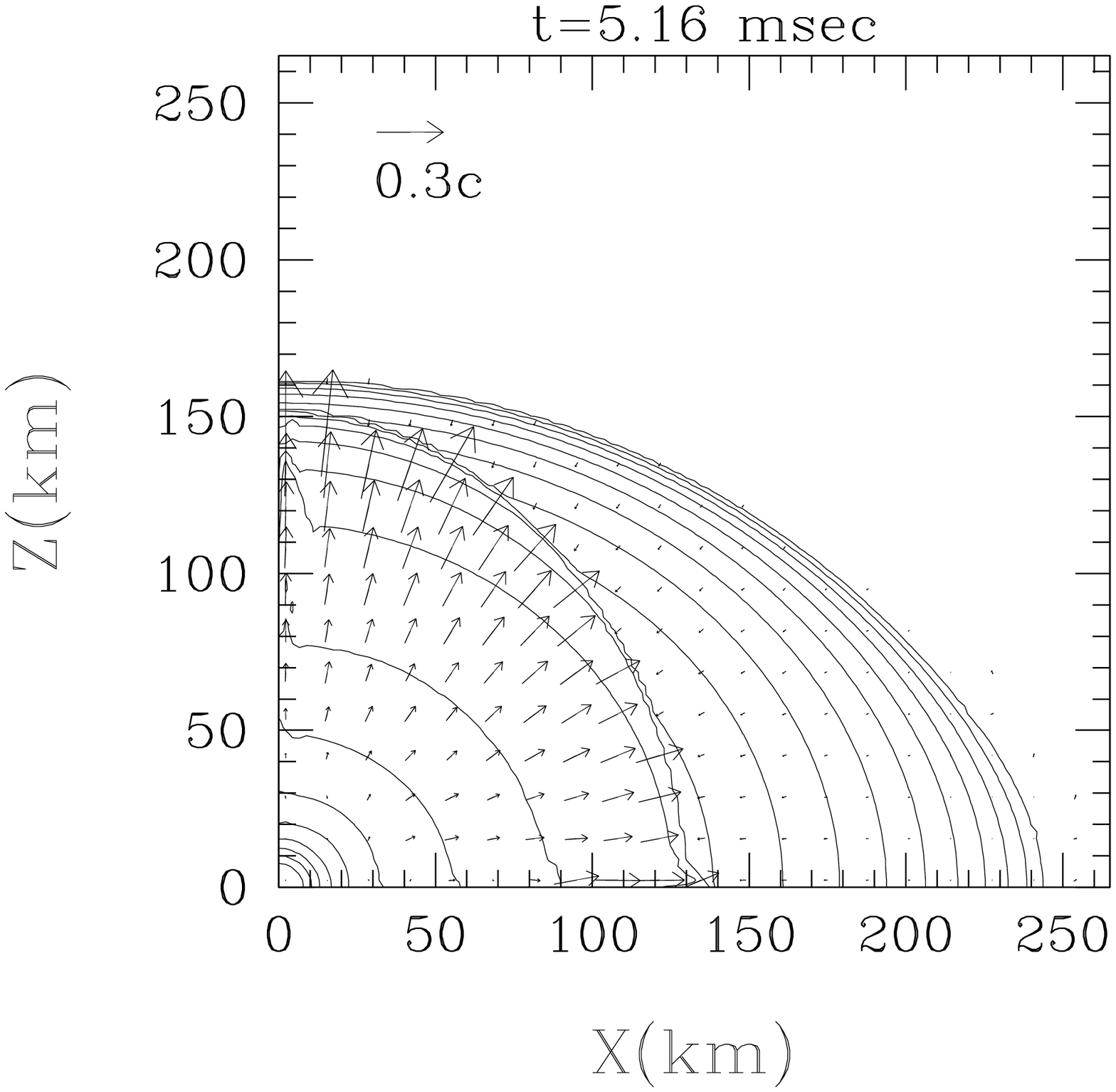}
\caption{
Density contour curves of $\rho$ and velocity fields of $v^A$
for model (C1a) at selected time steps.
The contour curves are drawn for $\rho/\rho_{\rm nuc}=10^{-0.5j}$, 
for $j=0,1,2,\cdots,20$. 
\label{FIG9}}
\end{center}

\end{figure}
\begin{figure}[htb]
\vspace*{-4mm}
\begin{center}
\epsfxsize=2.2in
\leavevmode
\epsffile{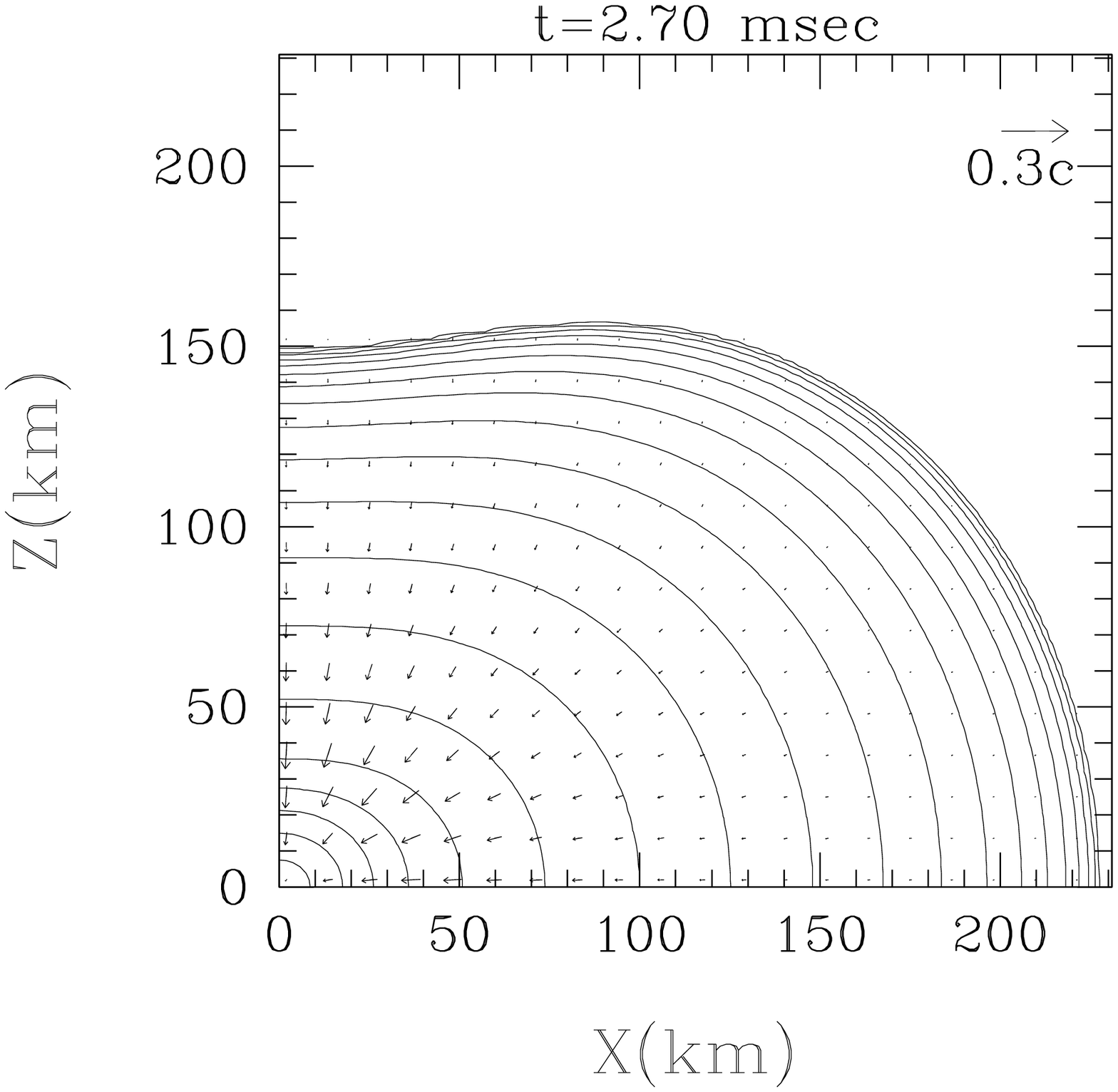}
\epsfxsize=2.2in
\leavevmode
\epsffile{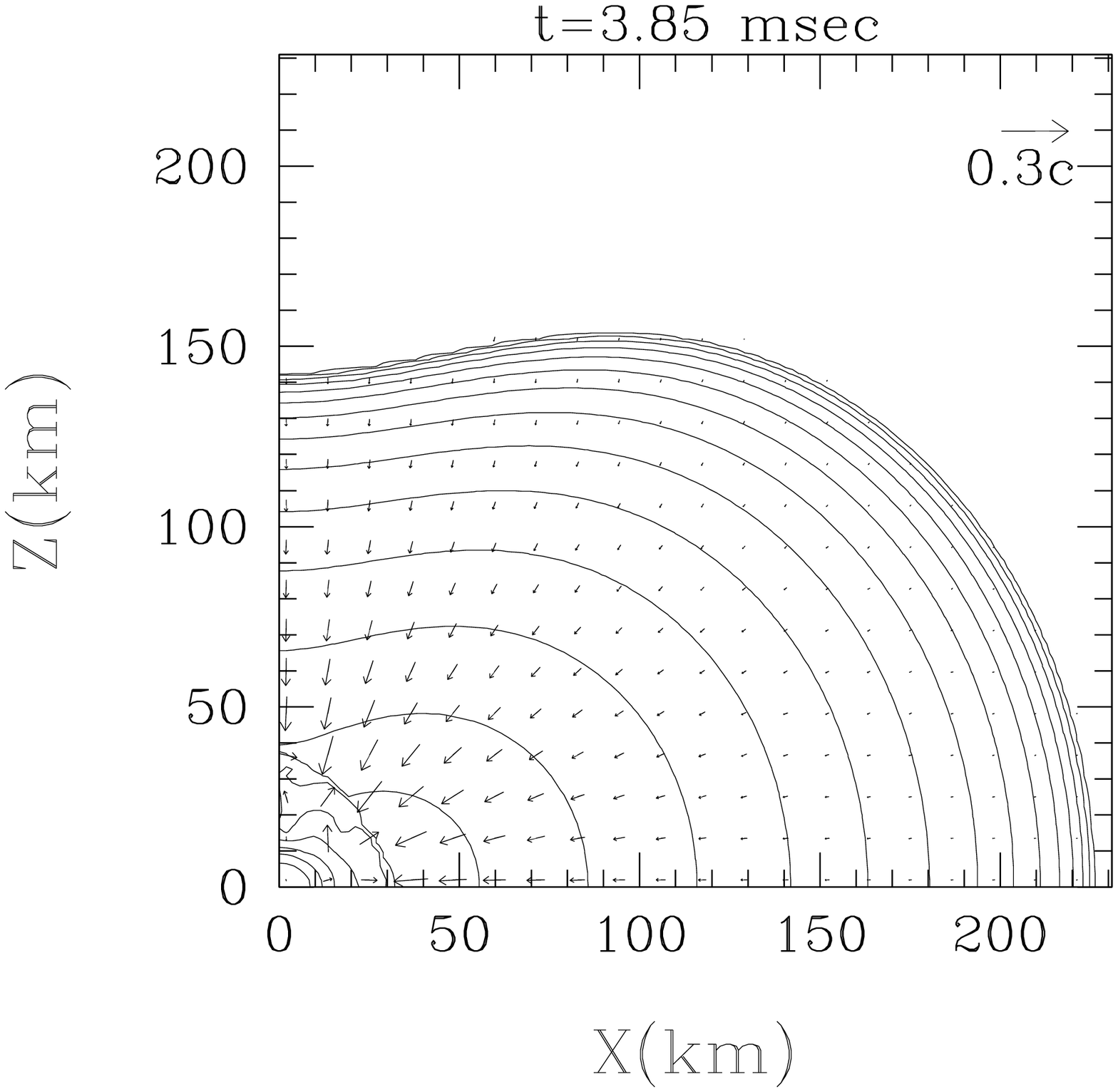}
\epsfxsize=2.2in
\leavevmode
\epsffile{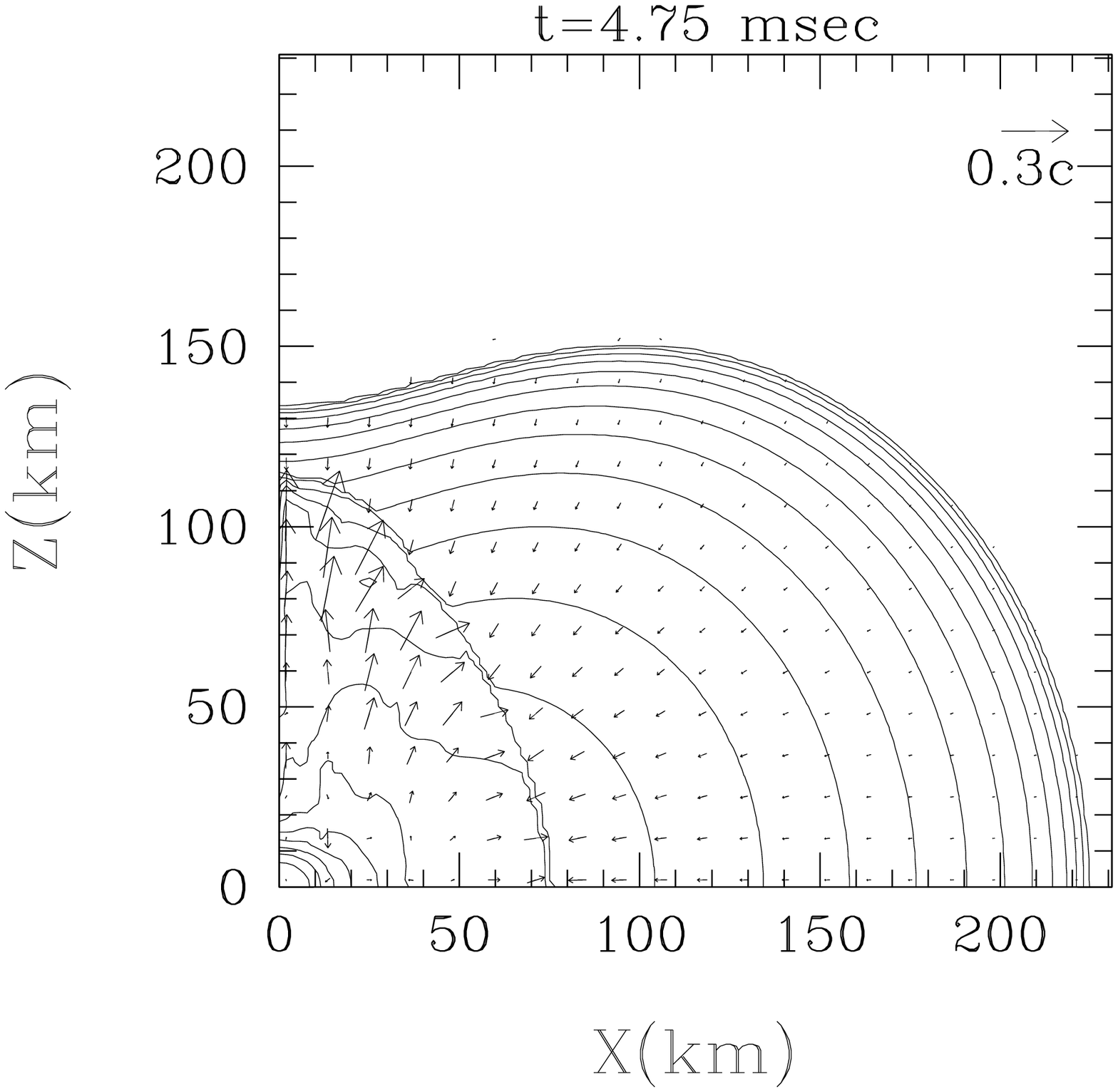}
\caption{
The same as Fig. \ref{FIG9} but for model (C2a). 
\label{FIG10}}
\end{center}
\end{figure}

\begin{figure}[htb]
\vspace*{-4mm}
\begin{center}
\epsfxsize=2.6in
\leavevmode
(a)\epsffile{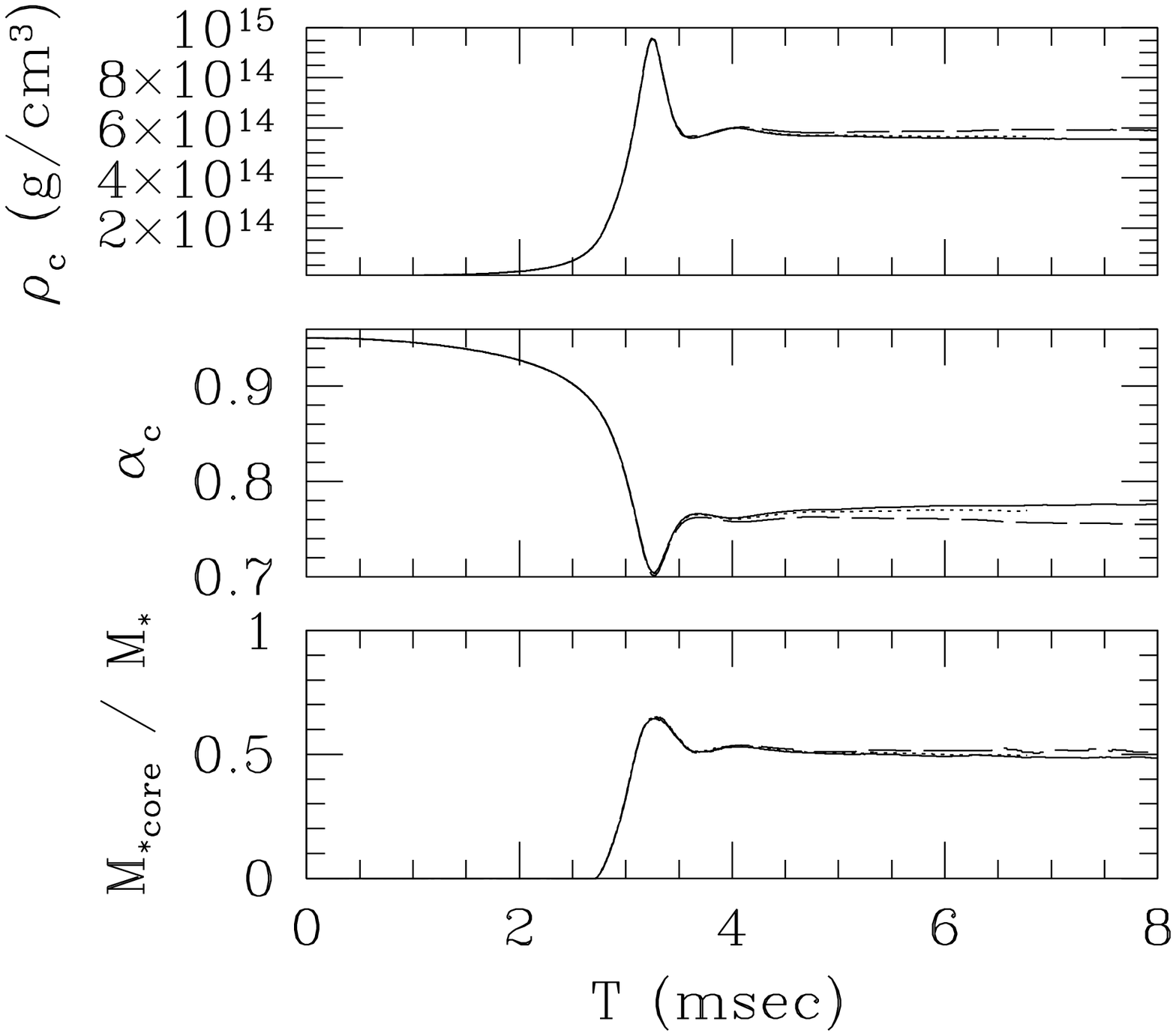}
\epsfxsize=2.6in
\leavevmode
~~~(b)\epsffile{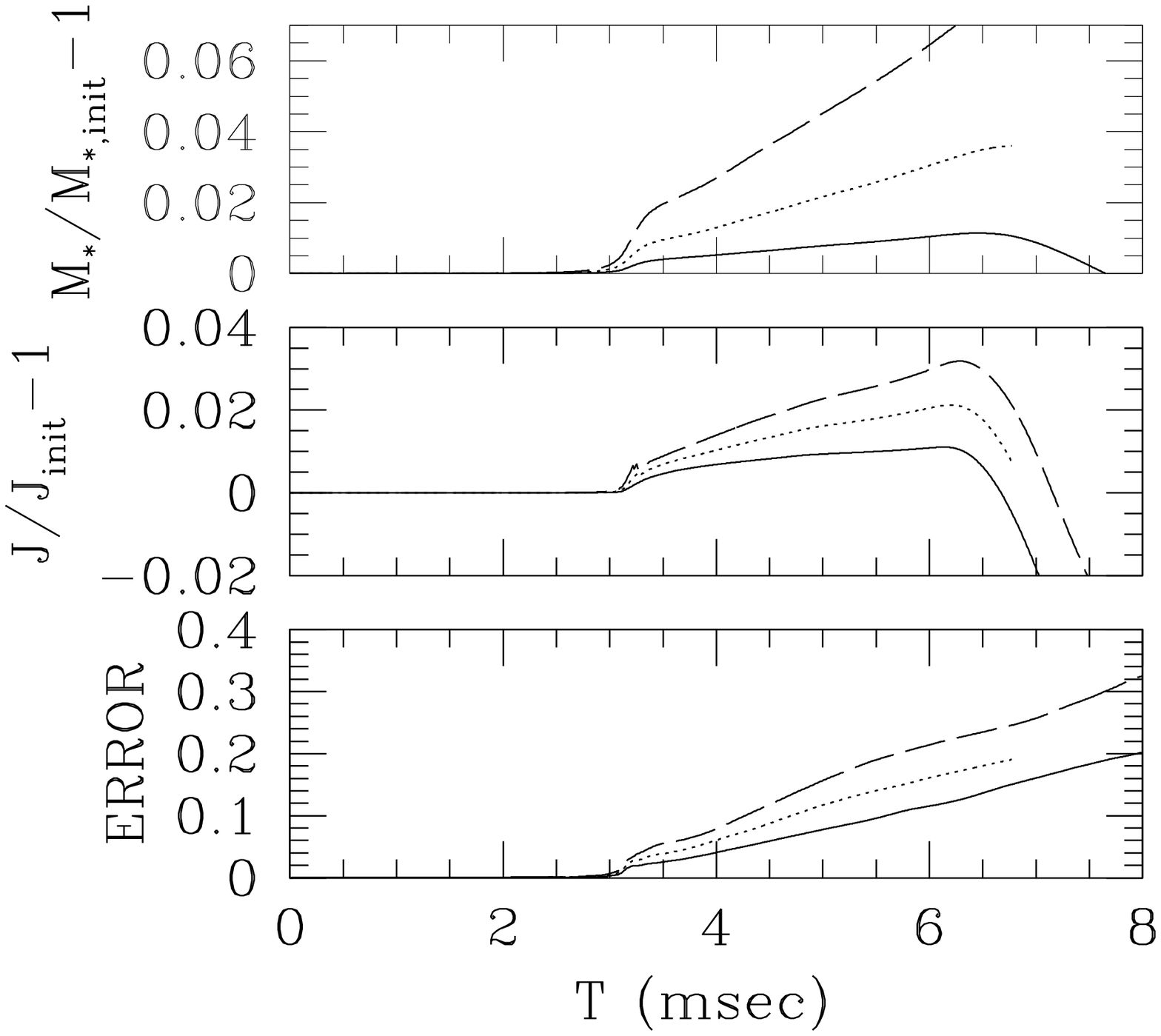}
\caption{
(a) Time evolution of central density, lapse function at origin, 
and fraction of baryon rest-mass for $\rho > \rho_{\rm nuc}$, and 
(b) time evolution of violation of rest-mass conservation,
angular momentum conservation, and averaged violation of the
Hamiltonian constraint for collapse of a star (C1a). 
In both figures, the solid, dotted, and dashed 
curves denote the results with $N=600$, 400, and 300. 
With these grid numbers,
%$\Delta x/M=0.2213$, 0.3319 and 0.4425.
$\Delta x=0.442$, 0.663, and 0.884 km.
For $t > 6.5$ msec, shocks reach outer boundaries
of the computational domain, and matter starts escaping. 
\label{FIG11}
}
\end{center}
%\vspace{-1mm}
\end{figure}

\begin{figure}[htb]
\vspace*{-4mm}
\begin{center}
\epsfxsize=2.6in
\leavevmode
(a)\epsffile{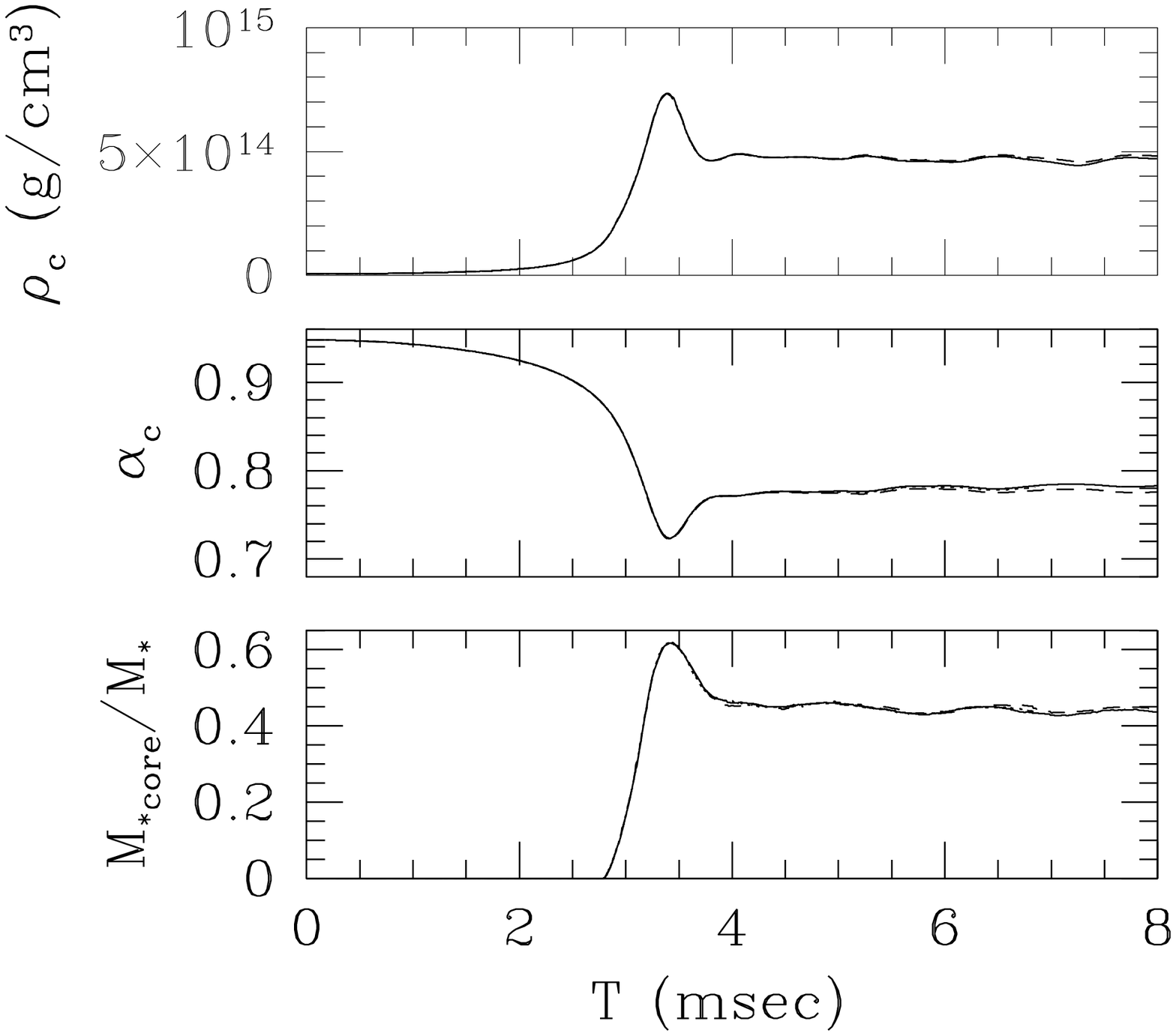}
\epsfxsize=2.6in
\leavevmode
~~~(b)\epsffile{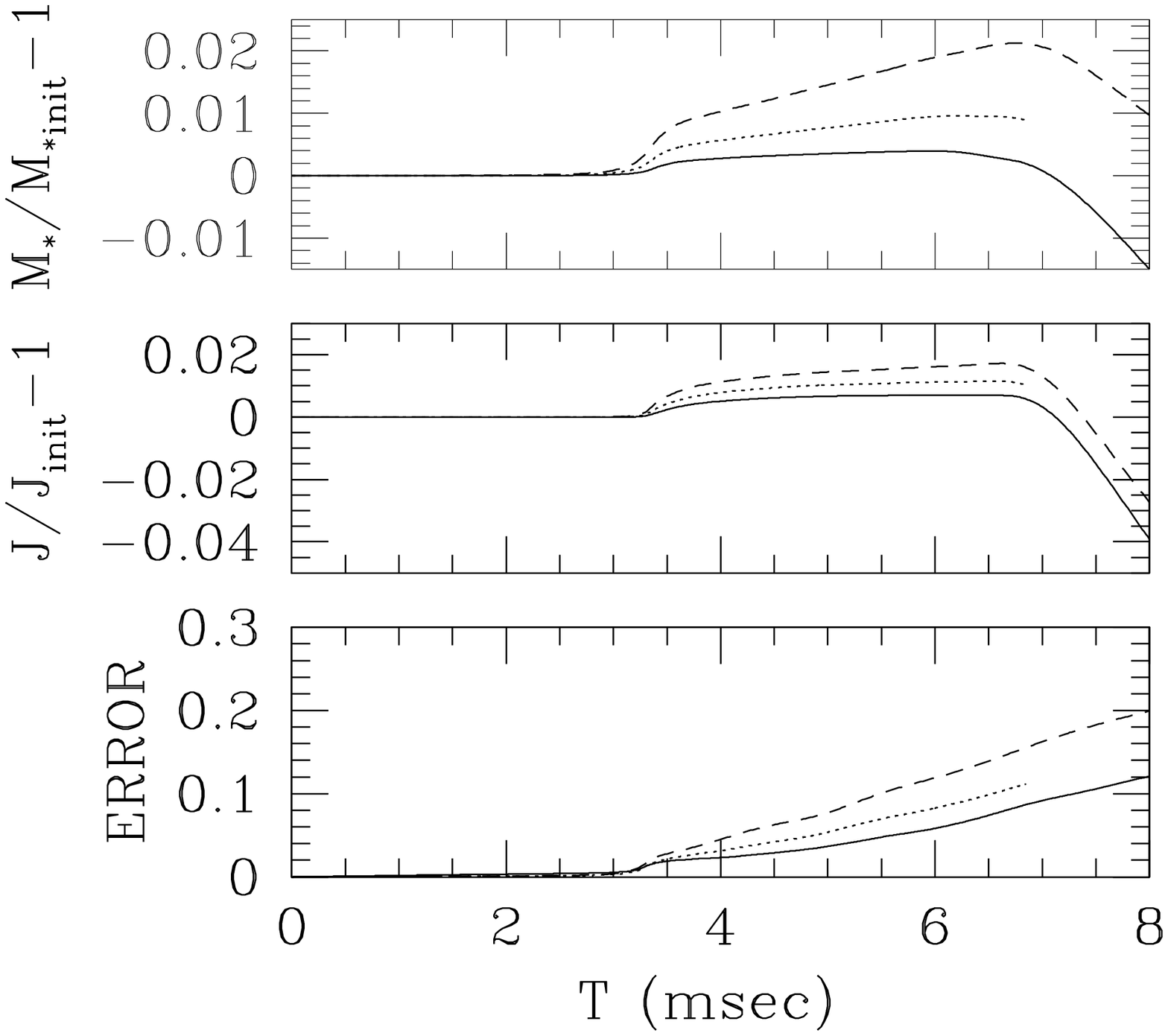}
\caption{
The same as Fig. \ref{FIG11} but 
for collapse of a star (C2a). 
In both figures, the solid, dotted, and dashed 
curves denote the results with $N=600$, 400, and 300. 
With these grid numbers,
%$\Delta x/M=$0.1777, 0.2665 and 0.3553. 
$\Delta x=$0.385, 0.577, and 0.770 km. 
For $t > 7$ msec, shocks reach outer boundaries 
of the computational domain, and matter starts escaping. 
\label{FIG12}}
\end{center}
%\vspace{-1mm}
\end{figure}

In Figs. \ref{FIG9} and \ref{FIG10},
we display snapshots of the density contour curves and
velocity fields at 
selected time steps for models (C1a) and (C2a) as examples. 
It is found that model (C1) has a spheroidal structure initially,
while model (C2) has a slightly toroidal shape due to the effect of 
differential rotation. In Figs. \ref{FIG11} and \ref{FIG12}, 
we also show the time evolution of several quantities for 
models (C1a) and (C2a).
Here $M_{*{\rm core}}$ denotes the total baryon rest-mass of
high-density matter with $\rho > \rho_{\rm nuc}$. 
As indicated in these figures, the collapses proceed in the following 
manner. In the early phase in which the central density is smaller 
than $\rho_{\rm nuc}$, the central region of the
star contracts approximately in 
a homologous manner because the adiabatic index is close to 4/3 
\cite{GoldW}. After the central density exceeds $\rho_{\rm nuc}$, 
a core is formed at the central region, and 
the mass gradually increases as a result of subsequent accretion. 
At the time when the central density becomes $\sim 4.5 \rho_{\rm nuc}$
for (C1a) and $\sim 3.5 \rho_{\rm nuc}$ for (C2a), 
the increase of the core mass stops, and 
strong shocks rapidly propagate outward
due to the restoring force of the core. 
After this moment, the core settles down toward an approximately stationary 
state, and finally the central density relaxes to 
$\sim 3 \rho_{\rm nuc}$ for (C1a) and $\sim 2.5 \rho_{\rm nuc}$ for (C2a). 
On the other hand, the shocks propagate outward, sweeping the 
infalling matter. In both cases (C1a) and (C2a), the 
baryon rest-mass of $\rho \geq \rho_{\rm nuc}$ is 
$\sim 0.65M_{\odot}$ after shock propagates far from the core. 
Since the angular momentum at $t=0$ for (C2a)
is larger than that for (C1a), 
the baryon rest-mass of the core and central density of the final
state of the core are slightly smaller for model (C2a). 
These facts indicate that the products after collapse depend basically on 
the equation of state, but are modified by the magnitude and
distribution of angular momentum 
that are initially retained by a precollapse star.
If the initial angular momentum is much larger than those
for (C1a) and (C2a), the collapse would be halted before 
the central density reaches $\rho_{\rm nuc}$, 
and as a result, a neutron star would not be formed \cite{HD}. 

Another noticeable feature is that for the differentially rotating 
initial condition, a small modulation is found 
in the time evolution of the central density and $\alpha_c$ 
after a quasistatic core is formed. 
This is likely because in the collapse with 
a differentially rotating initial condition, the collapsing
star deviates highly from spherical symmetry and 
shocks are formed in a nonspherical manner. As a result,
the nonspherical oscillation is excited in the formed core. 
Similar results are observed in \cite{HD}. 

In Figs. \ref{FIG11}(b) and \ref{FIG12}(b),
we display the violation of the baryon rest-mass conservation, 
angular momentum conservation, and averaged violation of 
the Hamiltonian constraint. First, we note that 
at $t \sim 6.5$ msec, the shocks reach the outer boundaries, and 
the matter escapes from the computational domain.  
This causes the rapid angular momentum decrease for $t \agt 6.5$ msec. 
Besides this, the errors converge to zero with improving the grid resolution, 
as we saw in Secs. IV A and IV B. 
However, there are several noticeable properties that have 
not been seen in the previous subsections. 
One is that the magnitude of the errors quickly increases 
when shocks are generated. The second is that the 
averaged violation of the Hamiltonian constraint converges
to zero at about first-order (not at second-order). 
We cannot explain the reason for them correctly. However, 
we deduce it as follows: 
At the shocks, the hydrodynamic quantities are discontinuous, and 
as a result, the second partial derivatives of the metric are 
discontinuous and the complete analyticity for the metric is violated.
This may change the global convergence property from second-order to 
first-order. 

The magnitude of the errors for (C2a) is smaller than that for
(C1a) with identical grid spacing. This is a consequence of 
the fact that the central density for (C2a)
after shock formation is smaller than that for (C1a), and
as a result, the grid resolution for (C2a) is better than that for (C1a). 

\begin{figure}[htb]
\vspace*{-4mm}
\begin{center}
\epsfxsize=2.6in
\leavevmode
(a) \epsffile{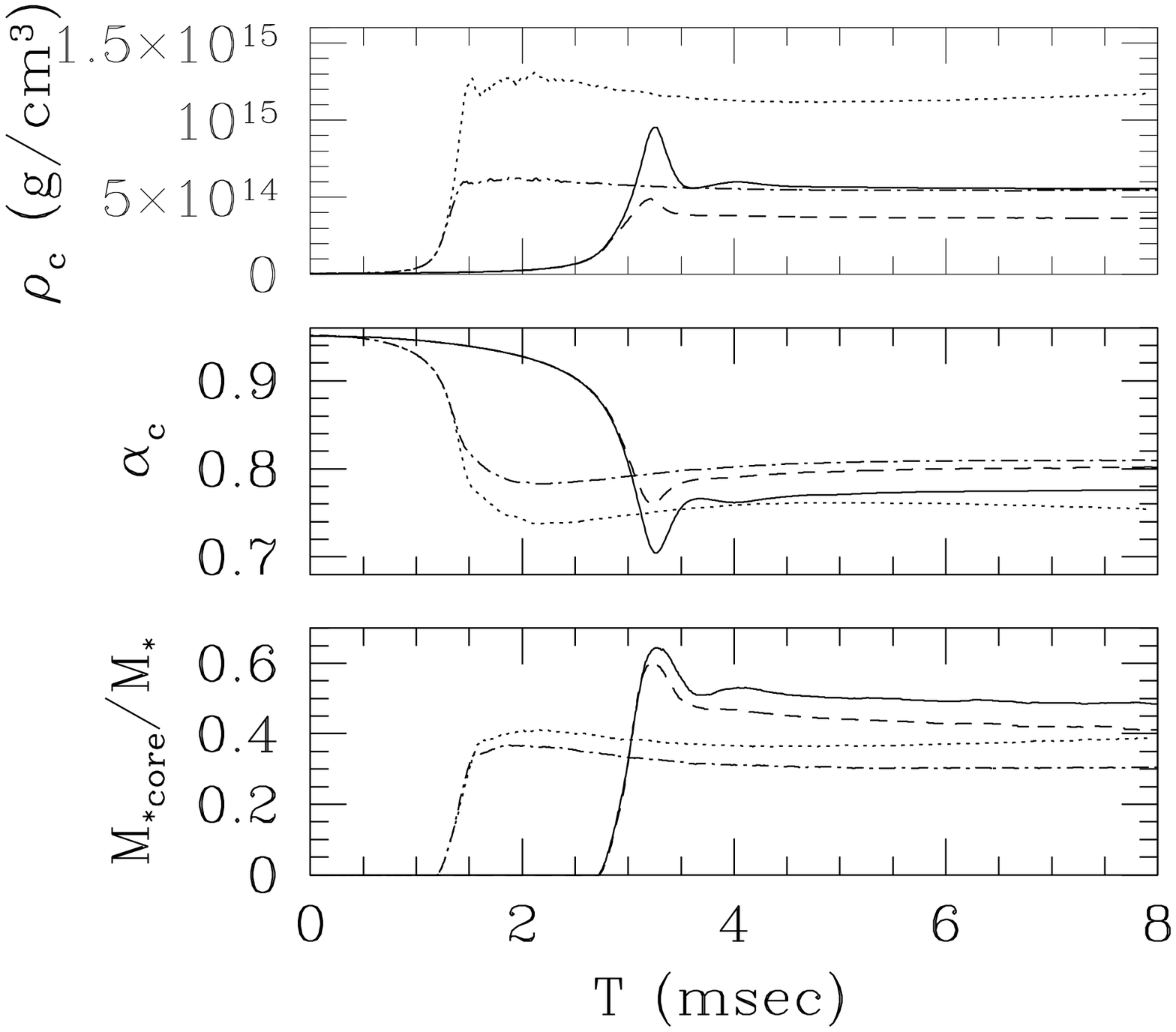}
\epsfxsize=2.6in
\leavevmode
~~~(b) \epsffile{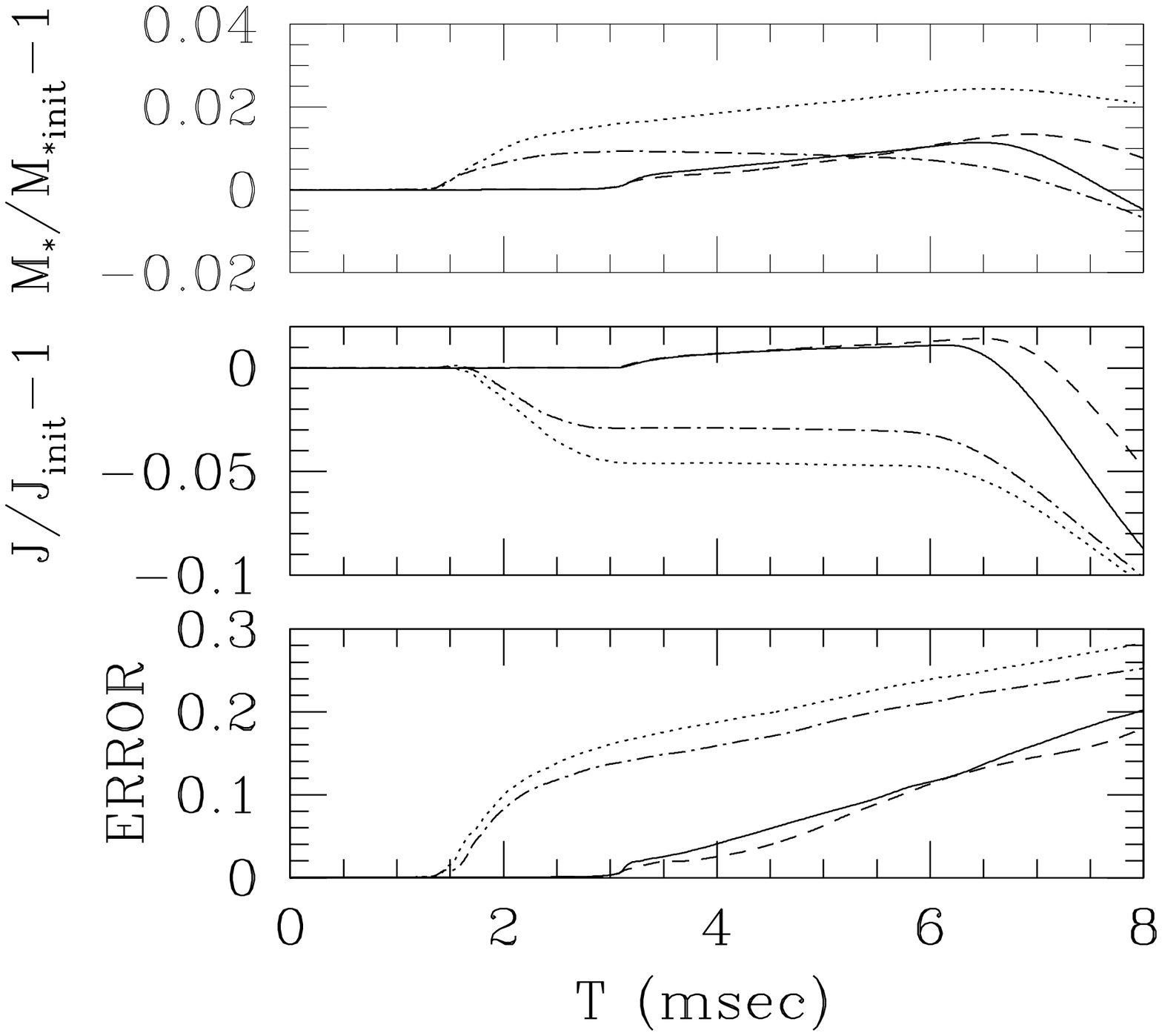}
\caption{
(a) Time evolution of central density, lapse function at origin, 
and fraction of baryon rest-mass for $\rho > \rho_{\rm nuc}$, and 
(b) time evolution of violation of rest-mass conservation,
angular momentum conservation, and averaged violation of the 
Hamiltonian constraint
for (C1a) (solid curves), (C1b) (dotted curves), (C1c) (dashed curves), 
and (C1d) (dotted-dashed curves). 
We adopt $N=600$, by which the equatorial radius is initially
covered by $N$ grid points for all the models. 
\label{FIG13}}
\end{center}
\end{figure}

To see the effects of the equations of state on the dynamics 
of collapse and on accumulation of the numerical errors, 
we display numerical results for (C1a), (C1b), (C1c), and (C1d)
in Fig. \ref{FIG13}. All the simulations were done 
with $N=600$, by which the equatorial radius is initially 
covered by $N$ grid points. 
Reflecting the difference of $\Gamma_1$ and $\Gamma_2$, the products
after the collapse and time evolution of the numerical errors 
are different. The following is a summary of the differences: 
(a) for $\Gamma_1=1.3$, the magnitude of the 
errors is larger than that for $\Gamma_1=1.325$, 
(b) for the identical value of
$\Gamma_1$, the central density and $M_{*{\rm core}}$
in the final state of the core are
smaller for the larger value of $\Gamma_2$, 
(c) $M_{*{\rm core}}$ is 
larger for the larger value of $\Gamma_1$ with the identical
value of $\Gamma_2$, 
(d) the magnitude of the error of the baryon rest-mass conservation always 
increases with time (never decreases in the absence of mass ejection from the
computational domain) irrespective of $\Gamma_1$ and $\Gamma_2$, and 
(e) the angular momentum increases due to 
the numerical error for $\Gamma_1=1.325$ but decreases for $\Gamma_1=1.3$. 
The reason for (a) is clear because for smaller values of $\Gamma_1$, the 
central density increases by a large factor, and as a result, 
the grid resolution becomes worse even in the identical grid spacing. 
The reason for (b) is also clear because with 
stiffer equations of state, the central density of neutron stars 
is smaller. The result (c) comes from the fact that with 
smaller values of $\Gamma_1$, shocks are stronger, and as a result, the 
amount of mass that is ejected outside the core is larger. 
However, the reason for (d) and (e) is not clear at all. 
These may be consequences of our choice of the finite 
differencing scheme for the hydrodynamic equations. 
Indeed, a similar tendency is found in the results of 
an approximate general relativistic simulation \cite{HD}. 

It is interesting to note that even for (C1b), in which 
the maximum allowed baryon rest-mass of neutron stars is 
very small ($\sim 1M_{\odot}$),
a low-mass neutron star of $M_{*{\rm core}} \sim
0.55 M_{\odot}$ is formed after the collapse. 
This is because the shocks explode the infalling matter sufficiently. 
Thus, even in the case that the stellar core mass before collapse 
is much larger than the maximum allowed mass of the neutron star for 
a given equation of state, a neutron star instead of 
a black hole could be formed at least temporarily, 
although such a neutron star could easily collapse to a 
black hole by subsequent accretion of matter or by a fall-back.

\begin{figure}[htb]
\vspace*{-4mm}
\begin{center}
\epsfxsize=2.4in
\leavevmode
(a)\epsffile{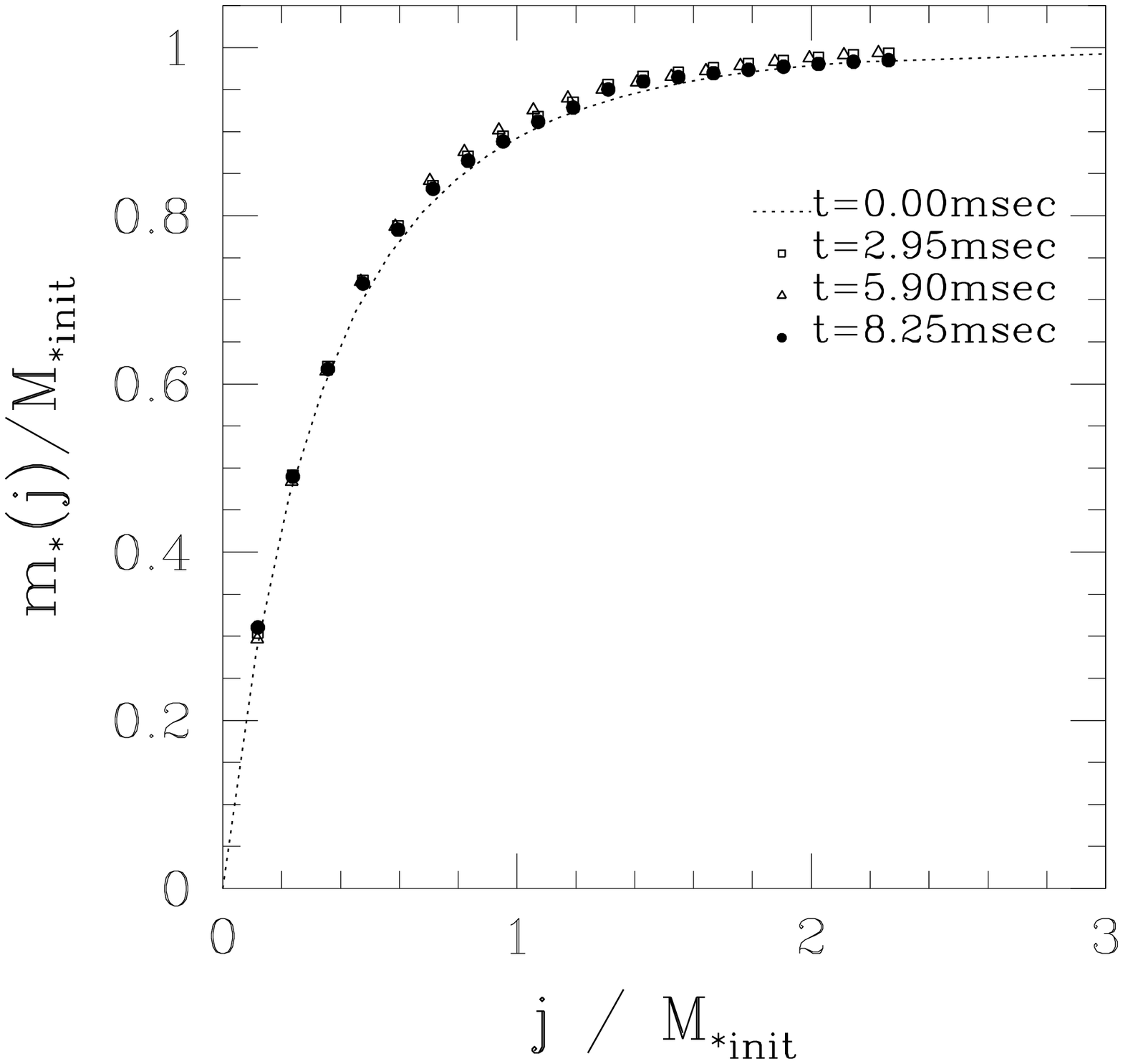}
\epsfxsize=2.4in
\leavevmode
~~~(b)\epsffile{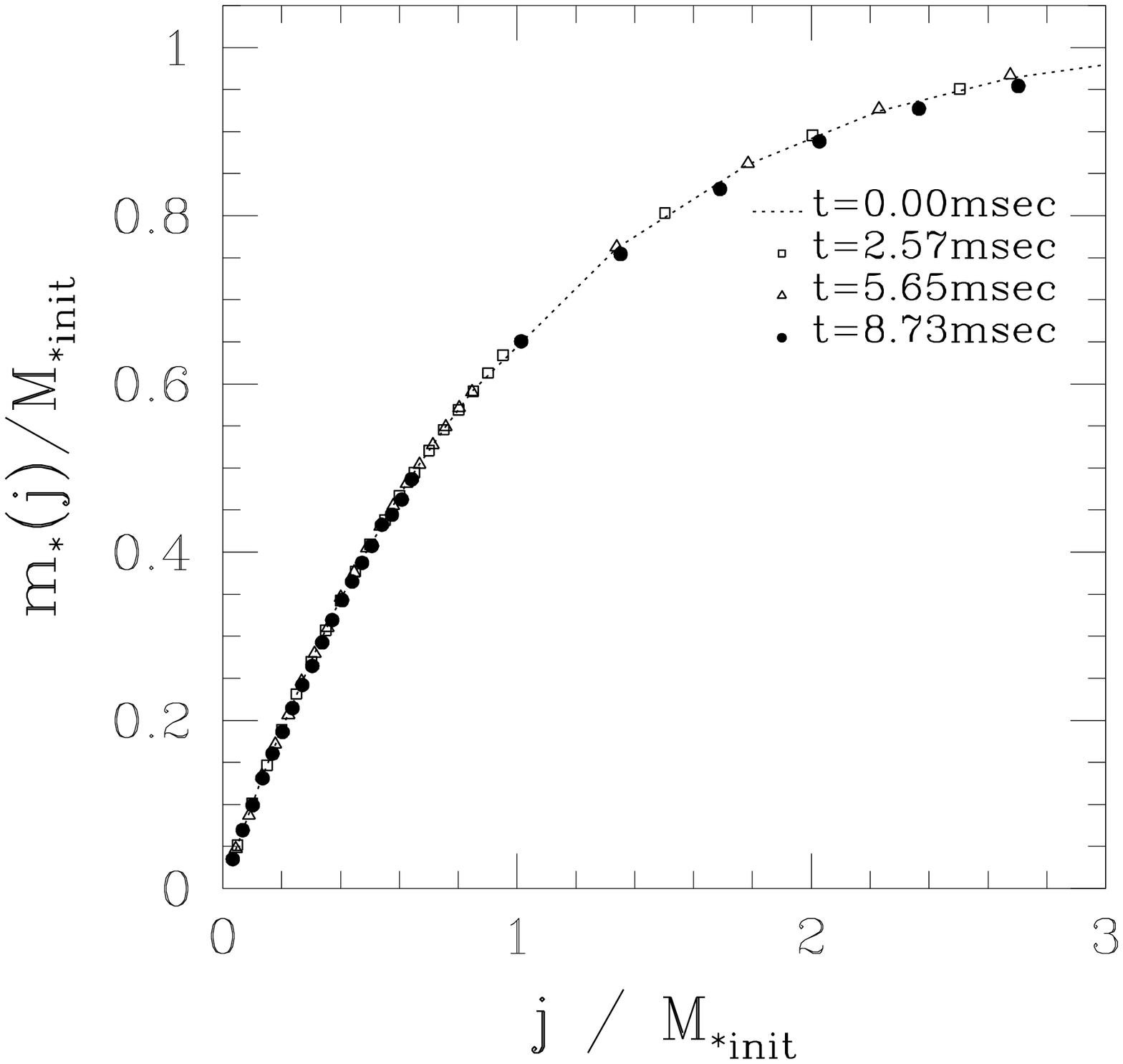}
\caption{
Specific angular momentum spectra at selected time steps
(a) for (C1d) and (b) for (C2a). 
\label{FIG14}}
\end{center}
%\vspace{-1mm}
\end{figure}

In Fig. \ref{FIG14}, the specific angular momentum spectra 
at selected time steps are shown for (C1d) and (C2a) as examples. 
In both cases, we take $N=600$. 
The figures demonstrate that the spectral shape is well 
conserved throughout the simulations irrespective of the 
initial angular velocity profile. Therefore, we conclude that 
angular momentum transfer due to numerical dissipation is 
sufficiently small in numerical computations.

\subsection{An example of stellar core collapse}

In this subsection, 
we present numerical results for a simulation of rotating
stellar core collapse that is started from a realistic 
initial condition with the central density $\sim 10^{10}~{\rm g/cm^3}$ 
and with $\Gamma=4/3$. 
We give a rigidly rotating equilibrium star that is 
close to the mass shedding limit as the initial condition. 
Several quantities of this rotating star are listed in Table V. 
Since it is rapidly rotating and its compactness is sufficiently small, 
this equilibrium star is dynamically stable against 
gravitational collapse even for 
the polytropic equation of state with $\Gamma=4/3$ \cite{BS}.
Thus, the collapse is triggered by the slight decrease of the
adiabatic constant from 4/3 to $\Gamma_1$.  
In the present simulation, we set $\Gamma_1=1.325$ and $\Gamma_2=2$. 

\begin{table}[t]
\begin{center}
\begin{tabular}{|c|c|c|c|c|c|c|} \hline
$\rho_c~({\rm g/cm^3})$ & $M_*(M_{\odot})$
& $M(M_{\odot})$ & $R$~(km) & $|T/W|$ & $J/M^2$ & $\alpha_c$ \\ \hline
$1.65 \times 10^{10}$  & 1.491 & 1.491 & 1910 &
$8.89 \times 10^{-3}$ & 1.136  & 0.993 \\ \hline
\end{tabular}
\caption{Central density, baryon rest-mass, ADM mass,
equatorial radius, ratio of the rotational kinetic energy to
potential energy, non-dimensional angular momentum parameter, and
central value of the lapse function 
of a rotating star chosen as an initial condition 
for a stellar core collapse simulation in Sec. IV D. 
}
\end{center}
\vspace{-5mm}
\end{table}

Since the characteristic length scale changes by a factor of $\sim 100$  
during the collapse, we performed the simulation changing the 
grid size and grid number as done in \cite{SS}. 
The grid size and computational domain were changed 
monitoring the value of the lapse function at the center $(\alpha_c)$, 
which approximately indicates the compactness of
the collapsing star. 
Whenever we carried out regridding, we made the grid spacing half
and used cubic interpolation \cite{recipe} 
for assigning the values of variables 
on the finer grids. The simulation was started 
with $N=500$, by which the equatorial radius is covered by 
480 grid points initially. At $t=0$, $\alpha_c \approx 0.993$. 
We carried out the first regridding when $\alpha_c$ was 0.975, at which
the mean radius of the collapsing star became $\sim 1/4$ of the initial one.
In this regridding, we chose $N=900$ and made the grid spacing half.
The next regridding was carried out when $\alpha_c=0.95$ and 0.90, and
we chose $N=1500$ and 2100, respectively. 
After $\alpha_c$ reached 0.90, we fixed $N$ and grid spacing. 
$L$ and $\Delta x$ in the final stage are about 1050 km and 0.5 km, 
respectively. For this simulation, 
the computational time was about 100 CPU hours for $\approx$
60000 time steps using eight processors of the FACOM VPP 5000 machine. 

Since the computational region was reduced whenever we carried out
the regridding, 
a small amount of mass that is outside the new computational domain 
was discarded. However, the magnitude of
the violation of mass and angular momentum
conservation is less than $\sim 0.5$\% and 2\%,
respectively [see Fig. \ref{FIG16}(b)]. This implies that 
the total amount of the discarded mass is comparable to that of
the numerical error associated with the finite differencing,
which does not much affect the evolution of the system, 
as indicated in Sec. IV C. 

\begin{figure}[t]
\vspace*{-4mm}
\begin{center}
\epsfxsize=2.2in
\leavevmode
\epsffile{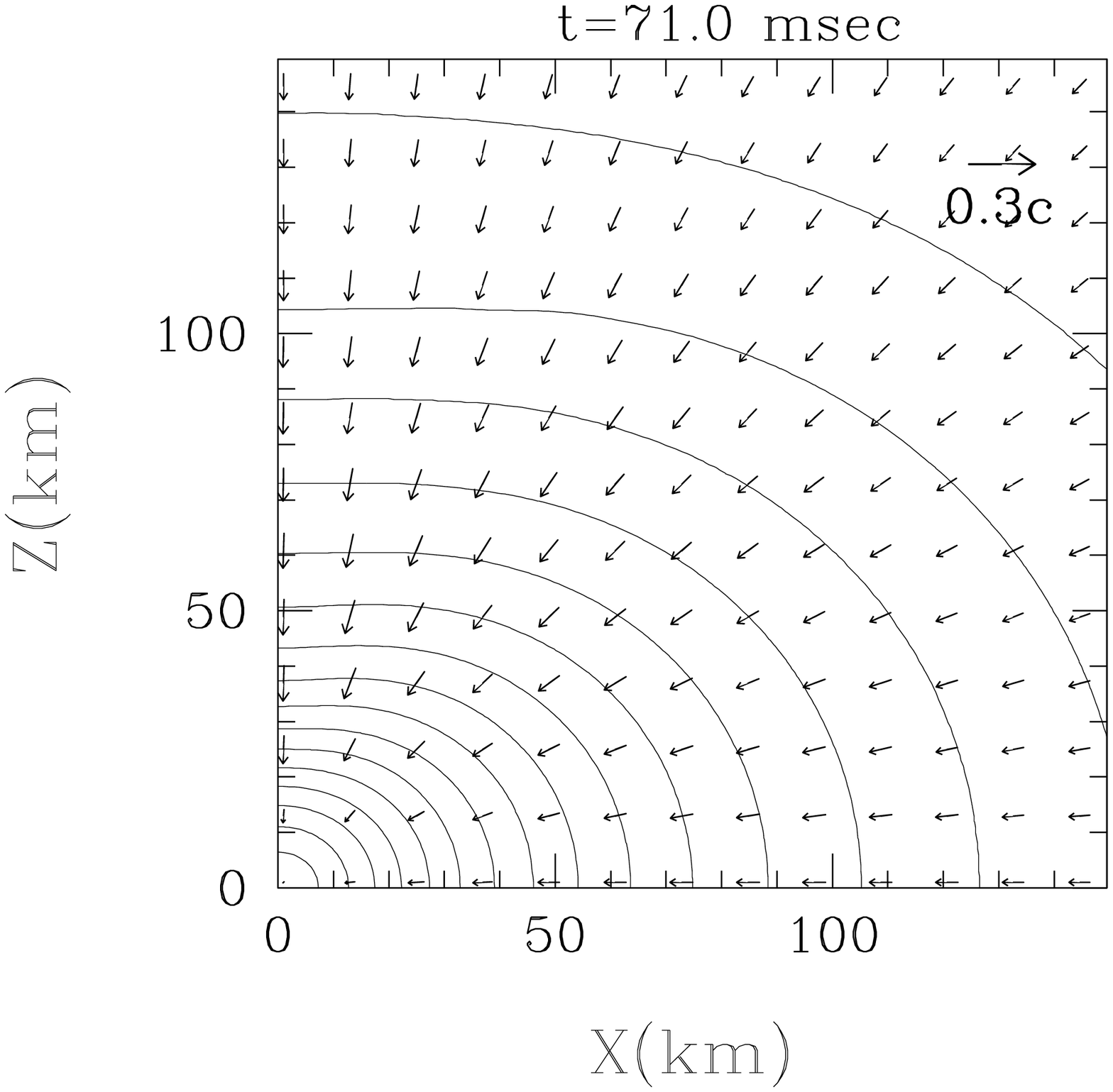}
\epsfxsize=2.2in
\leavevmode
\epsffile{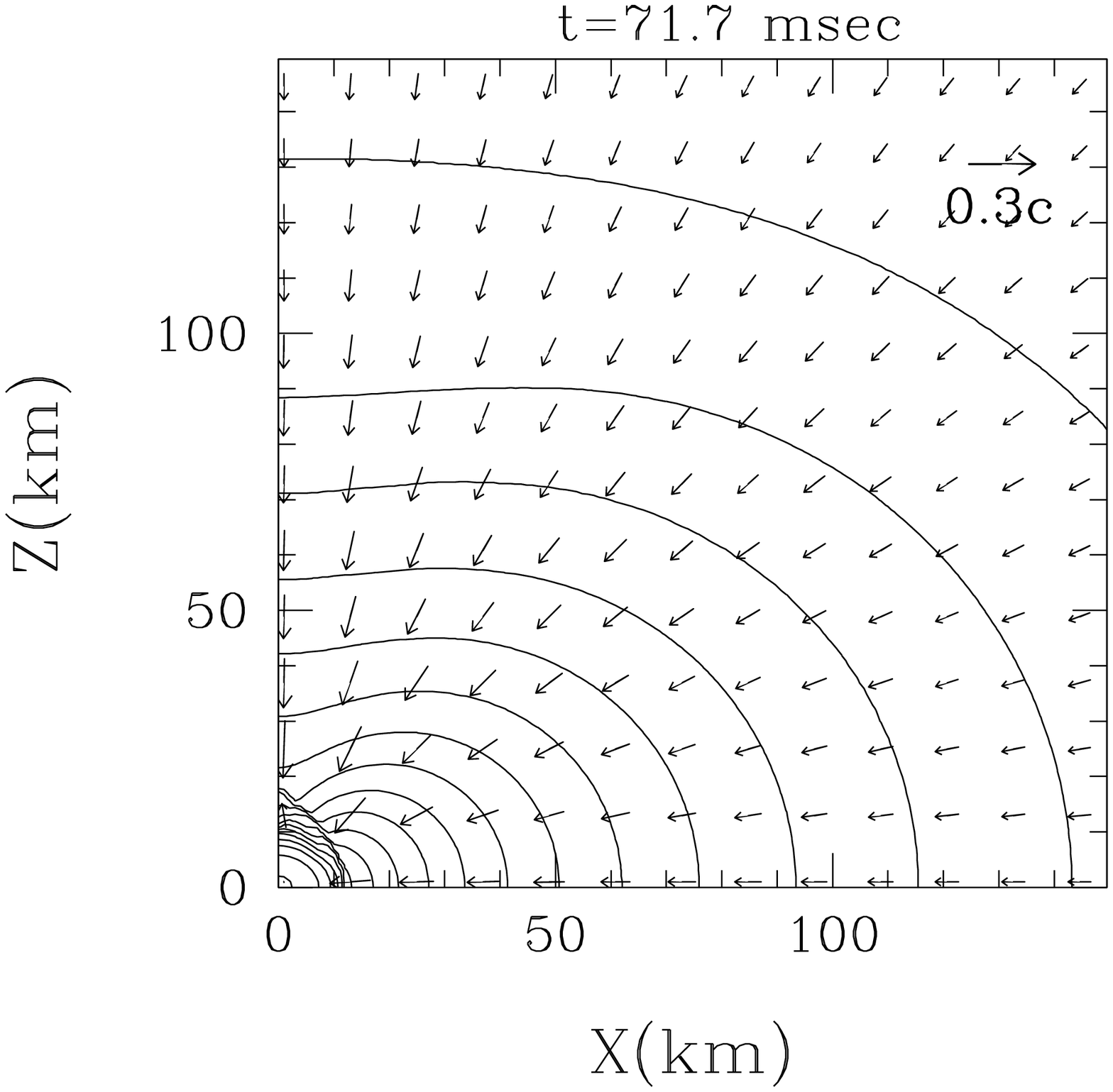}
\epsfxsize=2.2in
\leavevmode
\epsffile{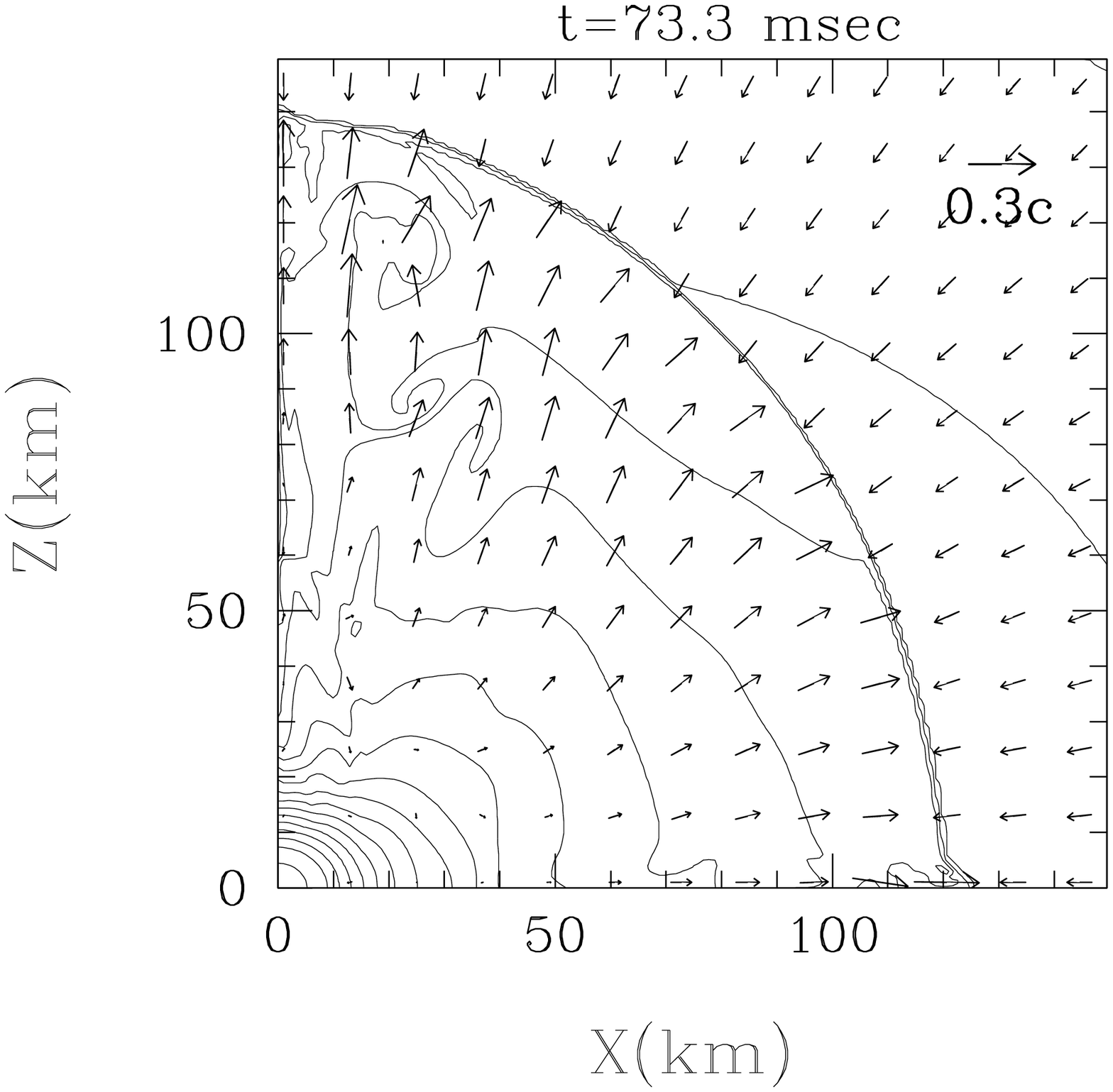}
\caption{
Density contour curves of $\rho$ and velocity fields of $v^A$
at selected time steps around which shocks are formed.
The contour curves are drawn for $\rho/\rho_{\rm nuc}=10^{-0.4j}$, 
for $j=0,1,2,\cdots,20$. 
\label{FIG15}}
\end{center}
\end{figure}

In Fig. \ref{FIG15}, we display the density contour curves 
and velocity fields at selected time steps around
which shocks are formed.  
The time of the shock formation is $\sim 71.7$ msec, which is 
in good agreement with that for model A1B3G1 in \cite{HD} with the 
correction factor which is associated with the dynamical time scale as
$(\rho_{c,{\rm init}}/10^{10}~{\rm g/cm^3})^{-1/2}$. 
(Note that in \cite{HD}, the central density of the initial condition
is $10^{10}~{\rm g/cm^3}$, while here $\rho_{c,{\rm init}} \approx
1.65 \times 10^{10}~{\rm g/cm^3}$.)
This coincidence suggests that the approximate general relativistic
approach adopted in \cite{HD} is indeed suited for 
study of axisymmetric stellar core collapse to neutron stars in
general relativity. 

After the shock formation, the shock fronts of 
prolate shape spread outward. The prolateness is produced by
the fact that the shocks are stronger for the $z$ direction due to
the absence of centrifugal force \cite{YS,Muller,HD}. 
In Fig. \ref{FIG16}(a), we display the time evolution of the 
central density and lapse function at the center. 
Global features are qualitatively the same as those for the simulation 
of model (C1a) presented in Sec. IV C. As in that case, 
the central density (lapse function) monotonically increases 
(decreases) until it exceeds $\rho_{\rm nuc}$. 
When it becomes $\sim 3.5\rho_{\rm nuc}$, the collapse is halted 
and the shocks start propagating outward, while the core 
gradually settles down toward a quasistationary state 
of $\rho_c \sim 2 \rho_{\rm nuc}$.

Although these qualitative features are the same as those for (C1a), there are 
a few quantitative differences between the results of two models. First, the 
maximum density and central density of the formed neutron star
found here are slightly smaller than those for (C1a). 
This is likely due to the fact that the effect of the centrifugal force 
plays a more important role than for (C1a). Second, 
the core does not relax to a static state soon, but oscillates 
approximately in a quasiperiodic manner for several periods. 
This oscillation is not conspicuous for model (C1a). 
These facts
imply that to obtain quantitative outputs of stellar core collapse, 
we should start the simulations from an initial condition 
of a realistic density profile, although 
qualitative global features of the collapse can be found even using 
a more compact initial condition. 
We note that the quasiperiodic oscillation found here 
is also observed in \cite{HD}. 
As reported in \cite{Muller,HD}, quasiperiodic gravitational waves 
are likely emitted associated with this oscillation. 
However, we have not tried to compute gravitational waves 
in the present work. As expected from the results in \cite{HD}, 
the amplitude of gravitational waves is not very large, 
so that it would not be technically easy to extract them from 
the metric in which gauge modes and numerical noises are included.
Developing a method for the wave extraction of a weak signal 
will be one of the challenging problems in the future. 

In the last figure of Fig. \ref{FIG16}(b), the time 
evolution of baryon rest-mass of the core 
(of density larger than $\rho_{\rm nuc}$) is shown.
As in the time evolution of $\rho_c$, it reaches the maximum
at which $\rho_c$ reaches the maximum value, and
then settles down toward a constant $\sim 0.55M_{\odot}$.
Thus, the temporary product after the collapse is a low mass neutron star 
in this simulation.

\begin{figure}[htb]
\vspace*{-4mm}
\begin{center}
\epsfxsize=2.6in
\leavevmode
(a)\epsffile{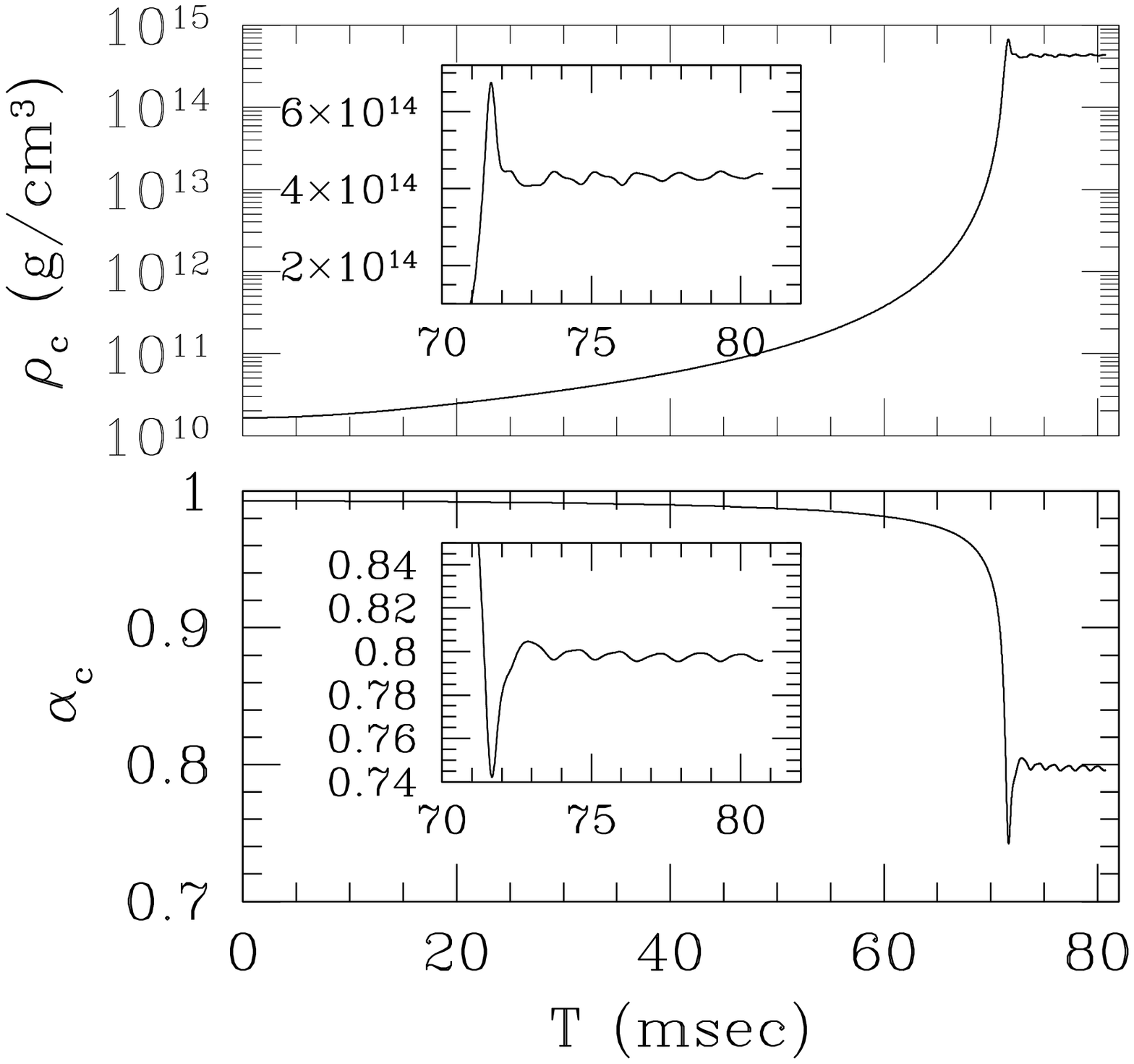}
\epsfxsize=2.6in
\leavevmode
~~~(b)\epsffile{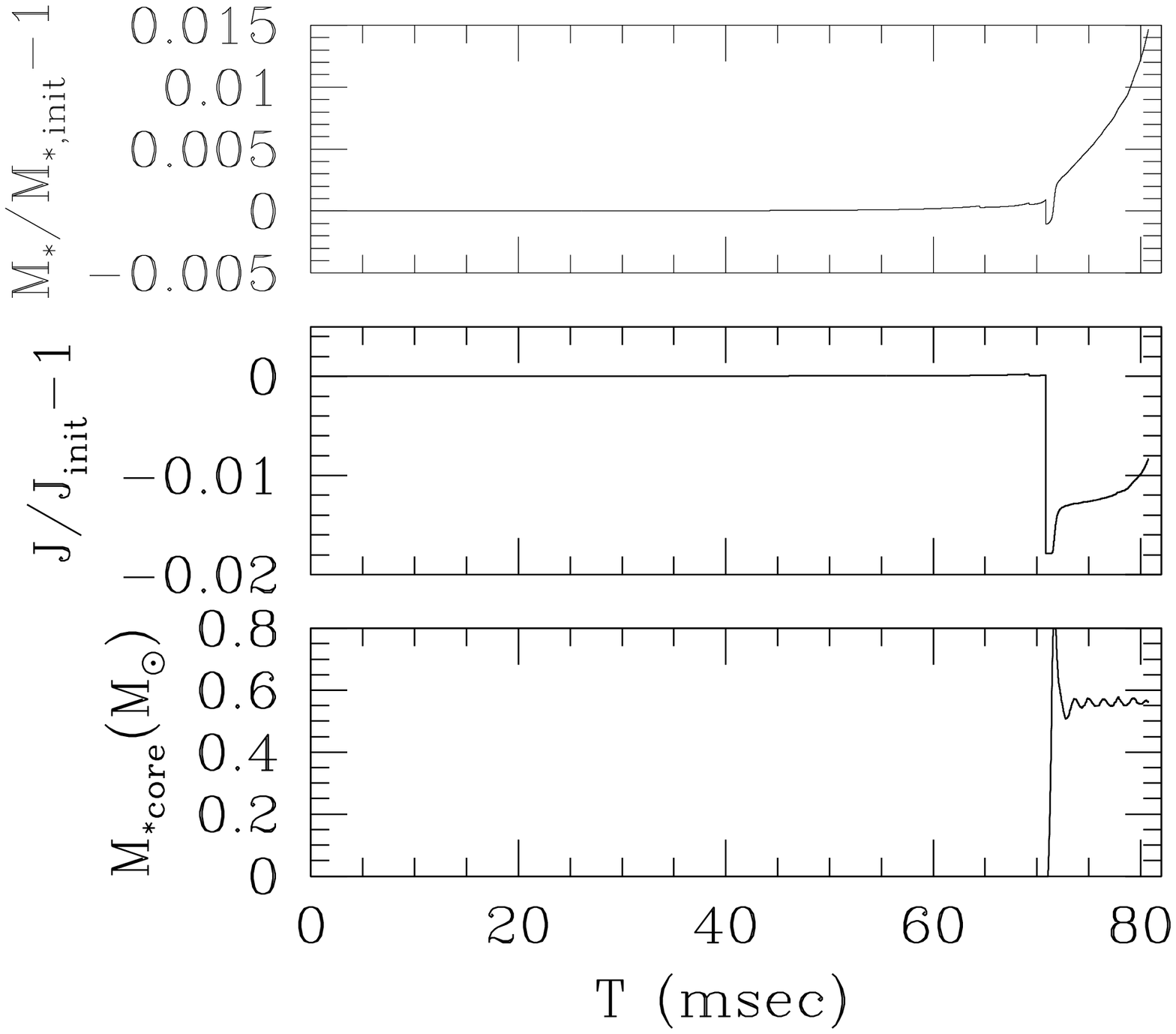}
\caption{(a) Time evolution of the central density and lapse
function at the center. (b) Time evolution of the
violation of rest-mass conservation,
angular momentum conservation, and the
time evolution of rest-mass of the core
with $\rho > \rho_{\rm nuc}$. 
\label{FIG16}}
\end{center}
%\vspace{-1mm}
\end{figure}

\section{summary}

We have presented numerical results obtained by an axisymmetric
general relativistic implementation, and demonstrated that 
with this new implementation, it is feasible 
to carry out long-term simulations for spherical and rapidly rotating 
neutron stars, and rotating 
stellar collapses to a neutron star and a Kerr black hole. 
It is shown that the simulations for stable neutron stars can be continued 
for $\agt 10$ dynamical time scales until the crash of computation 
with $N \sim 200$, 
even if the neutron stars are of $\sim 95 \%$ of the maximum allowed mass. 
The duration of the computation is 
long enough to obtain oscillation modes of neutron stars. 
The simulations are also feasible for collapse of rotating
neutron stars to Kerr black holes. 
We have illustrated that with our implementation, the 
mass of a Kerr black hole formed after the
collapse can be computed accurately. 
We have also demonstrated that it is 
feasible to perform the simulations of rotating stellar core collapse 
to a neutron star, adopting parametric 
equations of state that mimic realistic equations of state. 
This illustrates that the new implementation works well also 
for realistic equations of state that have not 
been adopted so far in fully general relativistic simulations. 
In conclusion, the axisymmetric numerical implementation 
presented here will be used for a wide variety of
astrophysical simulations such 
as rotating core collapse of a massive star to a neutron star or
a black hole, and accretion-induced collapse of a neutron star to
a black hole. As the next step, we plan to perform 
simulations for rotating core collapse to neutron stars, 
and compare the results with those in an approximate relativistic approach 
\cite{HD}.
We also consider that 
collapse of a massive stellar core to a black hole 
is one of the most interesting topics. 

Because of the assumption of axial symmetry, 
we were able to carry out a wide variety of tests and 
calibrations for our new hydrodynamic implementation 
with low computational costs. 
In previous works, e.g., \cite{gr3d,other}, tests of their hydrodynamic 
implementations picking up single stars have been done in three 
spatial dimensions. In those cases,
$N$ could be at most 100, since the computational costs were very 
high for $N > 100$. It is pragmatically 
very difficult to investigate the accuracy and convergence 
in a well-resolved simulation with $N \sim$ several hundreds 
under normal circumstances in which we can use at most $\sim 1000$
CPU hours per year. In the axisymmetric case, 
the simulation with $N \sim$ 
several hundreds is not very expensive, and thus for detailed tests of 
new hydrodynamic implementations in general relativity,
axisymmetric simulation has great advantages.
We expect that for the testing of new gauge conditions and 
wave extraction techniques, it would also play an important role. 

Finally, we note the following point. 
Although we focus only on the axisymmetric simulation 
in this paper, the present hydrodynamic implementation can be used 
for three-dimensional simulations with a slight modification, 
since the transport terms in the hydrodynamic equations are of the same form.
Actually, we have already implemented it and checked that it works. 
We expect that with the same grid resolution that we adopted in this paper, 
the same results will be obtained 
(although it takes a much longer time to carry out the long-term 
simulations). So far, we have performed simulations for a merger of 
binary neutron stars only using the $\Gamma$-law equation of 
state \cite{bina,bina2}. 
However, with the new hydrodynamic implementation reported 
in this paper, we will be able to adopt a variety of 
equations of state. We plan to perform the simulation for a merger of 
binary neutron stars adopting more realistic equations of state 
in the future. 

\begin{center}
{\bf Acknowledgments}
\end{center}

Part of this work was done when I visited the 
Theoretical Astrophysics Group of Caltech. 
I would like to thank Kip Thorne, Lee Lindblom, and Mark Scheel 
for their warm hospitality during my visit and for discussion. 
I also thank Toni Font for reading this manuscript and
giving helpful comments, Harald Dimmelmeier, 
Jose-Maria Ib\'a\~nez, Mark Miller, 
Eward M\"uller, Takashi Nakamura, and Nick Stergioulas for helpful
conversation, and Y. Sekiguchi for 
providing analytic solutions of the Riemann shock tube problem. 
This work is supported in part by Japanese Monbu-Kagakusho Grant
Nos. 13740143 and 14047207. 

\appendix

\section{Treatment for transport terms in the 
hydrodynamic equations} 

Equations (\ref{continuity1})--(\ref{energy1})
are of the forms 
\beqn
\pa_t Q_a + \pa_A F^A_a =S_a,  \label{eq21}
\eeqn
where $Q_a$ and $F^A_a$ for $a=1$--5 are defined as 
\beqn
&&Q_a=(\rho_*,~ J_x,~ J_y,~ J_z,~ E_*),\\ 
&&F^A_a=[\rho_* v^A,~
J_x v^A +P\alpha \sqrt{\gamma}\delta^{A}_{~x},~  
%J_y v^i +P\alpha\sqrt{\gamma}\delta^{A}_{~y}, 
J_y v^A,~ 
J_z v^A +P\alpha\sqrt{\gamma}\delta^{A}_{~z}, ~ 
E_* v^A + P \sqrt{\gamma}(v^A+\beta^A)]. 
\eeqn
Here, $J_i \equiv \rho_* \hat u_i$, $E_* \equiv \rho_* \hat e$, and 
$S_a$ in Eq. (\ref{eq21}) denote the right-hand sides of Eqs. 
(\ref{continuity1})--(\ref{energy1}). We note that 
$\gamma$ here is the determinant of the three-metric in 
the Cartesian coordinates. 

In numerical computation, we evaluate
the transport terms using the approximate Riemann solver, 
which relies on the characteristic decomposition 
of the equations (see e.g. \cite{Val2,fontrev} and references 
therein). To adopt this method,
we first need to compute the Jacobian matrix and then to carry out
the spectral decomposition of it. 

The Jacobian matrix for the $A$-th direction, $M^A_{ab}$, is defined by 
\beqn
M^A_{ab}={\pa F^A_a \over \pa Q_b}~~~(A=x~{\rm or}~z). 
\eeqn
Using this, Eq. (\ref{eq21}) may be expressed in the form
\beqn
\pa_t Q_a + \sum_{b=1}^5\sum_{A=x,z} M^A_{ab} \pa_A Q_b =S_a.  
\eeqn
Thus, $M^A_{ab}$ has information on the characteristic speed of the
fluid.

Following Font {\it et al}. \cite{Font,Val},
we calculate the Jacobian matrix from
\beqn
M^A_{ab}=\sum_{c=1}^5
{\pa F^A_a \over \pa q_c}
{\pa q_c \over \pa Q_b}
\equiv \sum_{c=1}^5 B^A_{ac} C_{bc}^{-1},
\eeqn
where
\beqn
&&B^A_{ac}\equiv {1 \over \sqrt{\gamma}}
{\pa F^A_a \over \pa q_c},\\
&&C_{bc}\equiv {1 \over \sqrt{\gamma}}{\pa Q_b \over \pa q_c}, 
\eeqn
and $q_c=(\rho, v^x, v^y, v^z, \varepsilon)$. 
Explicit forms for $C_{ab}$ and $B^x_{ab}$ in our notation are
\beqn
C_{ab}=\left[
\begin{array}{lllll}
w
& \displaystyle 
\rho w^3 {V_x \over \alpha^2}
& \displaystyle 
\rho w^3 {V_y \over \alpha^2}
& \displaystyle 
\rho w^3 {V_z \over \alpha^2} & 0 \\
\displaystyle 
{h_1 w^2 \over \alpha} V_x &
\displaystyle 
{\rho h w^2 \over \alpha} F_{xx} & 
\displaystyle 
{\rho h w^2 \over \alpha} F_{xy} &
\displaystyle 
{\rho h w^2 \over \alpha} F_{xz} & 
\displaystyle 
\rho h_2 {w^2 V_x \over \alpha} \\
\displaystyle 
{h_1 w^2 \over \alpha} V_y &
\displaystyle 
{\rho h w^2 \over \alpha}F_{xy} &
\displaystyle 
{\rho h w^2 \over \alpha}F_{yy} &
\displaystyle 
{\rho h w^2 \over \alpha}F_{yz} &
\displaystyle 
\rho h_2 {w^2 V_y \over \alpha} \\
\displaystyle 
{h_1 w^2 \over \alpha} V_z &
\displaystyle 
{\rho h w^2 \over \alpha}F_{xz} &
\displaystyle 
{\rho h w^2 \over \alpha}F_{yz} &
\displaystyle 
{\rho h w^2 \over \alpha}F_{zz} &
\displaystyle 
\rho h_2 {w^2 V_z \over \alpha} \\
\displaystyle 
h_1 w^2 -\chi
& \displaystyle 
2 {\rho h w^4 \over \alpha^2}V_x
& \displaystyle 
2 {\rho h w^4 \over \alpha^2}V_y
& \displaystyle 
2 {\rho h w^4 \over \alpha^2}V_z
& \displaystyle 
\rho h_2 w^2 - \rho \kappa
\end{array}
\right], 
\eeqn
and
\beqn
B^x_{ab}=\left[
\begin{array}{lllll}
w v^x
& \displaystyle 
\rho w \biggl(1+{w^2 V_x v^x\over \alpha^2}\biggr)
& \displaystyle 
\rho w^3 {V_y v^x \over \alpha^2}
& \displaystyle 
\rho w^3 {V_z v^x \over \alpha^2} & 0 \\
%%%%%%%%%%%%%%%%%%%%%%%%%%%%%%%%%%%%%%%%%%%%%%%%
\displaystyle 
{h_1 w^2 \over \alpha} V_x v^x + \alpha \chi
& \displaystyle 
{\rho h w^2 \over \alpha} (v^x F_{xx}+ V_x ) & 
\displaystyle 
{\rho h w^2 \over \alpha} v^x F_{xy} &
\displaystyle 
{\rho h w^2 \over \alpha} v^x F_{xz} &
\displaystyle 
\rho h_2 {w^2 V_x v^x \over \alpha} + \rho \kappa \alpha \\
%%%%%%%%%%%%%%%%%%%%%%%%%%%%%%%%%%%%%%%%%%%%%%%%
\displaystyle 
{h_1 w^2 \over \alpha} V_y v^x &
\displaystyle 
{\rho h w^2 \over \alpha} ( v^x F_{xy} + V_y )& 
\displaystyle 
{\rho h w^2 \over \alpha} v^x F_{yy} &
\displaystyle 
{\rho h w^2 \over \alpha} v^x F_{yz} &
\displaystyle 
\rho h_2 {w^2 V_y v^x \over \alpha} \\
%%%%%%%%%%%%%%%%%%%%%%%%%%%%%%%%%%%%%%%%%%%%%%%%
\displaystyle 
{h_1 w^2 \over \alpha} V_z v^x &
\displaystyle 
{\rho h w^2 \over \alpha} (v^x F_{xz} + V_z )& 
\displaystyle 
{\rho h w^2 \over \alpha} v^x F_{yz} & 
\displaystyle 
{\rho h w^2 \over \alpha} v^x F_{zz} &
\displaystyle 
\rho h_2 {w^2 V_z v^x \over \alpha} \\
\displaystyle 
h_1 w^2 v^x + \chi  \beta^x
& \displaystyle 
2 {\rho h w^4 \over \alpha^2}V_x v^x + \rho h w^2
& \displaystyle 
2 {\rho h w^4 \over \alpha^2}V_y v^x
& \displaystyle 
2 {\rho h w^4 \over \alpha^2}V_z v^x 
& \displaystyle 
\rho h_2 w^2 v^x + \rho \kappa \beta^x  
\end{array}
\right], 
\eeqn
where
\beqn
&& V_i \equiv \gamma_{ij}(v^j+\beta^j),\\
&& \chi\equiv {\pa P \over \pa \rho}\Big|_{\varepsilon}, \\ 
&& \kappa\equiv {1 \over \rho}{\pa P \over \pa \varep}\Big|_{\rho}, \\
&& F_{ij}=\gamma_{ij}+{2w^2 V_i V_j \over \alpha^2}, \\
&& h_1 \equiv 1+ \varep + \chi, \\
&& h_2 \equiv 1+ \kappa.
\eeqn
$B^z_{ab}$ is obtained by appropriate
exchanges of subscripts among $x, y, z$ of $B^x_{ab}$. 

The eigenvalues of the matrix $M^A_{ab}$, $\lambda^A$, correspond to
the characteristic speeds of the fluid in the $A$-th direction, and 
are derived from the equation 
\beq
\det(B^A_{ab}-\lambda^A C_{ab})=0.  
\eeq
The solutions are \cite{Val,Val2}
\beqn
\lambda^A=\lambda^A_{\pm},~~v^A~({\rm triple}),
\eeqn
where
\beqn
\lambda^A_{\pm}={1 \over \alpha^2 - V_k V^k c_s^2}
&& \biggl[v^A \alpha^2 (1-c_s^2) -\beta^A c_s^2(\alpha^2 - V_k V^k)
\nonumber \\
&&~~~~~~ \pm \alpha c_s \sqrt{(\alpha^2 -V_k V^k)\{ \gamma^{AA}
(\alpha^2 - V_k V^k c_s^2)-(1-c_s^2)V^A V^A \}} \biggr]\nonumber \\
&&~~~~~~ ({\rm no ~summation ~for}~ A)
\eeqn
and 
\beqn
&&c_s^2 = {1 \over h}\biggl( \chi + {P \over \rho}\kappa \biggr),\\
&&V^k =\gamma^{kl}V_l = v^k + \beta^k.  
\eeqn
Using the eigenvalues, the spectrum decomposition for 
$M^A_{ab}$ can be done in a straightforward manner as
\beqn
M^A_{ad}=\sum_{b,c} R^A_{ab} \Lambda^A_{bc} (R^A)^{-1}_{cd},
\eeqn
where 
$\Lambda^A_{bc}$ is the diagonal matrix composed of $\lambda^A$
in the following order: $[\lambda^A_+, v^A, v^A, v^A, \lambda^A_-]$. 
For convenience of the calculation of $R^A_{ab}$,
we define a matrix $T^A_{ab}$, which satisfies the relation as 
\beq
R^A_{ab}=\sum_{c=1}^5 C_{ac} T^A_{cb}. 
\eeq
Since $R^A_{ab}$ is calculated from the right eigenvectors of $M^A_{ab}$, 
$T^A_{ab}$ is composed of vectors $t_b^{(I)}$ that satisfy the equation 
\beq
\sum_{b=1}^5 (B^A_{ab}-\lambda^A C_{ab})(t^A_b)^{(I)}=0~~~
{\rm for}~~I=1\sim 5. 
\eeq
Then, $T^A_{ab}
=[(t^A_a)^{(1)},(t^A_a)^{(2)},(t^A_a)^{(3)},(t^A_a)^{(4)},(t^A_a)^{(5)}]$, 
and hence we obtain 
\beqn
&&T^x_{ab}=\left[
\begin{array}{lllll}
1 & -\kappa & 0 & 0 &1 \\
H^{xx}(\lambda^x_+) & 0 & 0 & 0 & H^{xx}(\lambda^x_-) \\
H^{xy}(\lambda^x_+) & 0 & \rho^{-1} & 0 & H^{xy}(\lambda^x_-) \\
H^{xz}(\lambda^x_+) & 0 & 0 & \rho^{-1} & H^{xz}(\lambda^x_-) \\
%%\displaystyle 
%%{-c_s^2(v^x-\lambda^x_+)\{\alpha^2\gamma^{xx}-V^x(\beta^x+\lambda^x_+)\}
%%\over \rho w^2(v^x-\lambda^x_+)^2} & 0 & 0 & 0 &
%%\displaystyle 
%%{-c_s^2(v^x-\lambda^x_-)\{\alpha^2\gamma^{xx}-V^x(\beta^x+\lambda^x_-)\}
%%\over \rho w^2(v^x-\lambda^x_-)^2} \\
%%\displaystyle 
%%{-c_s^2(v^x-\lambda^x_+)\{\alpha^2\gamma^{xy}-V^y(\beta^x+\lambda^x_+)\}
%%\over \rho w^2(v^x-\lambda^x_+)^2} & 0 & \displaystyle 
%%{1 \over \rho} & 0 &
%%\displaystyle 
%%{-c_s^2(v^x-\lambda^x_-)\{\alpha^2\gamma^{xy}-V^y(\beta^x+\lambda^x_-)\}
%%\over \rho w^2(v^x-\lambda^x_-)^2} \\
%%\displaystyle 
%%{-c_s^2(v^x-\lambda^x_+)\{\alpha^2\gamma^{xz}-V^z(\beta^x+\lambda^x_+)\}
%%\over \rho w^2(v^x-\lambda^x_+)^2} & 0 & 0 & \displaystyle 
%%{1 \over \rho} &
%%\displaystyle 
%%{-c_s^2(v^x-\lambda^x_-)\{\alpha^2\gamma^{xz}-V^z(\beta^x+\lambda^x_-)\}
%%\over \rho w^2(v^x-\lambda^x_-)^2} \\
\displaystyle 
{P \over \rho^2} & \displaystyle 
{\chi \over \rho} & 0 & 0 &
\displaystyle {P \over \rho^2}
\end{array}
\right], \\
&&~ \nonumber \\
%%%%%%%%%%%%%%%
%&&T^y_{ab}=\left[
%\begin{array}{lllll}
%1 & 0 & -\kappa & 0 &1 \\
%\displaystyle 
%{-c_s^2(v^y-\lambda^y_+)\{\alpha^2\gamma^{xy}-V^x(\beta^y+\lambda^y_+)\}
%\over \rho w^2(v^y-\lambda^y_+)^2} & \displaystyle 
%{1 \over \rho} & 0 & 0 &
%\displaystyle 
%{-c_s^2(v^y-\lambda^y_-)\{\alpha^2\gamma^{xy}-V^x(\beta^y+\lambda^y_-)\}
%\over \rho w^2(v^y-\lambda^y_-)^2} \\
%\displaystyle 
%{-c_s^2(v^y-\lambda^y_+)\{\alpha^2\gamma^{yy}-V^y(\beta^y+\lambda^y_+)\}
%\over \rho w^2(v^y-\lambda^y_+)^2} & 0 & 0 & 0 &
%\displaystyle 
%{-c_s^2(v^y-\lambda^y_-)\{\alpha^2\gamma^{yy}-V^y(\beta^y+\lambda^y_-)\}
%\over \rho w^2(v^y-\lambda^y_-)^2} \\
%\displaystyle 
%{-c_s^2(v^y-\lambda^y_+)\{\alpha^2\gamma^{yz}-V^z(\beta^y+\lambda^y_+)\}
%\over \rho w^2(v^y-\lambda^y_+)^2} & 0 & 0 & \displaystyle 
%{1 \over \rho} &
%\displaystyle 
%{-c_s^2(v^y-\lambda^y_-)\{\alpha^2\gamma^{yz}-V^z(\beta^y+\lambda^y_-)\}
%\over \rho w^2(v^y-\lambda^y_-)^2} \\
%\displaystyle 
%{-\chi+h c_s^2 \over \rho\kappa} & 
%0 & \displaystyle {\chi \over \rho}& 0 & 
%\displaystyle 
%{-\chi+h c_s^2 \over \rho\kappa}
%\end{array}
%\right], \\
%&&~ \nonumber \\
%%%%%%%%%%%%%%%%%%%%%%%%%%%%%%%%
&&T^z_{ab}=\left[
\begin{array}{lllll}
1 & 0 & 0 & -\kappa &1 \\
H^{zx}(\lambda^z_+) & \rho^{-1} & 0 & 0 & H^{zx}(\lambda^z_-) \\
H^{zy}(\lambda^z_+) & 0 & \rho^{-1} & 0 & H^{zy}(\lambda^z_-) \\
H^{zz}(\lambda^z_+) & 0 & 0 & 0 & H^{zz}(\lambda^z_-) \\
%%\displaystyle 
%%{-c_s^2(v^z-\lambda^z_+)\{\alpha^2\gamma^{xz}-V^x(\beta^z+\lambda^z_+)\}
%%\over \rho w^2(v^z-\lambda^z_+)^2} & \displaystyle 
%%{1 \over \rho} & 0 & 0 &
%%\displaystyle 
%%{-c_s^2(v^z-\lambda^z_-)\{\alpha^2\gamma^{xz}-V^x(\beta^z+\lambda^z_-)\}
%%\over \rho w^2(v^z-\lambda^z_-)^2} \\
%%\displaystyle 
%%{-c_s^2(v^z-\lambda^z_+)\{\alpha^2\gamma^{yz}-V^y(\beta^z+\lambda^z_+)\}
%%\over \rho w^2(v^z-\lambda^z_+)^2} & 0 & \displaystyle 
%%{1 \over \rho} & 0 &
%%\displaystyle 
%%{-c_s^2(v^z-\lambda^z_-)\{\alpha^2\gamma^{yz}-V^y(\beta^z+\lambda^z_-)\}
%%\over \rho w^2(v^z-\lambda^z_-)^2} \\
%%\displaystyle 
%%{-c_s^2(v^z-\lambda^z_+)\{\alpha^2\gamma^{zz}-V^z(\beta^z+\lambda^z_+)\}
%%\over \rho w^2(v^z-\lambda^z_+)^2} & 0 & 0 & 0 &
%%\displaystyle 
%%{-c_s^2(v^z-\lambda^z_-)\{\alpha^2\gamma^{zz}-V^z(\beta^z+\lambda^z_-)\}
%%\over \rho w^2(v^z-\lambda^z_-)^2} \\
\displaystyle {P \over \rho^2} & 
0 & 0 & \displaystyle {\chi \over \rho} & 
\displaystyle {P \over \rho^2}
\end{array}
\right], 
\eeqn
where
\beqn
&& H^{xk}(\lambda)=
{-c_s^2(v^x-\lambda)\{\alpha^2\gamma^{xk}-V^k(\beta^x+\lambda)\}
\over \rho w^2(v^x-\lambda)^2}, \\
&& H^{zk}(\lambda)=
{-c_s^2(v^z-\lambda)\{\alpha^2\gamma^{zk}-V^k(\beta^z+\lambda)\}
\over \rho w^2(v^z-\lambda)^2}.
\eeqn

After the above reconstruction of the fluid equation, the numerical fluxes
in the upwind scheme are computed from 
\beq
\hat F^A_a = {1 \over 2}\biggl[
F^A_a(Q^r_c)+F^A_a(Q^l_c)-\sum_{b=1}^5
(R|\Lambda|R^{-1})_{ab}(Q_b^l-Q_b^r)\biggr],
\eeq
where we omit the subscripts for $R_{ab}$ and $\Lambda_{ab}$. 
$Q^l_c$ and $Q^r_c$ denote 
$Q_c$ at the left and right sides of the corresponding interfaces, and
are evaluated using the third-order spatial interpolation.
At the interface between the $i$-th and $(i+1)$-th cells,
we define them according to  
\beqn
&&Q^l_c=Q_i+{\Delta_{i-1} \over 6}+{\Delta_{i} \over 3},\\
&&Q^r_c=Q_{i+1}-{\Delta_i \over 3}-{\Delta_{i+1} \over 6},
\eeqn
where $\Delta_i=Q_{i+1}-Q_i$. 
To suppress the oscillation near shock discontinuities, we 
modify the interpolation using the following min-mod limiter as \cite{JJJ} 
\beqn
&&Q_c^l=Q_i
+{\Phi(r^+_{i-1})\Delta_{i-1} \over 6}+{\Phi(r^-_{i})\Delta_{i} \over 3},\\
&&Q_c^r=Q_{i+1}
-{\Phi(r^+_{i})\Delta_i \over 3}-{\Phi(r^-_{i+1})\Delta_{i+1} \over 6},
\eeqn
where
\beqn
&&r^+_i=\Delta_{i+1}/\Delta_i,\\  
&&r^-_i=\Delta_{i-1}/\Delta_i,\\
&&\Phi(r)={\rm minmod}(1, br) ~~~~~(1\leq b \leq 4~{\rm for~TVD~condition}). 
\eeqn
For the simulations presented in this paper,
we choose $b=2$, since for $b \approx 1$,
the dissipation is so large that the envelope of neutron stars spreads
outward too quickly, while for $b=4$,
the oscillation around shock discontinuities is too serious. 

The values for components of matrices $R_{ab}$ and $\Lambda_{ab}$
at grid interfaces are computed using the Roe-type average such as
\cite{Roe,JJJ}
\beq
q_{i+1/2}={\sqrt{(\rho_*)_{i+1}}~q_{i+1} + \sqrt{(\rho_*)_i}~q_i
\over \sqrt{(\rho_*)_{i+1}} +\sqrt{(\rho_*)_{i}}}, 
\eeq
where we carry out the average for variables $\hat u_i$, 
$\kappa$, $\chi$, and $h$ (i.e., $q_i$ is one of these variables).
Other variables are computed from them. 
It should be noted that in the relativistic case,
the average is not uniquely specified in contrast with the
Newtonian case \cite{Roe}. However, numerical results of test computations
(see below) seem to indicate that this averaging is appropriate. 

To confirm that our hydrodynamic implementation can capture shocks 
accurately, we 
carried out the simulations for Riemann shock tube problems and
wall shock problems in the 1+1 special relativistic spacetime
with $(t,x)$ as the coordinates. 
In this test, we adopt the $\Gamma$-law equation of state. 
In both tests, we take $N=400$ with $\Delta x =1/N$.

In the Riemann shock-tube problem, we choose $\Gamma=5/3$. 
The parameters of the 
initial condition are chosen as $\rho=10$ and $P=13.3$ for $x<0$ and
$\rho=1$ and $P=10^{-6}$ for $x>0$ following previous papers
\cite{fontrev}. 
In the wall shock problem, we set the parameters 
as $v^x=0.9$, $\rho=1$, and $P=10^{-6}$ with $\Gamma=4/3$. 

\begin{figure}[htb]
\vspace*{-4mm}
\begin{center}
\epsfxsize=2.6in
\leavevmode
\epsffile{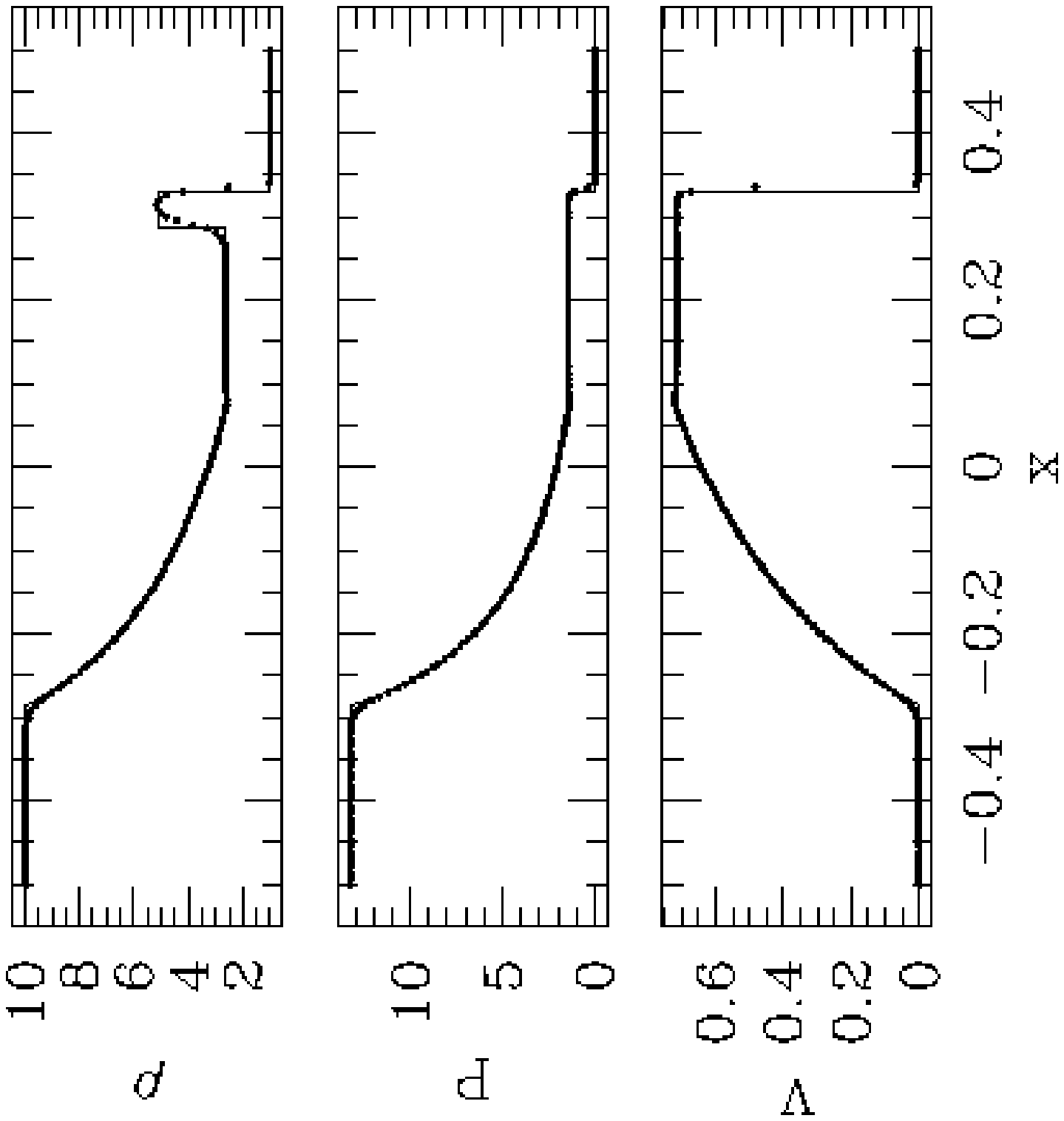}
\epsfxsize=2.6in
\leavevmode
~~\epsffile{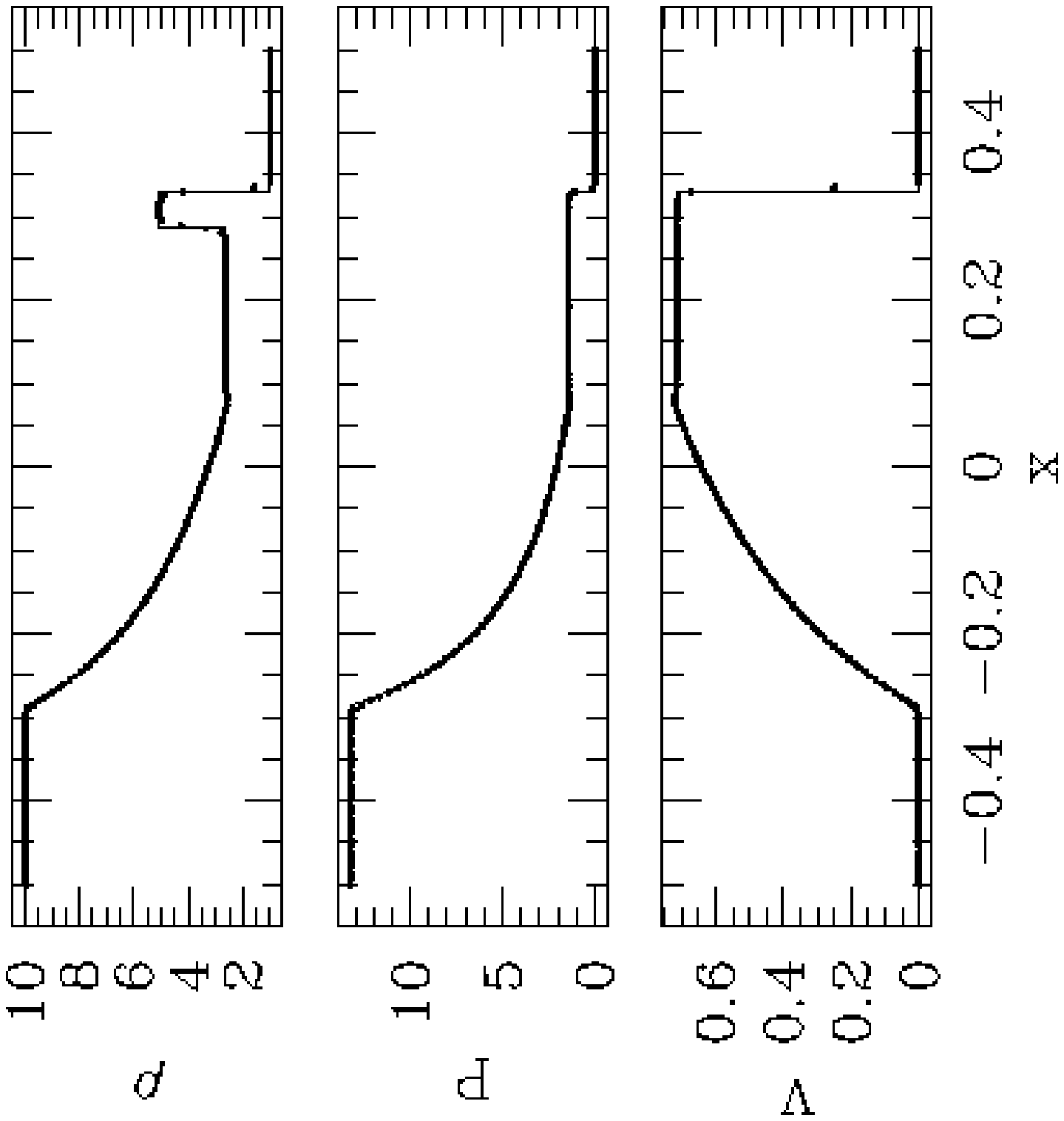}
\vspace*{-4mm}
\caption{Comparison of numerical solutions (filled circles) of 
a one-dimensional Riemann shock-tube problem 
with the analytical solution (solid curves) at $t=0.4$ 
for $b=1$ (left) and 2 (right). The grid number is 400
and the grid spacing is 0.0025. Only 200 data points are plotted. 
\label{FIGA1}
}
\end{center}
\end{figure}

\begin{figure}[htb]
\vspace*{-4mm}
\begin{center}
\epsfxsize=2.6in
\leavevmode
\epsffile{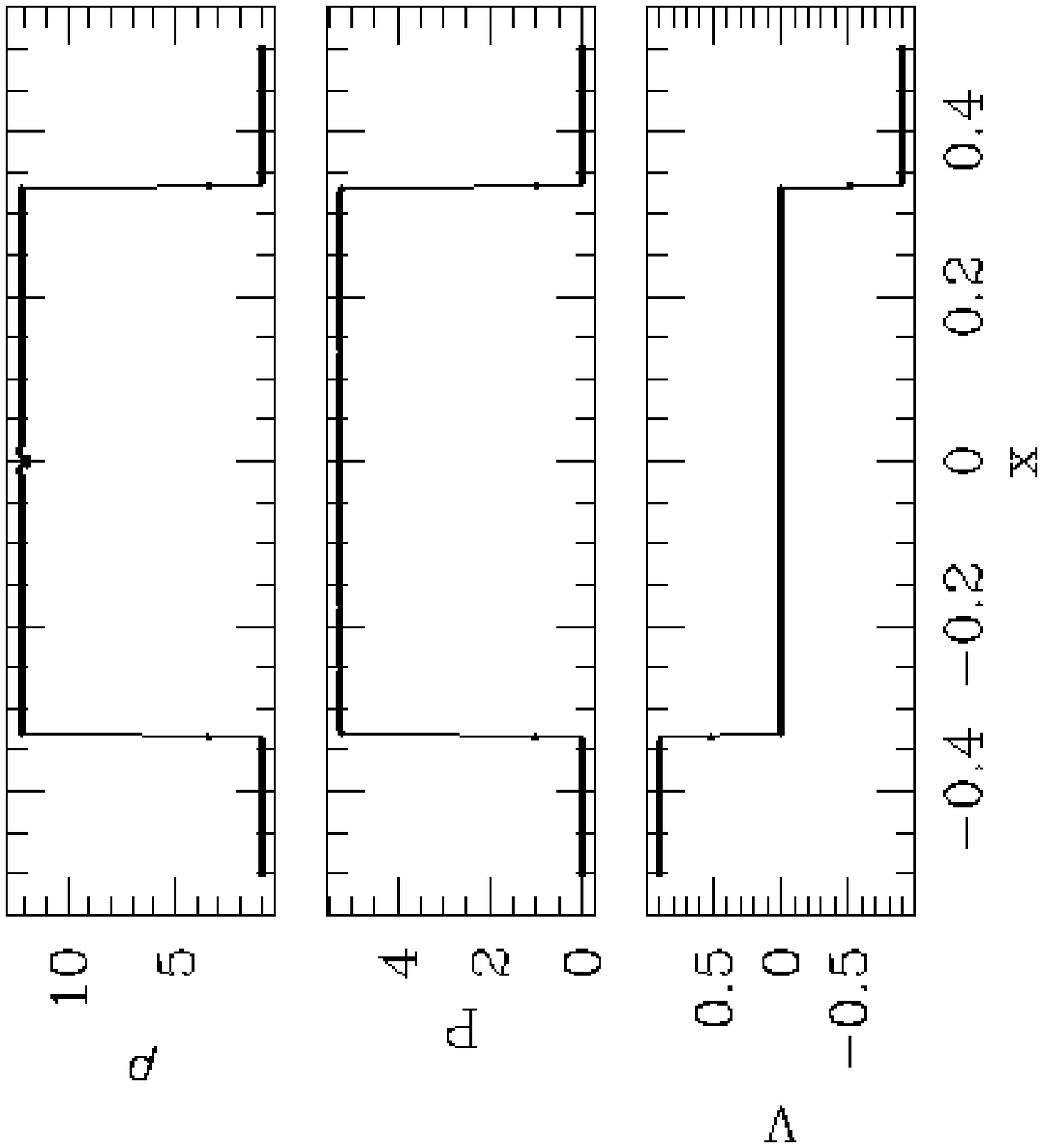}
\epsfxsize=2.6in
\leavevmode
~~\epsffile{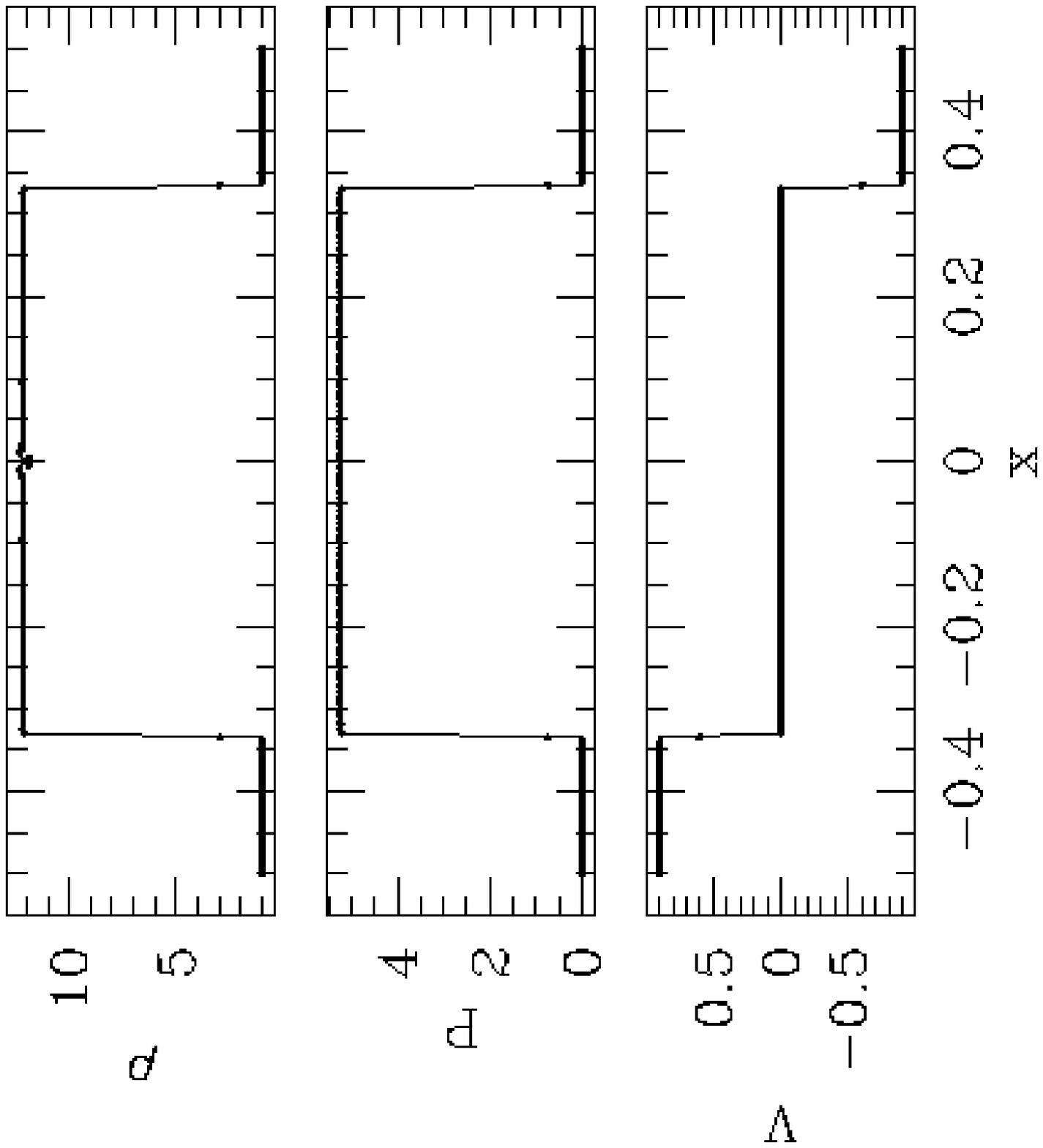}
\vspace*{-4mm}
\caption{
The same as Fig. \ref{FIGA1}, but for 
a one-dimensional wall shock problem with $\Gamma=4/3$ 
at $t=1.6$. 
\label{FIGA2}
}
\end{center}
\end{figure}

In Figs. \ref{FIGA1} and \ref{FIGA2}, 
we compare numerical results of the Riemann shock tube
problem and of the wall shock problem with analytical solution
(solid curves) for the choice $b=1$ and 2. As indicated in these figures,
numerical results agree well with analytic solutions,
in particular for $b=2$. 

\begin{thebibliography}{99}

\bibitem{Nakamura} T. Nakamura, Prog. Theor. Phys. 
{\bf 65}, 1876 (1981); {\bf 70}, 1144 (1983).

\bibitem{NOK} T. Nakamura, K. Oohara, and Y. Kojima, 
Prog. Theor. Phys. Suppl. {\bf 90}, 1 (1987). 

\bibitem{Maeda} K. Maeda, M. Sasaki, T. Nakamura and S. Miyama, 
Prog. Theor. Phys. {\bf 63}, 719 (1980). 

\bibitem{SP} R. F. Stark and T. Piran, Phys. Rev. Lett. 
{\bf 55}, 891 (1985); in {\it 
Dynamical Spacetimes and Numerical Relativity}, 
edited by J. M. Centrella (Cambridge University Press,
Cambridge, England), p. 40.

\bibitem{BP} J. M. Bardeen and T. Piran, Phys. Rep. {\bf 96}, 205 (1983). 

\bibitem{acst} Note, however, that collapse of a rotating cluster of 
collisionless matter is studied in the following reference: 
A. M. Abrahams, G. B. Cook, S. L. Shapiro, and S. A. Teukolsky, 
Phys. Rev. D {\bf 49}, 5153 (1994). 

\bibitem{hyper} S. E. Woosley, Astrophys. J. {\bf 405}, 273 (1993); 
B. Paczynski, Astrophys. J. Lett. {\bf 494}, L45 (1998);
A. I. MacFadyen and S. E. Woosley, Astrophys. J. {\bf 524}, 262 (1999);  
A. I. MacFadyen, S. E. Woosley, and A. Heger, Astrophys. J. {\bf 550}, 410
(2001). 

\bibitem{Newton} L. S. Finn and C. R. Evans, Astrophys. J. {\bf 351}, 
588 (1990).  

\bibitem{Newton2} R. M\"onchmeyer, G. Sch\"afer, E. M\"uller, and R. 
Kates, Astron. Astrophys. {\bf 246}, 417 (1991); 
E. M\"uller and H.-T. Janka, Astron. Astrophys. {\bf 103}, 358 (1997). 

\bibitem{Newton3}
S. Bonazzola and J.-A. Marck, Astron. Astrophys. {\bf 267}, 623
(1993).

\bibitem{YS}
S. Yamada and K. Sato, Astrophys. J. {\bf 434}, 268 (1994); 
{\bf 450}, 245 (1995). 


\bibitem{Muller} 
T. Zwerger and E. M\"uller, Astron. Astrophys.
{\bf 320}, 209 (1997); 
M. Rampp, E. M\"ulelr, and M. Ruffert, Astron. Astrophys.
{\bf 332}, 969 (1998). 

\bibitem{fryer} C. Fryer and A. Heger, Astrophys. J. {\bf 541}, 1033 (2000);
C. Fryer, D. E. Holz, and A. Heger, Astrophys. J. {\bf 565}, 430
(2002).

\bibitem{HD} H. Dimmelmeier, J. A. Font, and E. M\"uller,
Astron. Astrophys. {\bf 388}, 917 (2002); {\bf 393}, 523 (2002). 

\bibitem{David} But, with special choice of variables and 
gauge conditions, 
it may still be possible to perform long-term
simulations without artificial viscosity. 
See, e.g., D. Garfinkle and G. C. Duncan, 
Phys. Rev. D {\bf 63}, 044011 (2002). 

\bibitem{alcu} M. Alcubierre, S. Brandt, B. Br\"ugmann, 
D. Holz, E. Seidel, R. Takahashi, and J. Thornburg,
Int. J. Mod. Phys. D {\bf 10}, 273 (2001). 

\bibitem{gw3p2} M. Shibata, Prog. Theor. Phys. {\bf 101}, 1199 (1999). 

\bibitem{gr3d} M. Shibata, Phys. Rev. D {\bf 60}, 104052 (1999).

\bibitem{bina} M. Shibata and K. Uryu, Phys. Rev. D {\bf 61}, 064001
(2000). 

\bibitem{bina2} M. Shibata and K. Uryu, Prog. Theor. Phys.
{\bf 107}, 265 (2002). 

\bibitem{gr2d} M. Shibata, Prog. Theor. Phys. {\bf 104}, 325 (2000). 

\bibitem{SS} M. Shibata and S. L. Shapiro, Astrophys. J. Lett. {\bf 572},
L39 (2002). 

\bibitem{Font} J. A. Font, J.-Ma. Ib\'a\~nez, A. Marquina, and J. M. Marti,
Astron. Astrophys. {\bf 282}, 304 (1994).

\bibitem{Val} F. Banyuls,
J. A. Font, J.-Ma. Ib\'a\~nez, J. M. Marti, and J. A. Miralles, 
Astrophys. J. {\bf 476}, 221 (1997).

\bibitem{Val2} J.-Ma. Ib\'a\~nez
and J. Ma. Marti, J. Comput. Appl. Math. {\bf 109}, 
173 (1999): J.-Ma. Ib\'a\~nez, {\it et al.}, in {\it proceedings of Godonov 
Methods}, editted by E. F. Toro (Kluwer Academic, Dordrecht,
2001), p. 485. 

\bibitem{Iba} M. A. Aloy, J.-Ma. Ib\'a\~nez, J. M. Marti, and E. M\"uller, 
Astrophys. J., Suppl. Ser. {\bf 122}, 151 (1999). 

\bibitem{fontrev}
J. A. Font, Living Review Relativity {\bf 3}, 2 (2000); 
http://www.livingreviews.org/Articles/Volume2/2000-2font. 

\bibitem{other} J. A. Font {\it et al}., Phys. Rev. D {\bf 65},
084024 (2002). 

\bibitem{SBS} M. Shibata, T. W. Baumgarte, and 
S. L. Shapiro, Phys. Rev. D {\bf 61}, 044012 (2000). 

\bibitem{gr3drot} M. Shibata, T. W. Baumgarte, and 
S. L. Shapiro, Astrophys. J. {\bf 542}, 453 (2000). 

\bibitem{SN} M. Shibata and T. Nakamura, Phys. Rev. D {\bf 52},
5428 (1995). 

\bibitem{recipe} W. H. Press, B. P. Flannery, S. A. Teukolsky, and 
W. T. Vetterling, {\it Numerical Recipes} 
(Cambridge University Press, Cambridge, England, 1989). 

\bibitem{ST} For example, S. L. Shapiro and S. A. Teukolsky, {\em Black
Holes, White Dwarfs, and Neutron Stars} (Wiley Interscience, 
New York, 1983).


\bibitem{miller} M. Miller, W. Suen, and M. Tobias, Phys. Rev. D
{\bf 63}, 121501 (2001). 

\bibitem{chandra} S. Chandrasekhar, Astrophys. J. {\bf 140}, 417 (1964). 

\bibitem{nick} N. Stergioulas, personal communication:
Talk at 4th European Network Meeting, Palma, Spain (2002). 

\bibitem{SP2} R. F. Stark and T. Piran, 
Comput. Phys. Rep. {\bf 5}, 221 (1987). 

\bibitem{CST} G. Cook, S. L. Shapiro, and S. A. Teukolsky, 
Astrophys. J. {\bf 422}, 227 (1994). 

\bibitem{cox} J. P. Cox, {\em Theory of Stellar Pulsation}
(Princeton University Press, Princeton, NJ, 1980), p. 106. 

\bibitem{SNinit} S. E. Woosley and T. A. Weaver, Astrophys.
J., Suppl. Ser. {\bf 101}, 181 (1995). 

\bibitem{KEH} See, e.g., H. Komatsu, Y. Eriguchi, and 
I. Hachisu, Mon. Not. R. Astron. Soc., {\bf 237}, 355 (1989);
{\bf 239}, 153 (1989). 

\bibitem{Ster} See N. Stergioulas, Living Rev. Relativ, {\bf 1}, 8 (1998) 
for a historical review
about computation of relativistic rotating stars. 


\bibitem{BS} T. W. Baumgarte and S. L. Shapiro, 
Astrophys. J. {\bf 526}, 941 (1999).

\bibitem{GoldW} P. Goldreich and S. W. 
Weber, Astrophys. J. {\bf 238}, 991 (1980). 

\bibitem{Roe} P. L. Roe, J. Comput. Phys. {\bf 43}, 357 (1981).

\bibitem{JJJ} M. Yasuhara and H. Ohmiyaji, 
{\it Computational hydrodynamics} (in Japanese)
(University of Tokyo Press, Tokyo, 1992). 

\end{thebibliography}
\end{document}